\documentclass[aps,prb,preprint,superscriptaddress,longbibliography,floatfix]{revtex4-2}

\usepackage[version=3]{mhchem}
\usepackage{textcomp}
\usepackage{graphicx}
\usepackage{amsmath}
\usepackage{booktabs}
\usepackage{placeins}

\usepackage{caption}
\usepackage{subcaption}
\usepackage[hidelinks]{hyperref}

\begin{document}

\title{
\texorpdfstring{
Ultrafast Nonthermal Lattice Destabilization and Suppression of Polar Optical
Scattering in Electronically Excited \(\alpha\)-\ce{SiO2} from First-Principles
and Deep Neural Network Potential Modeling
}{
Ultrafast Nonthermal Lattice Destabilization and Suppression of Polar Optical
Scattering in Electronically Excited alpha-SiO2 from First-Principles
and Deep Neural Network Potential Modeling
}
}

\author{Iyyappa Rajan Panneerselvam}
\email{i.panneerselvam@qub.ac.uk}
\author{Mark Yeung}
\author{Charlotte Palmer}
\author{Brendan Dromey}
\email{b.dromey@qub.ac.uk}
\author{Lorenzo Stella}
\email{l.stella@qub.ac.uk}

\affiliation{Centre for Light-Matter Interactions, Department of Physics and Astronomy, Queen’s University Belfast, Belfast BT7 1NN, United Kingdom}

\begin{abstract}
We present a multiscale first-principles-to-machine-learning approach to investigate ultrafast lattice dynamics in electronically excited $\alpha$-\ce{SiO2}. \emph{Ab initio} molecular dynamics (AIMD) based on electronic temperature-dependent density functional theory (DFT) are used to train electronic temperature-dependent deep neural network potentials (DNNPs). The use of DNNPs enables atomistic modeling at near DFT accuracy of large $\alpha$-\ce{SiO2} cells with thousands of atoms. In particular, the use of DNNPs allowed us to obtain accurate phonon band structures and molecular dynamics (MD) of $\alpha$-\ce{SiO2} excited by a sudden increase of its electronic temperature. In agreement with previous studies, by increasing the electronic temperature, $T_e$, a pronounced lattice destabilization of $\alpha$-\ce{SiO2} is found, as evidenced by violations of elastic stability criteria, substantial volumetric expansion, a sharp reduction of the bulk modulus, and progressive weakening of \ce{Si-O} bonding due to anti-bonding state occupation. From the knowledge of the electronic and phonon band structures, we estimated the Fr\"ohlich coupling constant, which is found to decrease as $T_e$ increases, suggesting a cross-over to a non-polar phase of $\alpha$-\ce{SiO2} at elevated electronic temperature. This suggestion is corroborated by the Bader charge analysis. We also suggest that the polar optical phonon scattering channel should be strongly suppressed at $T_e>2$~eV. From the DNNP-MD simulations on large cells, we show that a well-defined thermal equilibrium --- as defined by the Maxwell-Boltzmann distribution --- is not achieved over the first few hundreds of femtoseconds. This behavior explains the non-monotonic equilibration of the kinetic temperature after a sudden raise of $T_e$. After $T_e$ is raised to $2.6$~eV, \ce{Si} and \ce{O} atoms first equilibrate separately at two different temperatures, suggesting an atomic fluid phase, in agreement with recent experimental and theoretical findings. 
\end{abstract}

\maketitle
\raggedbottom
\clearpage

\section{Introduction}
Ultrafast excitation of wide band gap insulators such as silicon dioxide (\ce{SiO2}), induced by intense laser pulses, as well as high-energy radiation sources including electrons, protons, heavy ions, or X-rays, has attracted significant interest, both experimentally and theoretically,~\cite{Kennedy2024,Coughlan2020,Stuart1996,Meftah1994,Kishimoto2004,Tsaturyan2022,Kachan2023,Ono2025a,Ono2025b} as it drives highly non-equilibrium states and response~\cite{Tsaturyan2022,Stoneham1994,Kishimoto2004,Corso2024}. Irradiation by charged ions, optical photons or x-rays preferentially couples with the electronic degrees of freedom, initially driving an irradiated insulator into an out-of-equilibrium state typically represented as ``hot electrons'' and ``cold lattice''. Further assuming that the electronic and lattice degrees of freedom quickly and separately thermalize among themselves, the subsequent ultrafast relaxation is interpreted as an energy exchanges between the lattice and the electrons. This is the physical picture underlying the so-called two-temperature model (TTM),  initially devised for metals \cite{Kaganov1957,Anisimov1974} and then successfully applied to semiconductor and insulators~\cite{Stella2021}.

Consistent with this picture, a recent experimental study~\cite{Kennedy2024} further demonstrated that ultrafast pulses of X-ray and protons can drive transient electronic restructuring in nanostructured \ce{SiO2}, resulting in sustained electronic excitation and delayed energy transfer to the lattice. Despite its indisputable success, the microscopic justification of the TTM is still debated, especially concerning the timescales of thermalization and energy exchanges, which are not necessarily well-separated.\cite{Waldecker16,Ono2019,Caruso2022,Singh2025} Even assuming that electrons fully equilibrate at a temperature, $T_e$, on a femtosecond timescale, while the lattice remains at its initial (\emph{e.g.}, room) temperature, ultrafast, non-thermal lattice instabilities can be triggered by a $T_e$ of the order of a few eV\cite{Medvedev2024a,Medvedev2024b}.

Non-thermal mechanisms have been first considered to explain pulsed laser annealing of semiconductors \cite{Stampfli1990, Stampfli1994, Bennemann2004} and subsequently extended to ion irradiation of insulators. \cite{Stampfli1996,Schiwietz2004} In the case of radiation damage, this mechanism is alternative to the well-known inelastic thermal spike model of ion track formation,\cite{Toulemonde1992,Meftah1994} which is based on the TTM. The TTM electron-ion (also referred to as electron-phonon) coupling constant inferred from the experimental track sizes is found to be systematically larger than both the theoretical values and the estimate from laser-driven experiments.\cite{Medvedev2024a} This long standing discrepancy that can be reconciled by inclusion of non-thermal mechanism to radiation damage\cite{Medvedev2024a,Medvedev2024b}.

The capabilities of ab initio molecular dynamics (AIMD) with electronic-temperature-dependent density functional theory (DFT)\cite{Kohn1965,Mermin1965,Alavi1994} to model electronically excited insulators have been demonstrated by Silvestrelli \emph{et al.}~\cite{Silvestrelli1996,Silvestrelli1997} in the case of laser irradiation of crystalline \ce{Si}. Most recently, AIMD has been applied to model non-thermal lattice instability of quartz \cite{Boero2005,Boero2006,Tsaturyan2024} and other wide bandgap insulators.\cite{Ono2025a} The same AIMD with electronic-temperature-dependent DFT approach has been also applied to the study of quartz and fused silica in the so-called warmed dense matter (WDM) regime \cite{Laudernet2004,Denoeud2014,Engelhorn2015,Green2018, Zhang2022b}.

More recently, Tsaturyan \emph{et al.}~\cite{Tsaturyan2022, Tsaturyan2024} investigated $\alpha$-\ce{SiO2} under ultrafast laser excitation, providing estimates  of transport parameters such as electronic specific heat, deformation potential, and electron-phonon coupling from first principles, over a wide range of electronic temperatures (up to $100{,}000~\text{K}$, equivalent to $\approx 8.6~\text{eV}$). Zhang \emph{et al.}~\cite{Zhang2022b} also reported about the properties \ce{SiO2} computed using AIMD at elevated temperatures and pressures characteristic of the WDM regime. In particular, they studied the bond weakening at large electronic temperature using a crystal orbital Hamilton population and atomic charge analyses. In their study, the conclusions about the fluid structure of the WDM phases is just limited by the size of the simulations cell that can be currently used in AIMD. Ono \emph{et al.}~\cite{Ono2025a} also modeled $\alpha$-\ce{SiO2} and other wide band gap insulator using finite-electronic-temperature AIMD. They showed a general softening of the phonon band structure of this class of materials, eventually yielding dynamical lattice instability at elevated electronic temperature.

As the conclusions of previous studies were often limited by the size of the AIMD simulation cells used~\cite{Zhang2022b}, we employ deep neural network potentials (DNNPs) trained on \emph{ab initio} data, enabling larger scale molecular dynamics (MD) simulations at near first-principles accuracy~\cite{Wang2018,Zeng2023}. In particular, we are able to provide more accurate assessment of the lattice temperature and correlations of the steady-state phase achieved after a sudden electronic excitation (modeled as an instantaneous increase of the electronic temperature~\cite{Silvestrelli1996,Silvestrelli1997}) of $\alpha$-\ce{SiO2} initially at room temperature. In this way it was also possible to obtain well-converged phonon band structures over a range of electronic temperature relevant relevant for experiments. 
Based on the results of the DNNP-MD simulations, we conclude that the lattice degrees of freedom of the  electronically excited phases do not immediately equilibrate, \emph{i.e.,} the atomic speeds are not distributed according to the Maxwell-Boltzmann distributed over the first 100 fs. This conclusion confirms some of the early criticisms to the TTM, especially the assumption that the lattice degrees of freedom fully and quickly thermalize before the heat-transfer from electron-ion coupling kicks in. Our AIMD simulations cannot model non-adiabatic electronic excitations and election-ion relaxation, which are beyond the scope of this work. Although we do not compute the electron-phonon coupling from first principles, as in \cite{Tsaturyan2024}, we have estimated the Fr\"{o}hlich coupling constant 
and found that it decreases as the electronic temperature increases, effectively suppressing the polar phonon scattering in electronically excited $\alpha$- \ce{SiO2}.

The rest of this paper is organized as follow:

\section{Computational Methods}

\subsection{Finite Electronic Temperature DFT Calculations}

First-principles electronic structure calculations were performed within the framework of DFT using the Vienna \textit{ab initio} Simulation Package (\textsc{vasp})~\cite{Kresse1996,KresseJoubert1999,KresseFurthmueller1996,KresseHafner1993} . The projector augmented-wave (PAW) method~\cite{Bloechl1994} was employed along with the Perdew-Burke-Ernzerhof (PBE) exchange-correlation functional.\cite{Perdew1996} To obtain more accurate estimates, \emph{e.g.}, of the band gap, the more accurate Heyd–Scuseria–Ernzerhof (HSE06) screened hybrid exchange–correlation functional~\cite{Heyd2003,Krukau2006} was used. A plane-wave kinetic energy cutoff of 600~eV was used in all first-principles calculations. 

The crystal structure of $\alpha$-quartz \ce{SiO2} (space group $P3_221$, No.~154) was initially obtained from the Materials Project (mp-6930).~\cite{MaterialsProjectSiO2}. It contains  a primitive unit cell with nine atoms, three \ce{Si} and six \ce{O}. A full structural optimization (both lattice parameters and atomic positions) was performed for the above unit cell using the conjugate gradient algorithm~\cite{Press1986}. The electronic self-consistency was considered converged when the total energy change between successive iterations was below $10^{-8}$~eV, and atomic positions were optimized until the forces on each atom are below 0.03~eV/\AA. The Brillouin zone was sampled using a $\Gamma$-centered $15\times15\times9$ $k$-point mesh for structural optimization, $9\times9\times5$ mesh for finite differences and DFPT calculations, and a single $\Gamma$-point for AIMD simulations performed in the NVT and NVE ensembles (also see Sec.~\ref{sec:AIMD}). A small (0.05~eV) Gaussian smearing was used at zero electronic temperature ($T_e=0$), while Fermi-Dirac smearing was used at finite temperature ($T_e>0$) to set the appropriate orbital occupation. The Mermin's formulation of DFT at finite electronic temperature was used for the electronic-temperature-dependent calculations.\cite{Mermin1965,Alavi1994}
Therefore, all reported electronic energies at finite $T_e$ are free energies rather than internal energies.

The optimized $\alpha$-SiO$_2$ crystal structure at $T_e=0$ was used as the reference for electronic structure calculations at $T_e>0$. At each electronic temperature, the atomic positions were further relaxed while keeping the cell shape and volume fixed. In this work, we considered $T_e=1.0$, $2.0$, $2.1$, $2.2$, and $2.6$~eV, corresponding to approximately $11{,}605$, $23{,}209$, $24{,}370$, $25{,}530$, and $30{,}172$~K, respectively. Hereafter, electronic temperatures are expressed in energy units (eV), unless otherwise stated. To ensure adequate sampling of thermally populated conduction states, the number of electronic bands was increased with electronic temperature: 120 bands were used for $T_e = 1.0$ and $2.0$~eV, and 150 bands for $T_e = 2.1$, $2.2$, and $2.6$~eV. 
 
\subsection{Out-of-equilibrium Equation of State}\label{sec:EoS}

An out-of-equilibrium equation of state (EOS) of $\alpha$-SiO$_2$ can be formally defined when the electronic temperature, $T_e$ is not equal to the lattice temperature, $T_L$. At $T_e\ge 0$ the electronic free energies, $F_e\left(V,T_e\right)$, were computed as a function of the cell volume, $V$, using the Mermin's functional. For each value of the volume and electronic temperature, the cell shape and atomic positions were relaxed to find the (possibly local) minimum of the electronic free energy. Vibrational (harmonic or quasi-harmonic) contributions to the free energy~\cite{Togo2023JPSJ,Togo2023Phonopy} are expected to be comparably negligible at $T_L=300$~K, and were not included, \emph{i.e.}, $F\left(V,T_e,T_L\right)\approx F_e\left(V,T_e\right)$.

The EoS is fitted to the third-order Birch-Murnaghan equation of state,
\begin{equation}
\label{eq:BM}
\begin{split}
F\left(V,T_e\right) &= F_0(T_e) +
\frac{9 V_0(T_e) B_0(T_e)}{16} \left\lbrace \left[ \left( \frac{V_0(T_e)}{V} \right)^{2/3} - 1 \right]^3 B_0^{\prime}(T_e) + \right.\\
&\left.\left[ \left( \frac{V_0(T_e)}{V} \right)^{2/3} - 1 \right]^2 \left[6 - 4 \left( \frac{V_0(T_e)}{V} \right)^{2/3} \right] \right\rbrace
\end{split}
\end{equation}
from which the equilibrium volume, $V_0(T_e)$ bulk modulus, $B_0(T_e)$, and the pressure derivative $B_0^\prime(T_e)$, can be obtained.

A volume-independent correction was applied to the PBE free energy,
\begin{equation}
F_e^{\mathrm{corr}}(V,T_e) = F_e^{\mathrm{PBE}}(V,T_e) + \Delta F_{\mathrm{HSE}}(V_0^{GS},T_e),
\end{equation}
where $\Delta F_{\mathrm{HSE}}(V_0^{GS},T_e) = F_{\mathrm{HSE}}(V_0^{GS},T_e) - F_{\mathrm{PBE}}(V_0^{GS},T_e)$ is the difference between the HSE06 and PBE energies computed at the ground-state (\emph{i.e.} $T_e$) equilibrium volume, $V_0^{GS}$.
This approximation is expected to be accurate in the case of a rapid, isochoric increase of $T_e$ and does not require computational demanding HSE06 calculations over the full volume range.

\subsection{\emph{Ab Initio} Molecular Dynamics Simulations}
\label{sec:AIMD}

\emph{Ab Initio} molecular dynamics (AIMD) simulations at finite electronic temperature\cite{Alavi1994} were performed using a 2$\times$2$\times$2 supercell of $\alpha$-\ce{SiO_2} unit cell containing 72 atoms, and the Brillouin zone sampled at the $\Gamma$-point, only\cite{Boero2005,Boero2006}. Following the computational approach outlined in Refs.~\cite{Silvestrelli1996,Silvestrelli1997}, the previously optimized ground state (\emph{i.e.}, $T_e$) structure was equilibrated in the canonical (NVT) ensemble for 4~ps at $T_L=298$~K using the Nos\'e--Hoover thermostat with a temperature damping time of 0.1~ps. Initial velocities were assigned from Maxwell-Boltzmann distribution at 298~K. Equilibration was assessed by monitoring the stabilization of the temperature, total energy, and potential energy during the NVT propagation. The AIMD time-step was set to 1~fs. After equilibration, the electronic temperature, $T_e$ was instantaneously raised and the AIMD was continued for 6 more ps in the microcanonical (NVE) ensemble, sampling the electronic free energy at fixed $T_e$.

\subsection{Deep neural network potentials construction and Molecular Dynamics}
\label{sec:DNNP}

To enable classical simulations of electronically excited $\alpha$-SiO$_2$, DNNPs were developed following recent applications of finite electronic temperature machine learning interatomic potentials~\cite{Plettenberg2023,Zeng2023Laser,Corradini2025}, using the DeePMD-kit package~\cite{Wang2018,Zeng2023}. Independent models were trained at $T_e=0$, $1.0$, $2.0$, $2.1$, $2.2$, and $2.6$~eV using the AIMD data from the 6~ps NVE propagations. For each condition, 6,000 frames were collected, including atomic coordinates, cell dimensions, total energies, atomic forces, and virial tensors. A random selection of 1000 frames was used for validation, and the remaining 5,000 frames for training. Data preprocessing was performed using the \texttt{dpdata} Python module~\cite{Zeng2025dpdata}. Each model used a hybrid descriptor composed of \texttt{se\_e2\_a} and \texttt{se\_e3} types, with a cutoff radius of 6.0~\AA{} and a smoothing cutoff of 0.5~\AA{}. The embedding network architecture included three hidden layers with 25, 50, and 100 neurons, while the fitting network consisted of three hidden layers of 240 neurons each. Training was conducted for 500,000 steps with an exponentially decaying learning rate, starting from 0.0005 and stopping at $3.51 \times 10^{-8}$. The loss function balanced energy, force, and virial terms, starting with prefactors of 0.02, 1000, and 10, respectively, and decaying to 1. Batch sizes of 2 and 1 were used for training and validation, respectively. Other hyperparameter choices closely followed the DeePMD-kit documentation guidelines.

To extend the analysis to larger time and length scales, classical molecular dynamics simulations were performed using the LAMMPS package~\cite{Thompson2022LAMMPS} using the electronic-temperature-dependent DNNPs. Each simulation began by constructing a $7\times7\times7$ supercell of $\alpha$-\ce{SiO2} (3087 atoms, corresponding to 1029 SiO$_2$ formula units) from the previously optimized ground state (\emph{i.e.}, $T_e=0$) unit cell. The supercell was first equilibrated in the canonical (NVT) ensemble for 4~ps at $T_L=298$~K using a Nos\'e--Hoover thermostat, with a time step of 1~fs and a thermostat damping parameter of 0.1~ps. Initial velocities were assigned from Maxwell-Boltzmann distribution at 298~K. After equilibration, the final atomic configuration was used to initiate microcanonical (NVE) ensemble simulations. Separate NVE simulations were performed for 20~ps using the appropriate using the electronic-temperature-dependent DNNP. Along each trajectory, atomic positions and velocities, together with the lattice temperature, potential energy, kinetic energy, total energy, and pressure, were recorded every 10~fs. 

\subsection{Post-processing and Analysis}
\label{sec:postprocessing}

Post-processing and analysis in this study were performed using a combination of established tools, including \textsc{vaspkit}~\cite{Wang2021VASPKIT} for analysis of DFT outputs, \textsc{sumo}~\cite{Ganose2018sumo} for phonon band structure visualization, \textsc{phonopy}~\cite{Togo2023Phonopy,Togo2023JPSJ} and \textsc{phonolammps} for phonon calculations, and \textsc{lobster}~\cite{Dronskowski1993COHP,Deringer2011COHP,Maintz2013LOBSTER,Maintz2016LOBSTER,Nelson2020LOBSTER} for chemical bonding analysis. Bader charges were calculated by partitioning the VASP charge density using the grid-based algorithm implemented in the Henkelman-group code~\cite{Tang2009Bader}.
The velocity autocorrelation functions (VACF) and vibrational density of states (VDOS) were calculated from the MD velocity trajectories using a custom post-processing code.

\section{Results and Discussion}

\subsection{Elastic Stability Criteria Under Electronic Excitation}\label{sec:stability}

We assessed the elastic stability of the optimized $\alpha$-\ce{SiO2} structure at different electronic temperatures using the criteria defined for trigonal crystal systems~\cite{Mouhat2014Elastic}. While elastic criteria provide necessary conditions for crystalline structural stability, a complete assessment also includes dynamical stability, defined by the absence of imaginary phonon branches. Phonon band structure analysis and dynamical stability are discussed in Sec.~\ref{sec:phonons}. 

For the trigonal class ($P3_221$), the following conditions on the elastic stiffness constants ($C_{ij}$) must be satisfied to ensure mechanical stability: (i) $C_{11} - C_{12} > 0$, (ii) $C_{13}^2 < 0.5 \, C_{33} (C_{11} + C_{12})$, (iii) $C_{14}^2 < 0.5 \, C_{44} (C_{11} - C_{12})$, and (iv) $C_{44} > 0$. The calculated stability criteria, and the minimum eigenvalue of the elastic stiffness matrix at different electronic temperatures are summarized in Table~1 of the Supporting Information. The $\alpha$-\ce{SiO2} structures satisfy all of these stability conditions at $T_e=0$, $1.0$, $2.0$, $2.1$~eV. At $T_e=2.2$ and $2.6$~eV, the trigonal structure becomes elastically unstable as evidenced by negative eigenvalues in the stiffness matrix and the violation of conditions (i), (iii) and/or (iv). In particular, at $T_e=2.2$~eV, the conditions $C_{14}^2 < 0.5 \, C_{44}(C_{11} - C_{12})$ and $C_{44} > 0$, and at $T_e=2.6$~eV, the conditions $C_{11} - C_{12} > 0$ and $C_{44} > 0$ are no longer satisfied. These results suggest a threshold electronic temperature for elastic instability between $T_e=2.1$ and $2.2$~eV, which is interpreted as a progressive loss of shear rigidity due to the weakening of the \ce{Si-O} bonding at elevated electronic temperature\cite{Cabral2019,Choping2024}.Condition~(iii), involving $C_{14}$ and $C_{44}$, becomes violated, indicates that resistance to shear distortions involving off-diagonal components is lost~\cite{Mouhat2014Elastic,Wang1997Elastic}. 
The failure of condition~(iv) ($C_{44} > 0$) at $T_e=2.6$~eV indicates that the lattice can no longer sustain shear deformation even along principal planes --- a sign of incipient amorphization or melting. The eventual failure of condition~(i) ($C_{11} - C_{12} > 0$) at $T_e=2.6$~eV implies longitudinal softening, also consistent with a complete collapse of the trigonal crystalline structure at elevated electronic temperature.

Alongside the elastic stability criteria, the Debye temperature, $\Theta_D$, offers a vibrational perspective on lattice stiffness, as it is closely linked to the elastic response of the material~\cite{Anderson1963Debye}. The Debye temperature was calculated from the elastic constants using the approach implemented in \textsc{VASPKIT}~\cite{Wang2021VASPKIT}. In the ground state, the calculated Debye temperature of $\alpha$-\ce{SiO2} is approximately 558~K, consistent with its stiff covalent framework. As the electronic temperature increases, $\Theta_D$ decreases steadily, reaching 518~K at $T_e=1.0$~eV, 284~K at $T_e=2.0$~eV and 220~K at $T_e=2.1$~eV. This decline aligns with a global softening of phonon modes due to the weakening of force constants under electronic excitation, see Sec.\ref{sec:phonons}. At $T_e=2.2$~eV, $\Theta_D$ drops further to 167~K, and by $T_e=2.6$~eV, it becomes undefined, reflecting a breakdown of vibrational coherence~\cite{Recoules2006,Denoeud2014,Engelhorn2015} and the failure of the quasiharmonic approximation~\cite{Togo2023Phonopy,Togo2023JPSJ}.

\subsection{Equation of State and Volume Response Under Electronic Excitation}

In Sec.~\ref{sec:stability}, we showed evidence that the threshold electronic temperature required to maintain lattice stability of the optimized $\alpha$-\ce{SiO2} structure is $T_e=2.1$ and $T_e=2.2$~eV. 
In this Section, we quantify the equation of state (EOS) of $\alpha$-\ce{SiO2} under electronic excitation by computing energy-volume curves for electronic temperatures upto $T_e=2.2$~eV, while keeping the lattice temperature ``cold'', see Sec.\ref{sec:EoS}. The curves are then fitted to Eq.~(\ref{eq:BM}).

\begin{figure}[!htbp]
\centering
\includegraphics[width=0.85\linewidth]{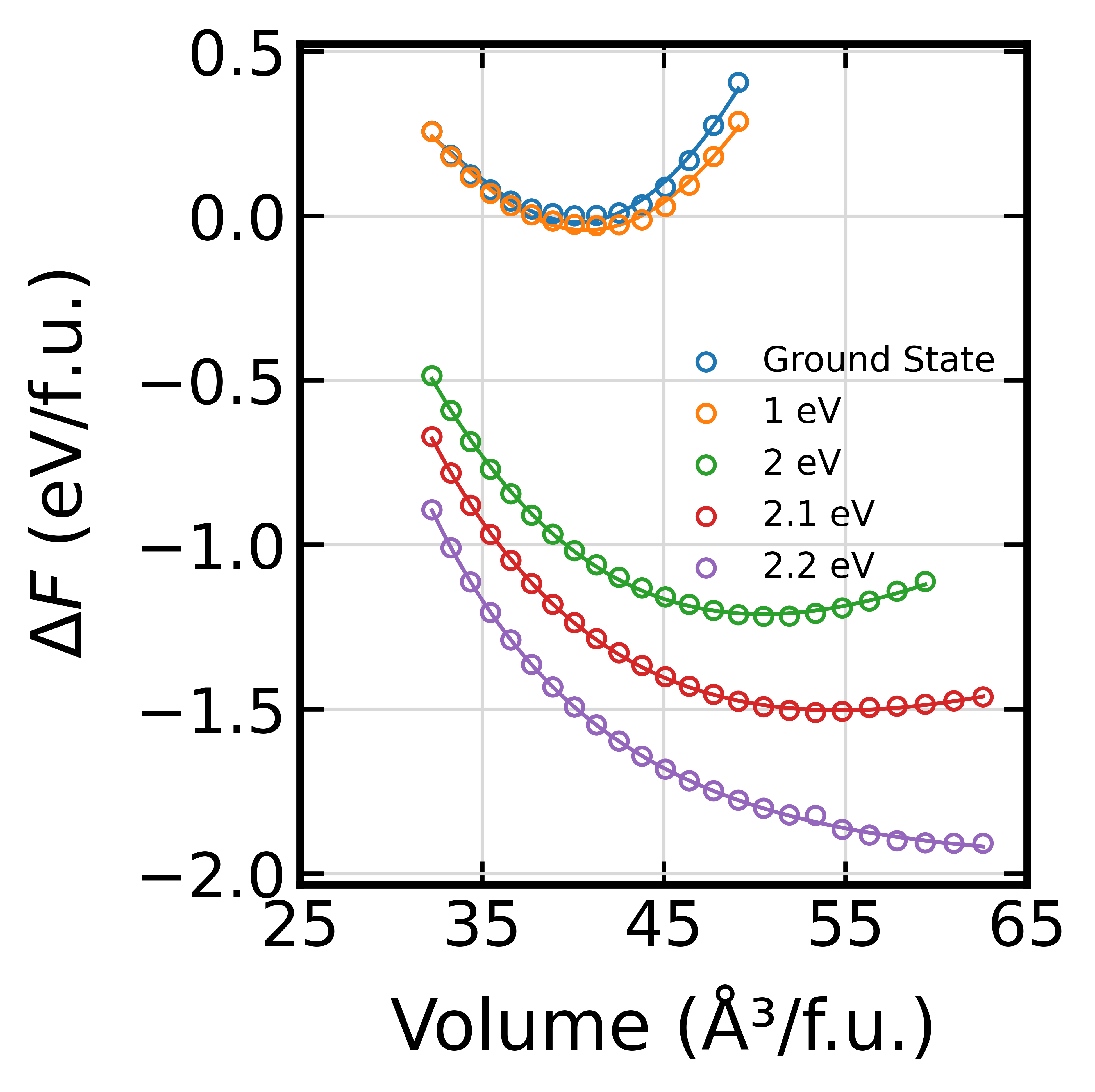}
\caption{Calculated electronic free energy-volume curves of $\alpha$-SiO$_2$ in the ground state and under electronic excitation ($1.0$-$2.2$~eV). The free energies were evaluated within the Mermin finite-temperature DFT formalism. To incorporate hybrid-functional accuracy, a rigid energy shift corresponding to the ground-state HSE06 energy was applied to the PBE free-energy data. All curves are referenced to the minimum of the ground-state curve, which is set to zero. The resulting curves are fitted using the third-order Birch-Murnaghan equation of state.The equilibrium volumes obtained from the EOS fits are 
$V_0^{GS} = 40.25$~\AA$^3$/f.u. (ground state), 
$40.70$~\AA$^3$/f.u. ($1.0$~eV), 
$50.44$~\AA$^3$/f.u. ($2.0$~eV), and 
$54.67$~\AA$^3$/f.u. ($2.1$~eV). For $T_e = 2.2$~eV, the fitted minimum lies outside the sampled volume range.}
\label{fig:EOS}
\end{figure}

Fig.~\ref{fig:EOS} shows the EoS of $\alpha$-\ce{SiO2} for the ground state (\emph{i.e.}, at $T_e=0$~eV) and at $T_e\le2.2$~eV, while corresponding fitting parameters are reported in Table 2 of the Supporting Information. Since above $T_e=2.2$~eV, the trigonal structure of $\alpha$-\ce{SiO2} exhibits clear signatures of elastic (see Sec.~\ref{sec:stability}) and dynamical instability (see Sec.~\ref{sec:phonons}), and obtaining a reliable EoS is challenging. 
In Fig.~\ref{fig:EOS}, differences in free energy are referenced to the minimum of the ground-state curve. 

With increasing electronic temperature, the equilibrium volume $V_0$ exhibits a systematic expansion, reflecting the buildup of electronic pressure arising from the occupation of antibonding states. The volume per formula unit first increases moderately at $1.0$~eV and then more substantially for $T_e\ge 2.0$~eV, in agreement with the evidence of \ce{Si-O} bond weakening gathered in Sec.\ref{sec:stability}.
Such a bond weakening can be ascribed to an increasing populations of antibonding orbitals as $T_e$ increases, as suggested in previous studies.\cite{Silvestrelli1996,Silvestrelli1997,Zhang2022b}
Simultaneously, the bulk modulus $B_0$ shows a pronounced decrease with electronic temperature, falling from 68.1~GPa in the ground state to just over 2.0~GPa at $T_e=2.2$~eV. This indicates a substantial reduction in the lattice resistance to volumetric compression. These trends are consistent with the softening of the elastic constants and the suppression of the Debye temperature discussed in Sec.~\ref{sec:stability}. 

\FloatBarrier

\subsection{Electronic Band structure, Bonding Evolution (pCOHP Analysis), and atomic charges Under Electronic Excitation}
\label{Sec:pCOHP}

To elucidate the electronic origins of the bond weakening and structural destabilization, we next examine the evolution of the electronic band structure and chemical bonding under electronic excitation. The HSE06 hybrid-functional band structures of $\alpha$-\ce{SiO2} at the ground state (\emph{i.e.}, $T_e=0$~eV) and at electronic temperatures up to $T_e=2.2$~eV are shown in Fig.~S1 of the Supporting Information. In its ground state, $\alpha$-\ce{SiO2} exhibits an indirect gap of $7.54$~eV and a direct gap of $7.84$~eV. These values are consistent with previous HSE06 result reported by Xiao \textit{et al.}~\cite{XiaoBandgap}. For comparison, $GW_0$ calculations yield $9.4$~eV~\cite{XiaoBandgap}, while calculations using the modified Becke-Johnson (mBJ) exchange potential gives an optical band gap of $9.2$~eV~\cite{OtobeBandgap,BeckeJohnson2006,TranBlaha2009}. The difference reflects the quasiparticle corrections captured by $GW$ approximation, which are not fully accounted for by hybrid DFT functionals.

\begin{figure}[!htbp]
\centering
\begin{subfigure}[t]{0.48\linewidth}
\centering
\includegraphics[height=5.2cm]{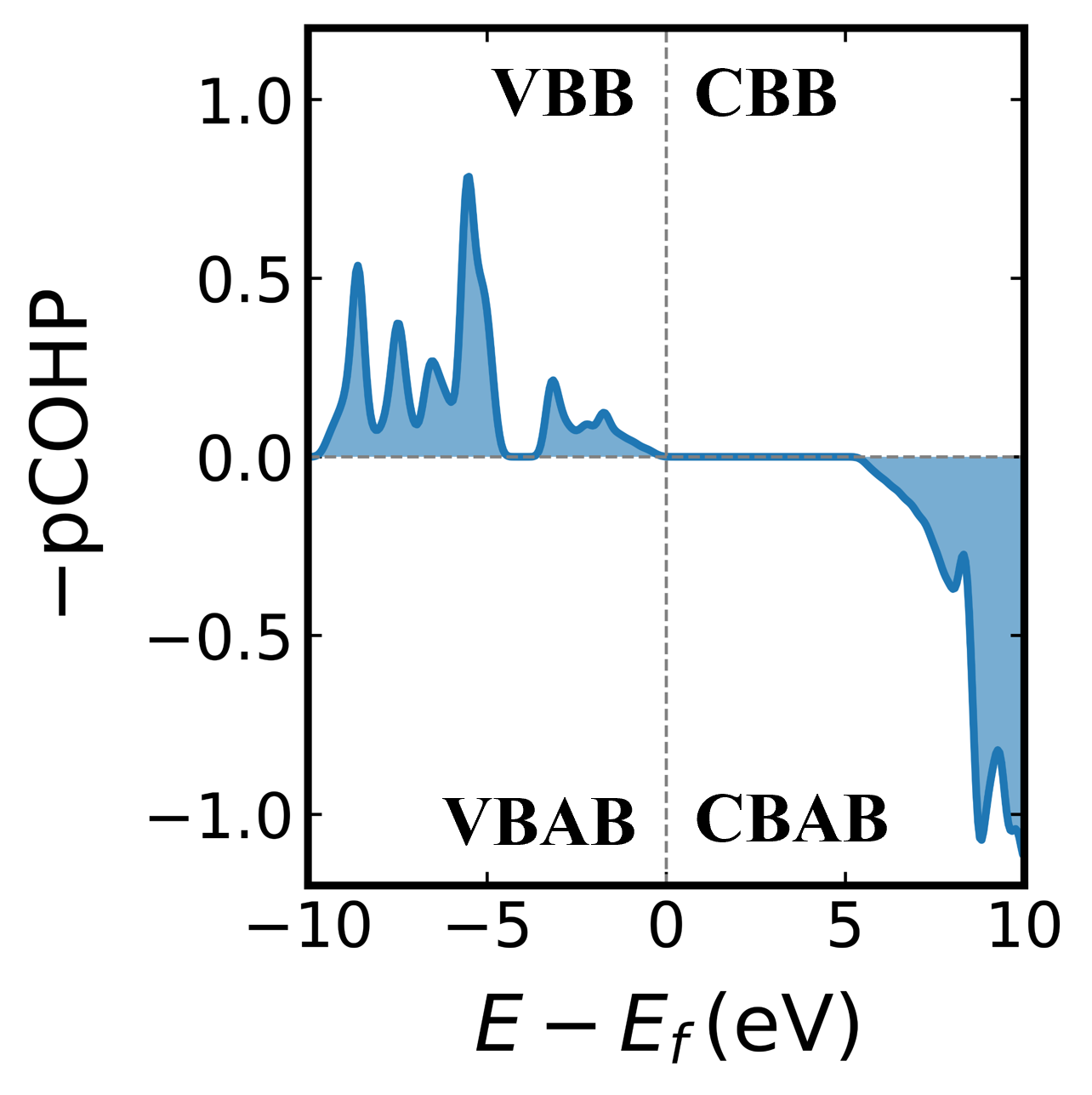}
\caption{Ground state}
\end{subfigure}\hfill
\begin{subfigure}[t]{0.48\linewidth}
\centering
\includegraphics[height=5.2cm]{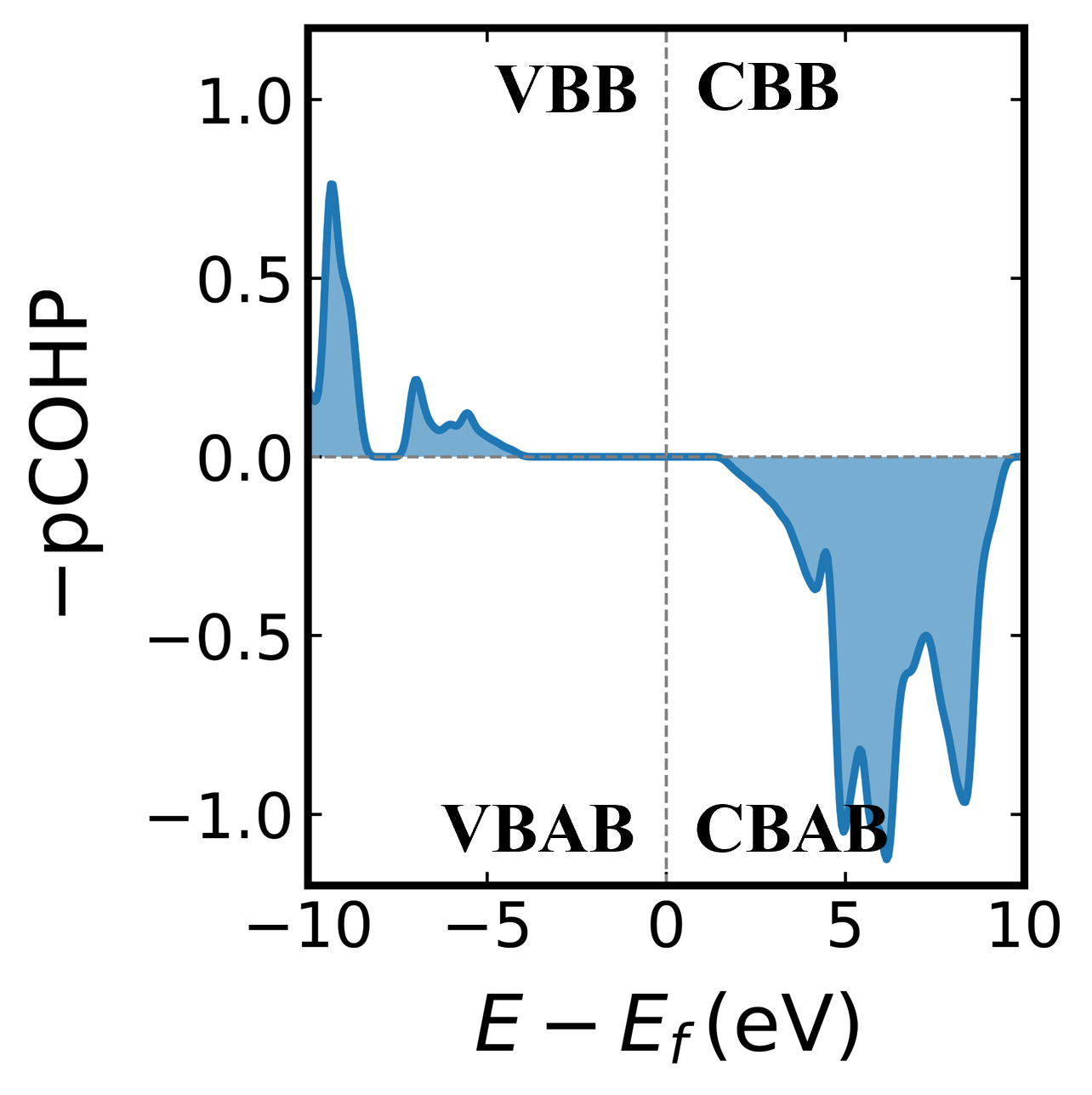}
\caption{$T_e = 1.0$~eV}
\end{subfigure}

\vspace{0.2cm}

\begin{subfigure}[t]{0.48\linewidth}
\centering
\includegraphics[height=5.2cm]{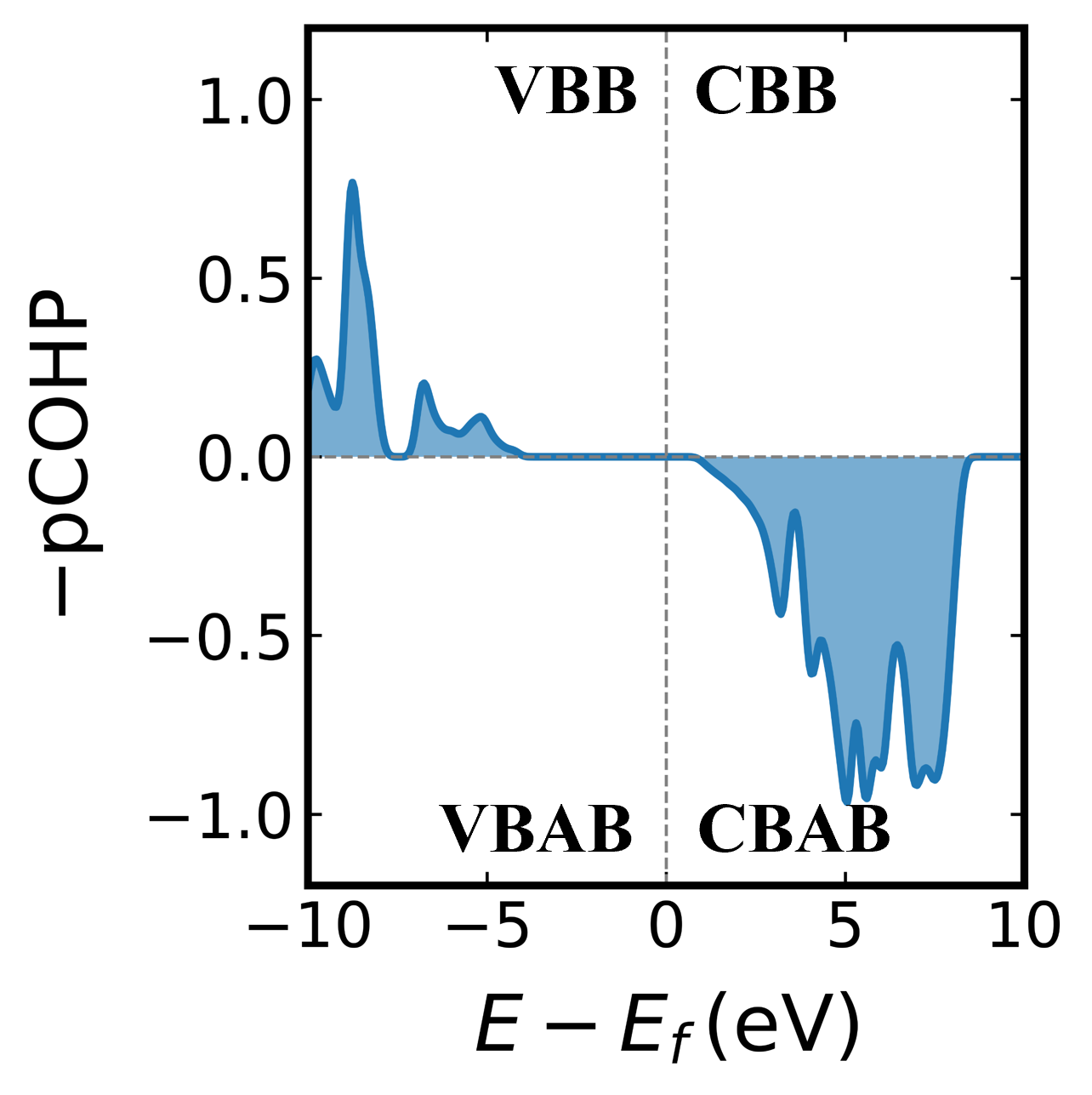}
\caption{$T_e = 2.0$~eV}
\end{subfigure}\hfill
\begin{subfigure}[t]{0.48\linewidth}
\centering
\includegraphics[height=5.2cm]{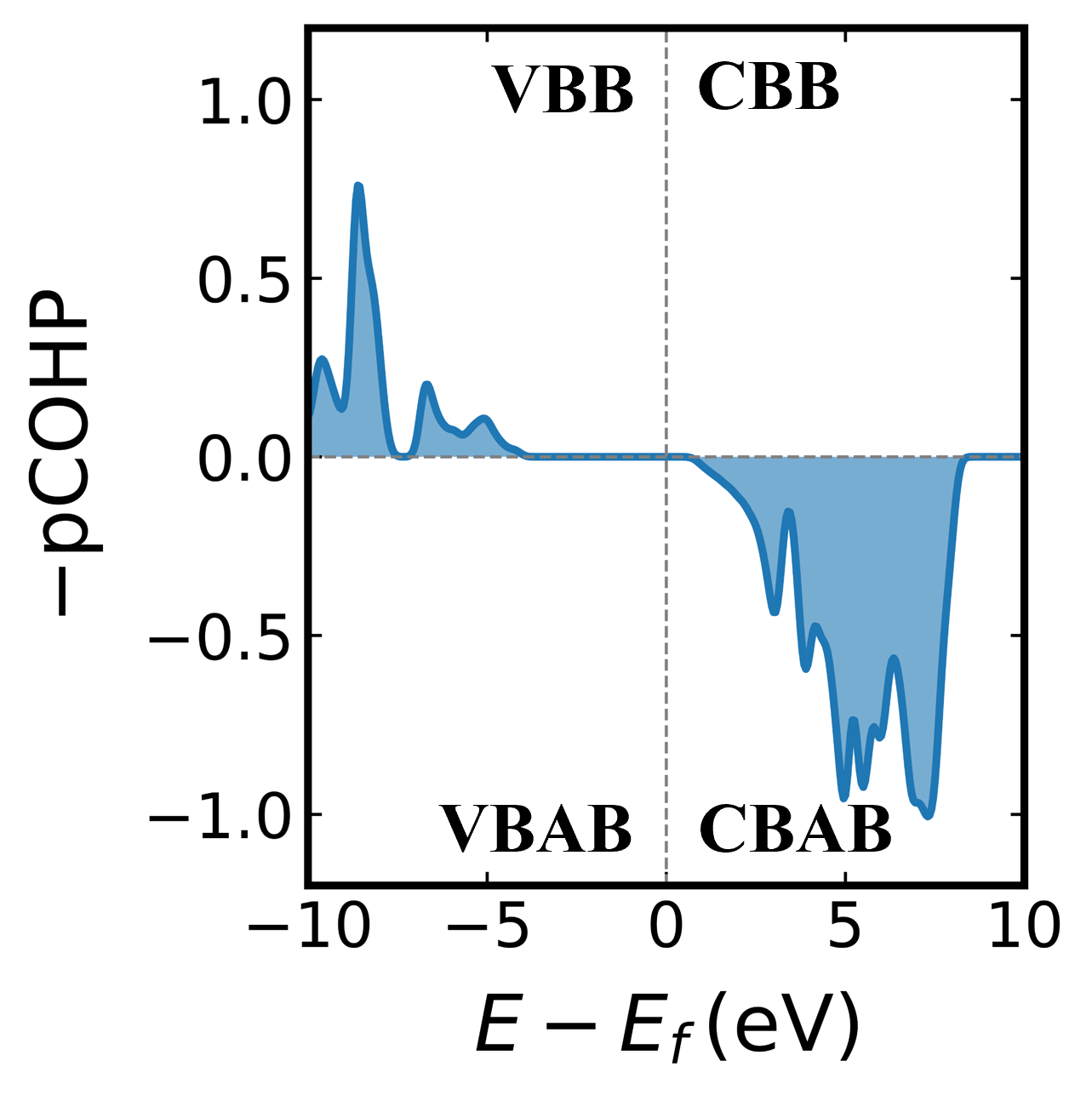}
\caption{$T_e = 2.1$~eV}
\end{subfigure}

\vspace{0.2cm}

\begin{subfigure}[t]{0.48\linewidth}
\centering
\includegraphics[height=5.2cm]{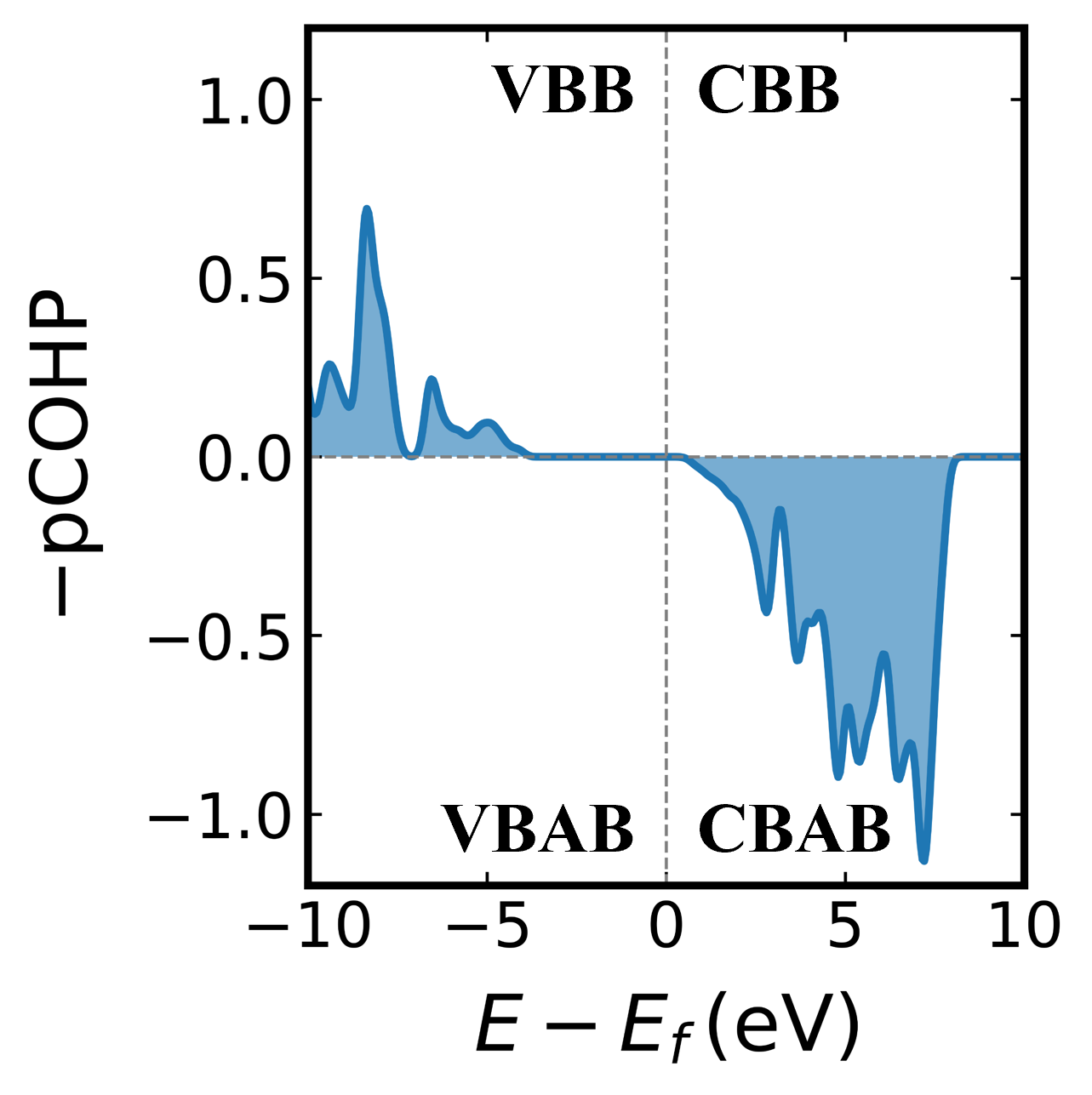}
\caption{$T_e = 2.2$~eV}
\end{subfigure}

\caption{PBE-based projected Crystal Orbital Hamilton Population ($-\mathrm{pCOHP}$) curves for representative Si--O bonds in $\alpha$-SiO$_2$ at the ground state and under increasing electronic temperatures. Following standard convention, $-\mathrm{pCOHP}$ is plotted such that positive and negative values correspond to bonding and antibonding interactions, respectively. Here, VBB and VBAB denote valence band bonding and antibonding states, while CBB and CBAB represent conduction band bonding and antibonding states.}
\label{fig:pcohp}
\end{figure}

In agreement with previous studies~\cite{Kachan2023,Tsaturyan2024}, we found that the band gap decreases systematically with increasing electronic temperature. However, a finite gap persists indicating band-gap narrowing rather than a true insulator-to-metal transition, at least for the value of $T_e$ considered in this work. The indirect (direct) band gaps decrease to $7.50$ ($7.80$)~eV at $T_e = 1.0$~eV, $6.55$ ($6.72$)~eV at $T_e = 2.0$~eV, $6.27$ ($6.40$)~eV at $T_e = 2.1$~eV, and $5.96$ ($6.05$)~eV at $T_e = 2.2$~eV.

To investigate the changes in chemical bonding driven by electronic excitation, we performed a PBE-based projected Crystal Orbital Hamilton Population (pCOHP) analysis for representative Si–O bonds in $\alpha$-\ce{SiO2}. The corresponding pCOHP curves for the ground state and $T_e=1.0$, $2.0$, $2.1$, and $2.2$~eV are shown in Fig.~\ref{fig:pcohp}(a)–(e), respectively. The pCOHP formalism partitions the electronic density of states into bonding and antibonding contributions, providing an orbital-resolved measure of bond strength. Following common practice, we plot $\mathrm{-pCOHP}$, such that positive and negative values correspond to bonding and antibonding interactions, respectively. In the ground state, the bonding states (VBB: valence band bonding) are well separated from the antibonding states (CBAB: conduction band antibonding), and the Fermi level lies in a region with negligible antibonding occupation. The bonding peaks are strong and sharp, confirm robust covalent \ce{Si-O} bonding. 

As the electronic temperature increases, two clear trends emerge. First, the magnitude of bonding peaks in the valence band near to the Fermi level decreases, indicating a weakening of the occupied bonding states. Second, and more critically, there is a gradual increase in occupation of CBAB states close to the Fermi level. This shift is consistent with electrons being excited into antibonding orbitals, thereby destabilizing the lattice and reducing bond strength.\cite{Silvestrelli1996,Silvestrelli1997,Zhang2022b} 
By $T_e=2.2$~eV, a pronounced occupation of CBAB states is evident, suggesting that a significant portion of the electronic population resides in antibonding orbitals. This correlates well with the reported elastic instability and lattice softening. To quantify the overall bond strength, we analyzed the integrated projected Crystal Orbital Hamilton Population ($\mathrm{-IpCOHP}$) values for \ce{Si-O} bonds across different electronic temperatures. In the ground state, $\mathrm{-IpCOHP}$ is approximately $4.0$~eV per bond, reflecting strong covalent bonding. With increasing electronic excitation, the magnitude of the $\mathrm{-IpCOHP}$ values decreases systematically, taking values of $3.95$, $3.40$, $3.30$, and $3.10$~eV at electronic temperatures of $T_e=1.0$, $2.0$, $2.1$, and $2.2$~eV, respectively. This progressive reduction indicates a weakening of \ce{Si-O} bonds due to electronic population of antibonding states. At $T_e=2.6$~eV, the $\mathrm{-IpCOHP}$ drops sharply to just $0.6$~eV, showing a collapse of the \ce{Si-O} bonding. 

A qualitatively similar decrease in \ce{Si-O} bond strength with increasing temperature was reported for liquid silica under extreme pressure, where the \(-\mathrm{IpCOHP}\) decreases as the system approaches the bonded to atomic regime~\cite{Zhang2022b}. Finally, Bader charge analysis~\cite{Tang2009Bader} yields average net charges of \(+3.226\,e/-1.613\,e\), \(+3.215\,e/-1.608\,e\), \(+2.975\,e/-1.487\,e\), \(+2.911\,e/-1.456\,e\), \(+2.816\,e/-1.408\,e\), and \(+1.508\,e/-0.754\,e\) on Si/O for the ground state and at \(T_e=1.0\), \(2.0\), \(2.1\), \(2.2\), and \(2.6\)~eV, respectively. The progressive reduction in charge separation indicates decreasing ionicity of the \ce{Si-O} bond with increasing electronic temperature.

\subsection{Phonon Band structure Under Electronic Excitation}
\label{sec:phonons}

While AIMD provides accurate insights into atomic-scale dynamics and vibrational properties, its practical application is constrained by significant computational demands. In particular, modeling electronically excited states at finite electronic temperatures requires a large number of partially occupied bands according to the Fermi-Dirac distribution. To overcome these limitations and use larger simulation cells to improve the thermodynamic sampling, we employed a machine-learned interatomic potential trained within the Deep Neural Network Potential (DNNP) framework. Six DNNPs were trained using AIMD data at $T_e=0.0$, $1.0$, $2.0$, $2.1$, $2.2$, and $2.6$~eV, see Sec.~\ref{sec:DNNP}. The performance of each model was validated against DFT reference energies and forces obtained from independent AIMD configurations.
The energy and force prediction accuracy is illustrated by validation plots in Figures~S2 and S3, while the corresponding quantitative errors, expressed in terms of mean absolute error (MAE) and root mean square error (RMSE), are summarized in Tables~S3 of the Supporting Information.

Using the electronic-temperature-dependent DNNPs and their integration with the \textsc{lammps} and \textsc{phononlammps} packages, we computed the phonon band structures of $\alpha$-\ce{SiO2} at $T_e=0.0$, $1.0$, $2.0$, $2.1$, $2.2$, and $2.6$~eV. Both the DFT and DNNP phonon band structures were calculated using $4\times4\times4$ supercells containing 576 atoms. Convergence studies showed that a $2\times2\times2$ supercell produces spurious imaginary acoustic frequencies near the $\Gamma$ point, indicating an insufficient real-space range of the second-order force constants. The sensitivity of $\alpha$-quartz phonons to supercell size has also been previously reported~\cite{Mizokami2018}. To ensure a consistent comparison between the DFT and DNNP phonon band structures, non-analytical term correction (NAC) were omitted.

\begin{figure}[!htbp]  
\centering
\includegraphics[width=0.85\linewidth]{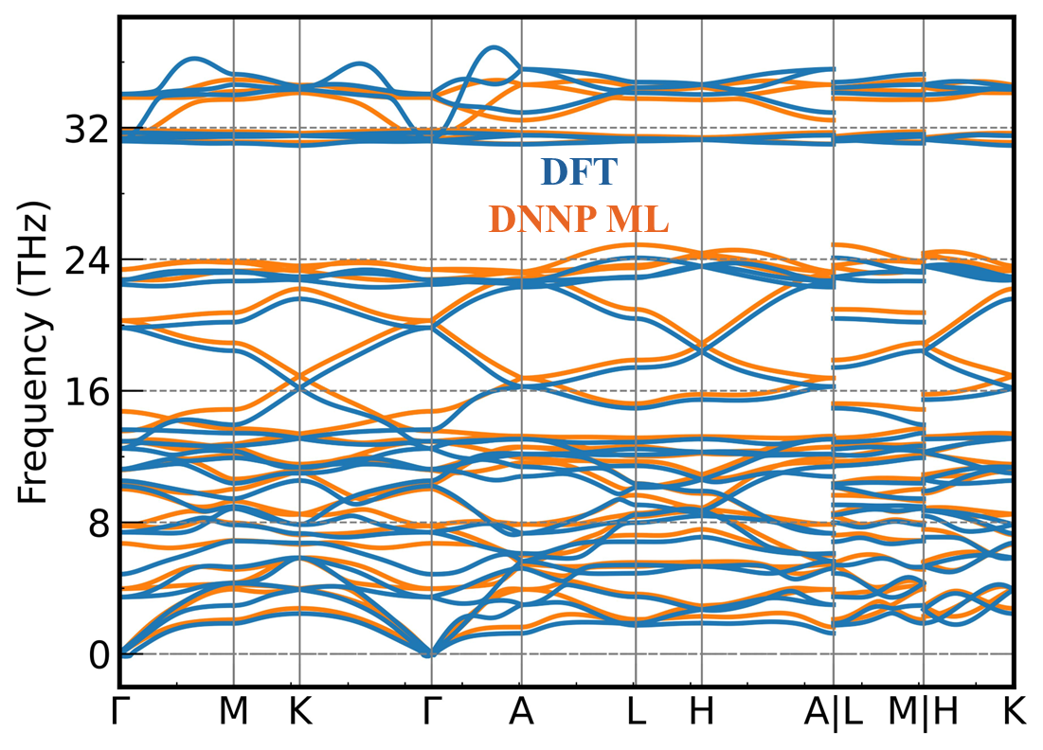}
\caption{Comparison of phonon band dispersions in $\alpha$-SiO$_2$ obtained from DFT (VASP + Phonopy) and DNNP (LAMMPS + Phonolammps). Both approaches employ equivalent $4\times4\times4$ supercells. The DNNP phonons are computed using a potential trained on AIMD data at 298~K. Overall agreement is good, although small discrepancies are observed in intermediate optical modes.}
\label{fig:phonon}
\end{figure}

To validate the reliability of the DNNP for vibrational properties, we compared the phonon band structure of ground-state (\emph{i.e.}, $T_e=0$~eV)$ \alpha$-SiO$_2$ computed using the finite displacement method with \textsc{vasp} and \textsc{phonopy} against the DNNP-based phonons obtained via the DNNPs. As shown in Fig.~\ref{fig:phonon}, the overall agreement between the two approaches is acceptable across the Brillouin zone. Observed deviations are likely due to the use of notionally 298 K AIMD data for the ground state DNNP, which introduces effective anharmonic contributions into the learned potential. Additionally, deviations may arise from the fact that phonon frequencies depend on second order force constants, which are not explicitly fitted during DNNP training.

The accuracy of our calculated DNNP phonon band structure is quantified by a root-mean-square error (RMSE) of 0.44~THz and a mean absolute error (MAE) of 0.35~THz across all phonon branches. A mode-resolved analysis further yields RMSE/MAE values of 0.37/0.29~THz for acoustic modes and 0.44/0.35~THz for optical modes, indicating slightly larger deviations for high-frequency vibrations. These results can be placed in context by comparing against recent benchmarks of universal machine-learned interatomic potentials~\cite{Loew2025}. In that study, models such as M3GNet and CHGNet exhibit errors in maximum phonon frequencies corresponding to approximately 1.85-2.04~THz, while MACE-MP-0 and SevenNet-0 achieve improved accuracies of $\sim$1.27~THz and $\sim$0.83~THz, respectively. The most accurate model, MatterSim-v1, reaches $\sim$0.35~THz. In comparison, our DNNPs significantly outperforms widely used models such as M3GNet and CHGNet, and also shows improved accuracy relative to MACE-MP-0 and SevenNet-0, while approaching the performance of state-of-the-art models such as MatterSim-v1. Notably, the overall error (0.44~THz) remains below the variation introduced by different exchange-correlation functionals ($\sim$0.7~THz), indicating that the predicted phonon frequencies lie within the intrinsic uncertainty due to the use of approximate DFT functionals. Consistent with this quantitative agreement, the overall dispersion trends, including the correct degeneracies at high-symmetry points and the absence of imaginary modes, are well reproduced, despite comparably minor deviations in selected optical branches.

\begin{figure}[!htbp]
\centering

\begin{subfigure}{0.48\linewidth}
\centering
\includegraphics[width=\linewidth]{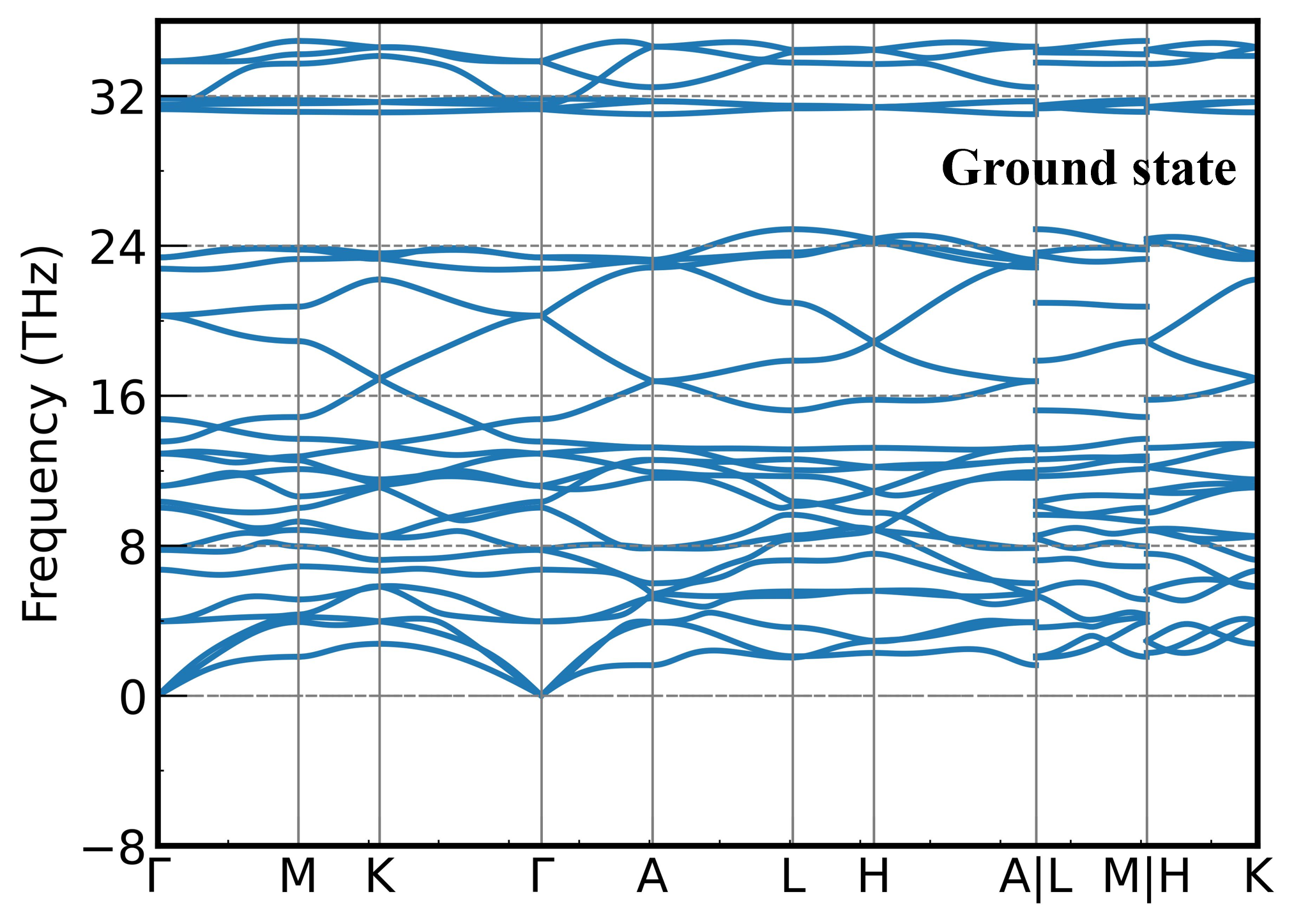}
\caption{Ground state}
\end{subfigure}\hfill
\begin{subfigure}{0.48\linewidth}
\centering
\includegraphics[width=\linewidth]{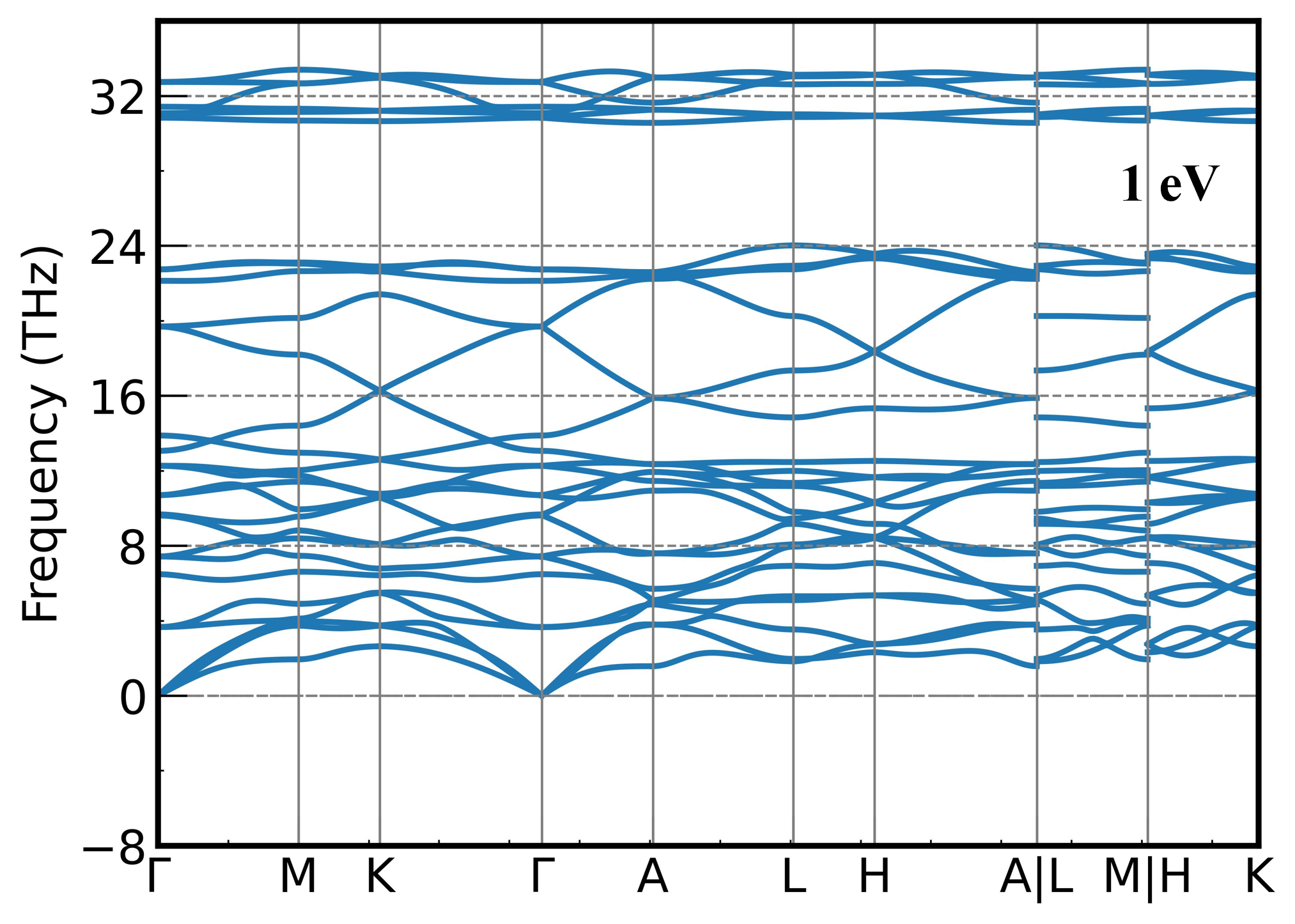}
\caption{$T_e = 1.0$~eV}
\end{subfigure}

\vspace{0.25cm}

\begin{subfigure}{0.48\linewidth}
\centering
\includegraphics[width=\linewidth]{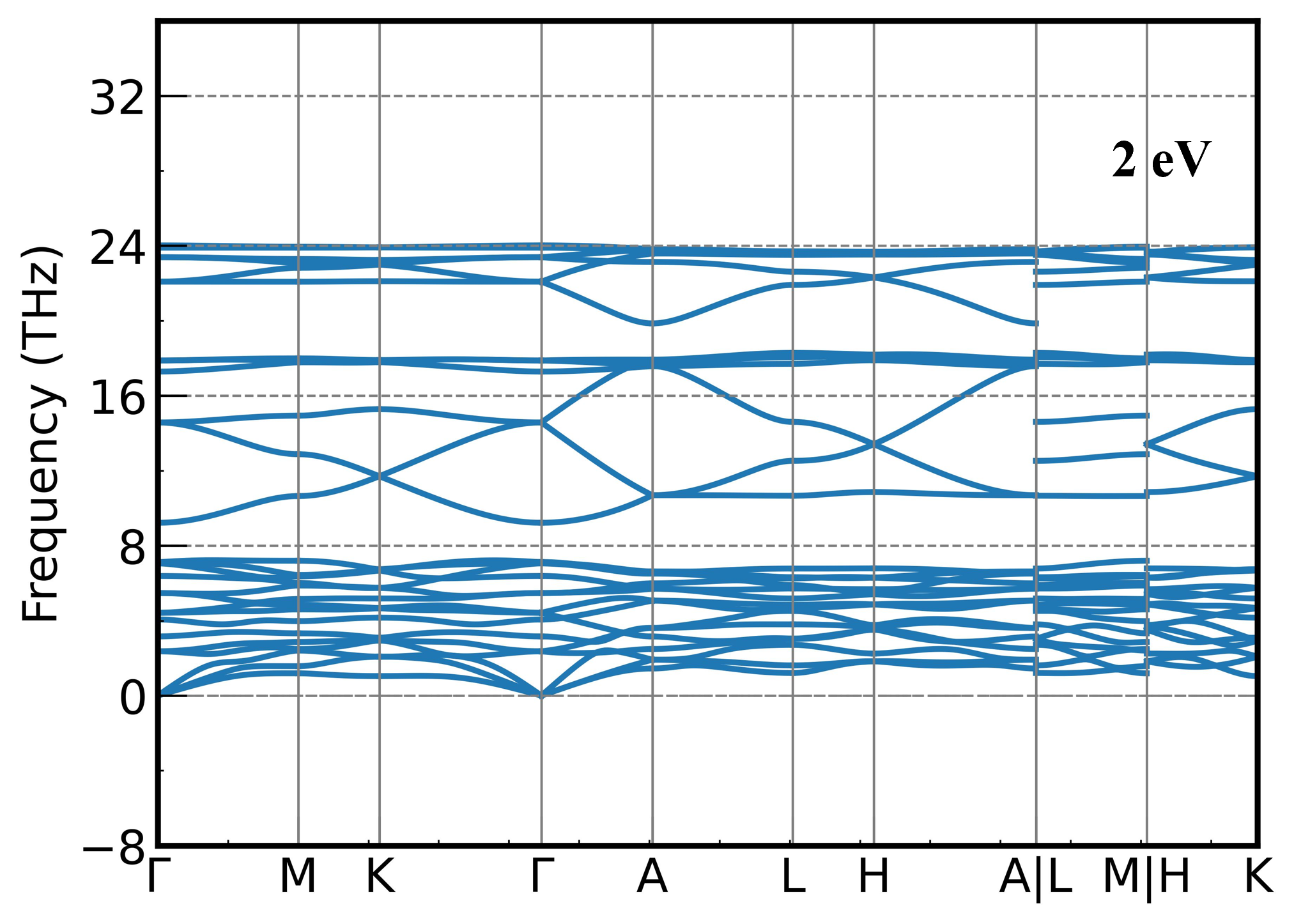}
\caption{$T_e = 2.0$~eV}
\end{subfigure}\hfill
\begin{subfigure}{0.48\linewidth}
\centering
\includegraphics[width=\linewidth]{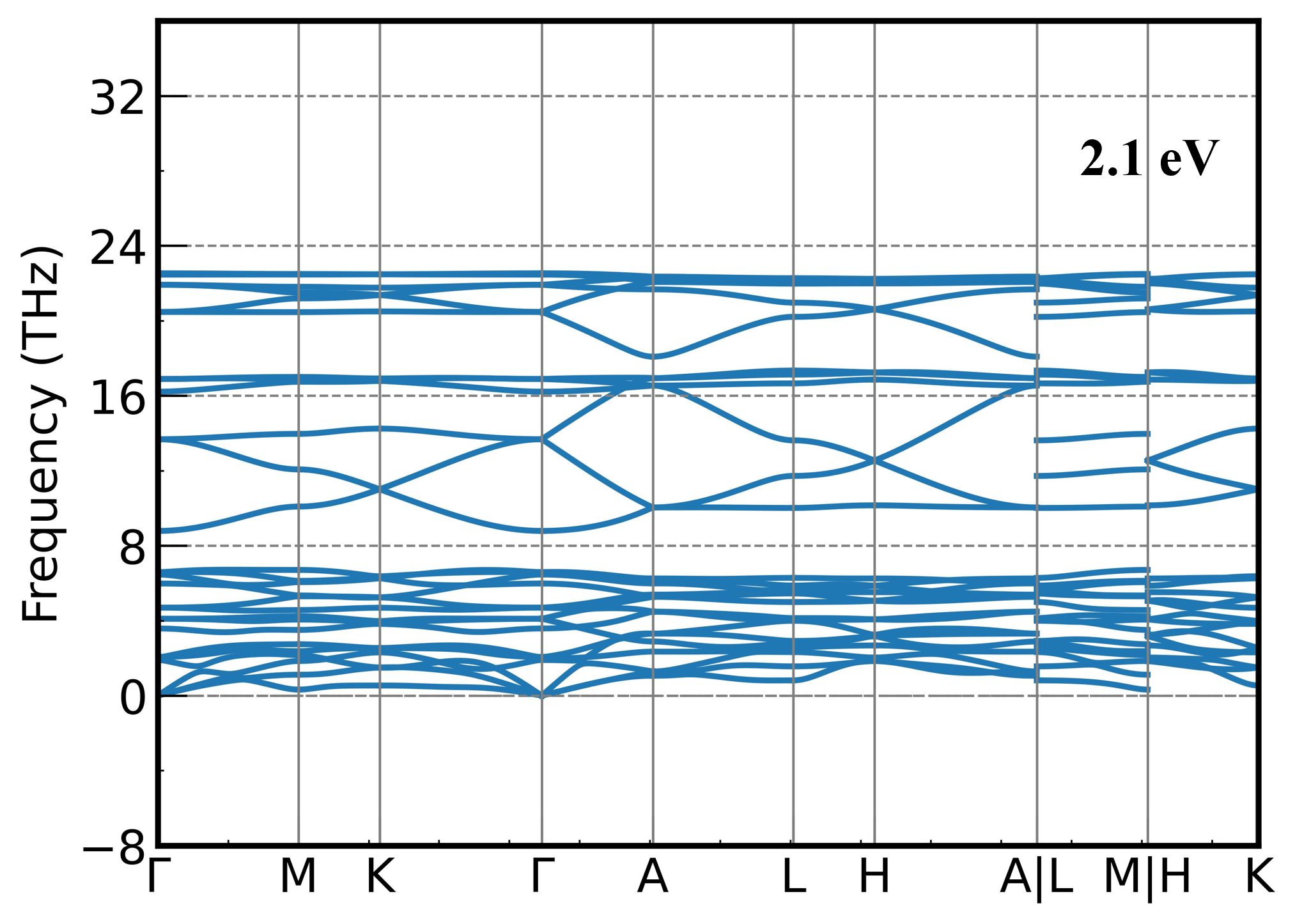}
\caption{$T_e = 2.1$~eV}
\end{subfigure}

\vspace{0.25cm}

\begin{subfigure}{0.48\linewidth}
\centering
\includegraphics[width=\linewidth]{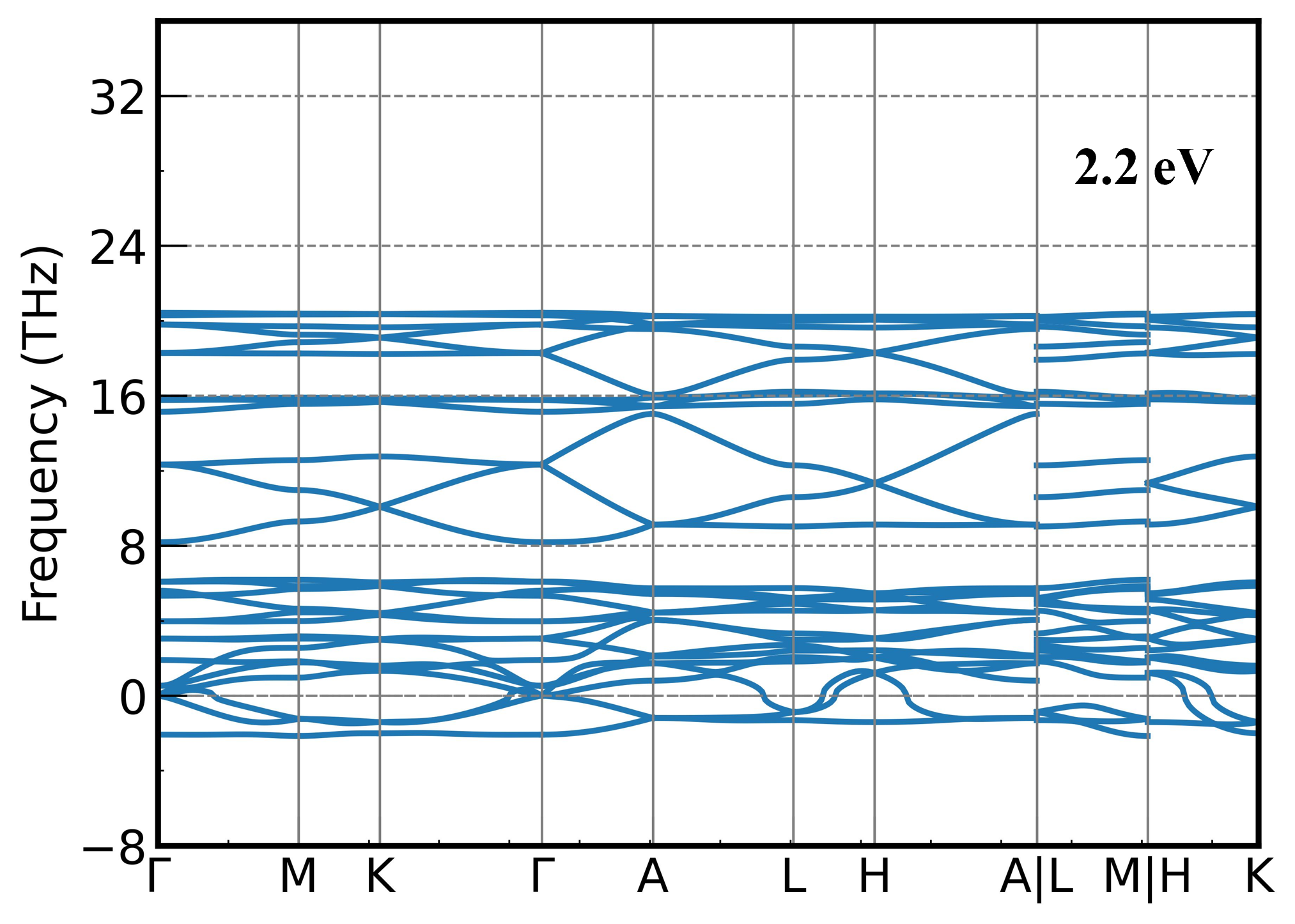}
\caption{$T_e = 2.2$~eV}
\end{subfigure}\hfill
\begin{subfigure}{0.48\linewidth}
\centering
\includegraphics[width=\linewidth]{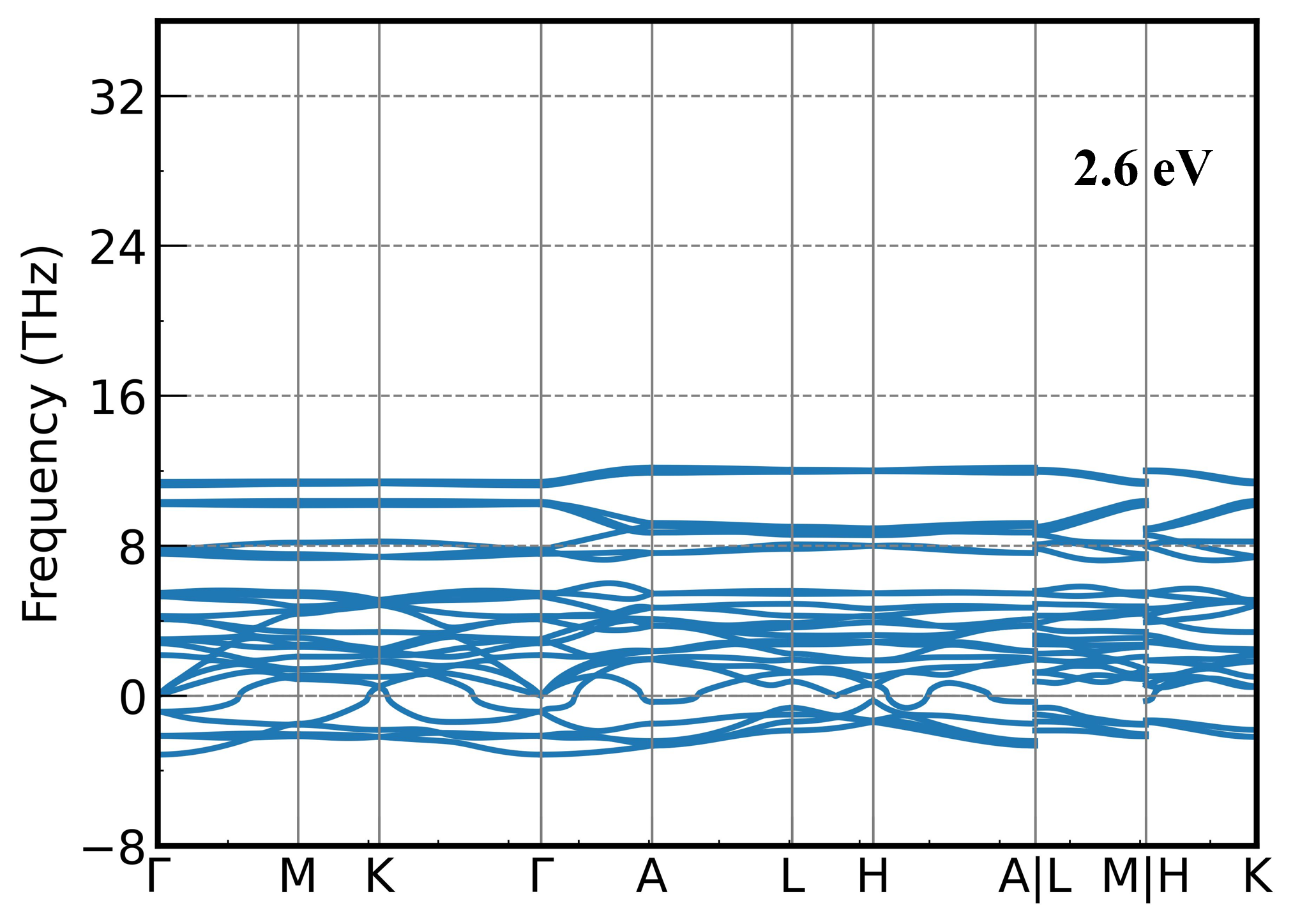}
\caption{$T_e = 2.6$~eV}
\end{subfigure}

\caption{Phonon dispersion relations of $\alpha$-SiO$_2$ computed using the DNNP-\textsc{phonolammps} framework at the ground state and under increasing electronic temperatures. All phonon calculations were performed at fixed volume using the corresponding optimized structures at each electronic temperature.}
\label{fig:phonon_excited}
\end{figure}

At higher electronic temperatures, DFT-based phonon calculations of a $4\times4\times4$ supercell become prohibitively expensive because of the large number of partially occupied bands required. Therefore, we computed the phonon band structures at different electronic temperatures using the DNNPs and \textsc{phonolammps} framework, as shown in Fig.~\ref{fig:phonon_excited}. At the ground state and $T_e = 1.0$~eV, the phonon bands are well-defined and all frequencies are real, and indication of full dynamical stability. The acoustic modes disperse smoothly from zero frequency at the $\Gamma$ point, and the optical branches retain their high-frequency character, consistent with the strong covalent \ce{Si-O} bonding network. At $T_e=2.0$ and $2.1$~eV, a noticeable softening occurs across both acoustic and optical branches, with reduced phonon bandwidths and lower maximum frequencies. While no imaginary modes are observed, yet, the compression of optical branches in the phonon spectrum suggests an incipient lattice structure softening and instability. At $T_e = 2.2$~eV, the first signatures of dynamical instability emerge, with a more pronounced softening of the optical branches and the appearance of imaginary frequencies  (conventionally represented as negative values). In fact, we have verified that at $T_e=2.2$~eV, the symmetry constrained trigonal structure becomes a saddle point on the free-energy surface. By performing a simulated annealing, the structure can be allowed to undergo a small symmetry breaking distortion and to relax into a dynamically stable local minimum.

By $T_e=2.6$~eV, several phonon branches exhibit imaginary frequencies, indicating the breakdown of harmonic dynamical stability and the onset of an electronically driven lattice instability. These phonon instabilities are ``diagnostics'' for electronically driven lattice destabilization which identify unstable directions on the fixed $T_e$ DNNP free energy surface. The progression of the phonon band structures from stable to unstable behavior with increasing electronic temperature mirrors the elastic softening trends observed in the elastic constants and EOS analysis. Recent first-principles calculations similarly reported progressive phonon softening and imaginary modes in $\alpha$-SiO$_2$ at elevated electronic temperatures~\cite{Ono2025b}. Our DNNPs reproduce this trend while providing finer $T_e$ resolution and reducing finite-supercell artifacts near the instability threshold.

\subsection{Fr\"ohlich Coupling Constant at Finite Electronic Temperature}

The electron-phonon coupling constant, $G_\mathrm{e-ph}$ of electronically excited $\alpha$-\ce{SiO2} has been recently computed using the Eliashberg spectral function formalism, showing a monotonic increase of $G_\mathrm{e-ph}$ as a function of the particle-hole density\cite{Tsaturyan2024}. The calculation of $G_\mathrm{e-ph}$ through the Eliashberg spectral function has been originally proposed by Allen~\cite{Allen1987} for metals, and its application to insulators not straightforward\cite{Medvedev2023}. An approach based on the Boltzmann equation formalism can account for both polar and nonpolar electron-phonon scattering mechanisms, and yield accurate estimate of $G_\mathrm{e-ph}$, yet in a less direct way\cite{Brouwer2014,Brouwer2017}.
Although the calculation of the electron-phonon coupling constant is beyond the scope of this work, 
from the phonon band structures reported in Sec.~\ref{sec:phonons}, the so-called Fr\"ohlich coupling constant can be easily determined.

The Fr\"ohlich model describes the long-range coupling between charge carriers and longitudinal optical (LO) phonons in polar materials, arising from the macroscopic electric field generated by polar lattice vibrations~\cite{Frohlich1954}. Within the Fr\"ohlich model, the dimensionless coupling constant 
\begin{equation}
\label{eq:alpha}
\alpha =
\frac{e^2}{4\pi\varepsilon_{\mathrm{0}}\hbar}
\sqrt{
\frac{m^\ast}{2\hbar\omega_{\mathrm{LO}}}
}
\left(
\frac{1}{\varepsilon_\infty}
-
\frac{1}{\varepsilon_{\mathrm{s}}}
\right)
\end{equation}
characterizes the strength of the long-range polar interaction and is commonly associated with polaronic properties such as carrier mass enhancement. In Eq.~\ref{eq:alpha}, $\varepsilon_{\mathrm{0}}$ is the vacuum permittivity, $m^\ast$ is the electron effective mass, $\omega_{\mathrm{LO}}=2\pi f_{\mathrm{LO}}$ is the angular frequency of the longitudinal optical phonon, while $\varepsilon_{\mathrm{s}}$ and $\varepsilon_\infty$ are the static and optical dielectric constants, respectively. These parameters were extracted consistently at each electronic temperature, with dielectric properties and effective masses obtained from DFT calculations --- here using the HSE06 functional --- and the LO phonon frequencies derived from phonon band structures computed using the DNNPs.

\begin{table}[!htbp]
\centering
\caption{Effective mass $m^\ast$ measured in unit of the bare electron mass, $m_e$, longitudinal optical phonon frequency $f_{\mathrm{LO}}$, dielectric constants $\varepsilon_s$ and $\varepsilon_\infty$, their ratio $\varepsilon_{\mathrm{s}}/\varepsilon_\infty$, and Fröhlich coupling constant $\alpha$ computed at different electronic temperatures $T_e$.}
\label{tab:SiO2_frohlich}
\begin{tabular}{ccccccc}
\toprule
$T_e$ (eV) &
$m^{*}$ ($m_e$) &
$f_{\mathrm{LO}}$ (THz) &
$\varepsilon_{s}$ &
$\varepsilon_{\infty}$ &
$\varepsilon_{\mathrm{s}}/\varepsilon_{\infty}$ &
$\alpha$ \\
\midrule
Ground state & 0.508 & 34.04 & 4.5890   & 2.4235 & 1.8935 & 1.3643 \\
1.0          & 0.505 & 33.07 & 8.0948   & 5.8728 & 1.3784 & 0.3313 \\
2.0          & 0.471 & 23.65 & 52.4733  & 48.8367 & 1.0745 & 0.0115 \\
2.1          & 0.468 & 22.02 & 68.2650  & 63.2207 & 1.0798 & 0.0098 \\
2.2          & 0.466 & 19.95 & 103.1635 & 95.3546 & 1.0819 & 0.0070 \\
\bottomrule
\end{tabular}
\end{table}

The ground state electron effective mass, high energy polar LO phonon frequency, and inverse-dielectric factor are comparable with the corresponding parameters considered by Porod and Ferry~\cite{Porod1985}.

From Table~\ref{tab:SiO2_frohlich} it can be seen that the effective mass, $m^\ast$, is a decreasing function of $T_e$, at variance of what reported by \cite{Tsaturyan2024}. However, in our study the atomic positions have been relaxed at each $T_e$. The softening of $f_{\mathrm{LO}}$ -- and of the phonon band structure, in general -- has been already reported.\cite{Tsaturyan2024,Ono2025b} It is worth noting that the observed softening of $f_{\mathrm{LO}}$ would, in isolation, tend to increase $\alpha$. The softening of $f_{\mathrm{LO}}$ is more than compensated by the decrease in the dielectric enhancement factor,  $(\varepsilon_{\infty}^{-1}-\varepsilon_{s}^{-1})$, which controls the strength of the long-range polar interaction. Consequently, the Fr\"ohlich coupling is strongly reduced, by more than two orders of magnitude, as $T_e$ increases, suggesting a suppression of the long-range polar optical phonon scattering channel at elevated electronic temperatures. As the ratio $\varepsilon_{\mathrm{s}}/\varepsilon_{\infty}$ tends to unity (see column six of Table~\ref{tab:SiO2_frohlich}), one can conclude that $\alpha$-\ce{SiO2} at elevated electronic temperature becomes essentially non-polar. This observation agrees with the Bader charge analysis reported in Sec.~\ref{Sec:pCOHP}

The values of $\omega_{\mathrm{LO}}$ reported in Table~\ref{tab:SiO2_frohlich} do not include the NAC. Although partially inconsistent, this approximation only introduces minor errors in estimates of $\alpha$-\ce{SiO2}, while simplifying the calculations. 
For the ground state, the uncorrected value of $f_{\mathrm{LO}}$ is $34.04$~THz. With NAC, the frequency increases to $35.93$~THz along the $\Gamma$-M direction and to $36.17$~THz along the $\Gamma$-A direction.
The corresponding Fr\"ohlich coupling constant decreases from $1.3643$ to $1.3280$ and $1.3236$, respectively.
Because of the general softening of the phonon band structure demonstrated in Sec.~\ref{sec:phonons}, the LO-TO splitting is expected to decrease as $T_e$ is increased, and the error from neglecting the NAC to decrease accordingly.
In general, the values of $\alpha$ reported in Table~\ref{tab:SiO2_frohlich} can be taken as upper limits.

\begin{figure}[!htbp]
\centering

\begin{subfigure}{0.48\linewidth}
\centering
\includegraphics[width=\linewidth]{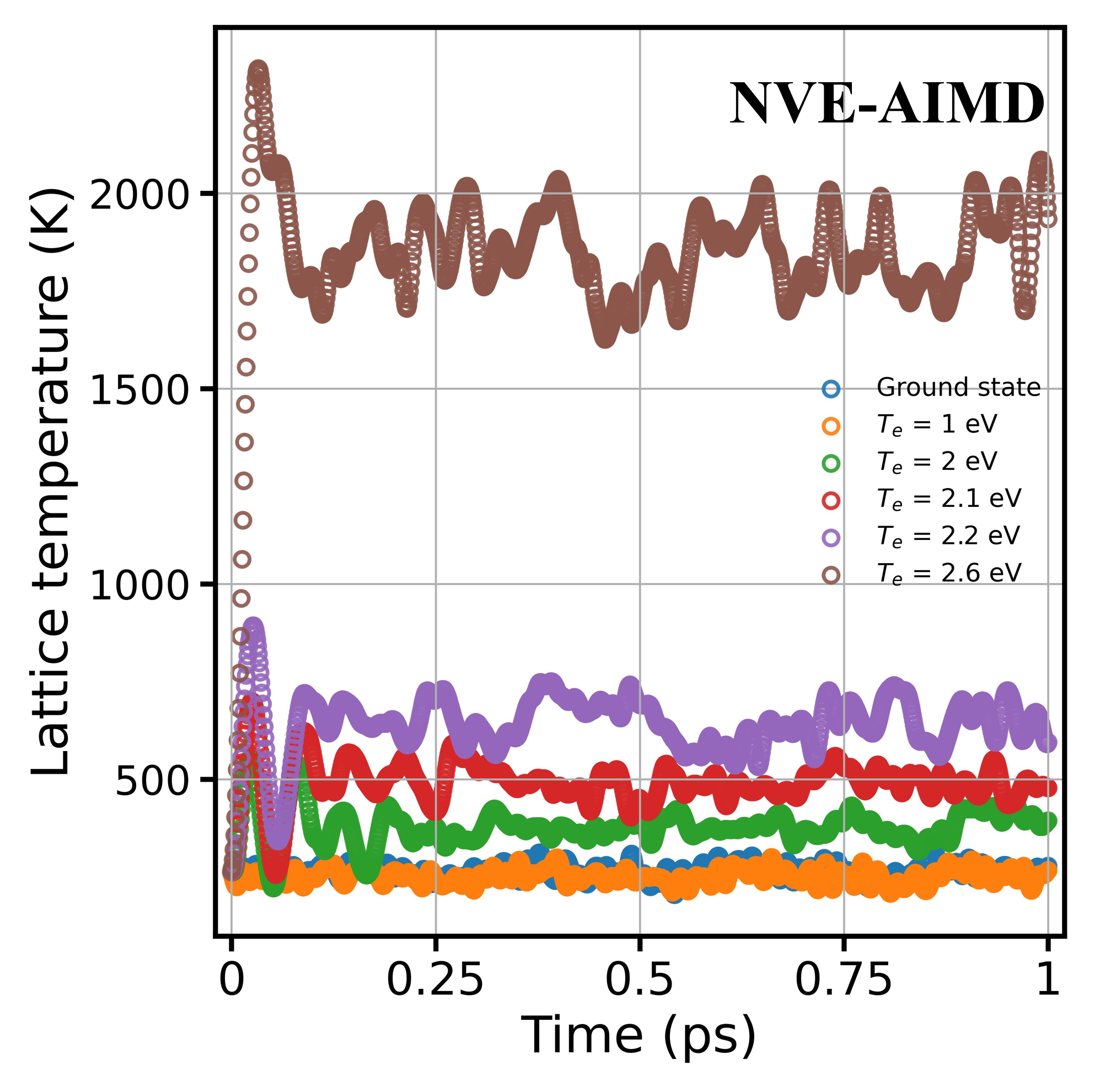}
\caption{NVE-AIMD}
\end{subfigure}\hfill
\begin{subfigure}{0.48\linewidth}
\centering
\includegraphics[width=\linewidth]{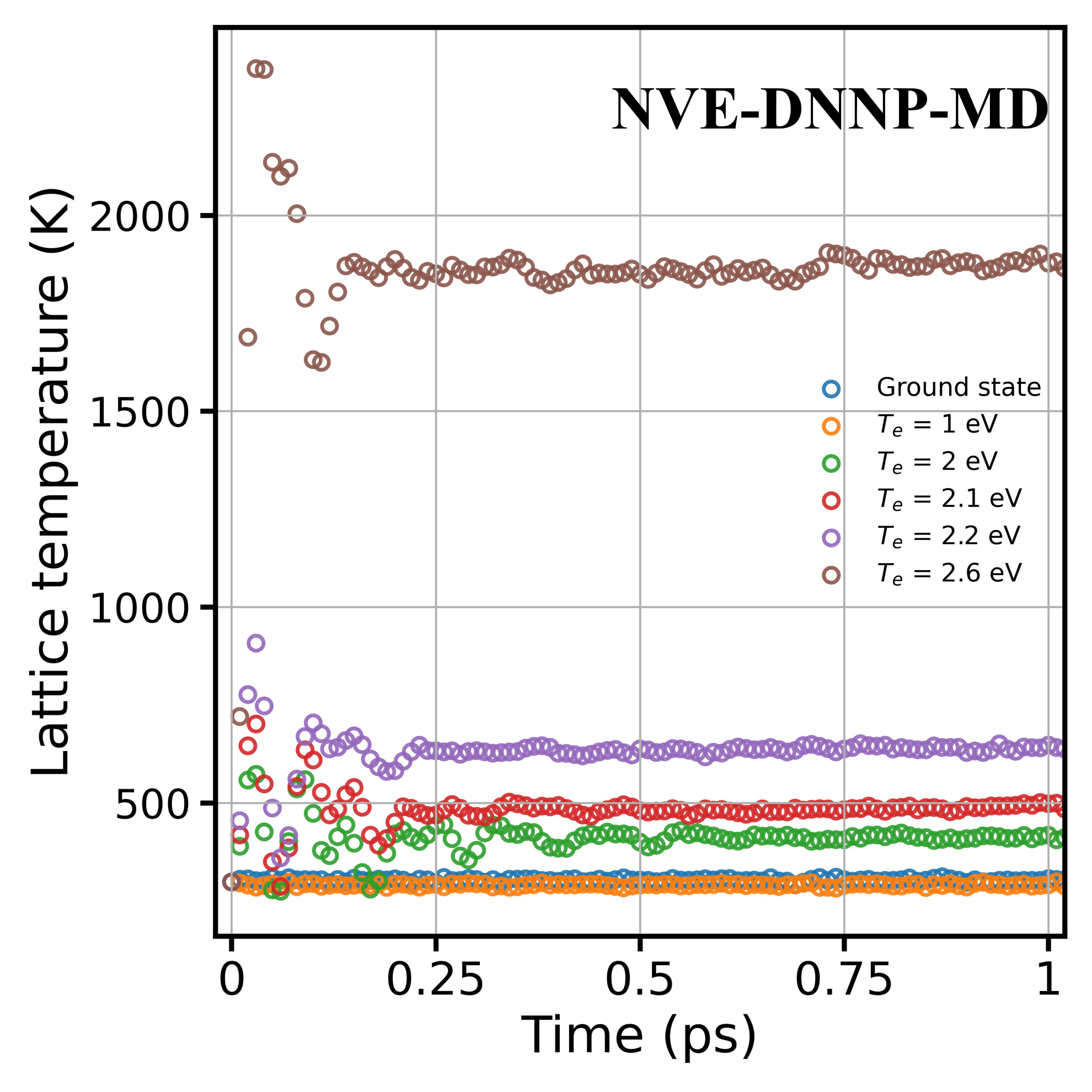}
\caption{NVE-DNNP-MD}
\end{subfigure}

\caption{Lattice temperature evolution in $\alpha$-SiO$_2$ for the ground state and under electronic excitation ($T_e = 1.0$--$2.6$~eV). Results from NVE-AIMD (a) and NVE-DNNP-MD (b) simulations are shown only up to 1~ps to highlight the initial lattice relaxation following electronic excitation.}
\label{fig:lattice_temperature}
\end{figure}

\subsection{Lattice Temperature Evolution after a sudden increase of the electronic temperature}
\label{sec:lattice_temperature}

A sudden electronic excitation as described by an instantaneous increase of the electronic temperature, $T_e$, modifies the interactomic forces and can drive a ultrafast non-thermal melting of the lattice.\cite{Silvestrelli1996,Silvestrelli1997,Medvedev2015} In Fig.~\ref{fig:lattice_temperature}, we report the time-dependent lattice temperature, $T_L(t)$,  defined so that 
\begin{equation}
\label{eq:kinetic}
   E_K(t) = \left(\frac{3N-3}{2}\right)k_B T_L(t)\;, 
\end{equation}
where $N$ is the number of atoms in the simulation cell and $E_K(t)$ the average kinetic energy measured at time $t>0$, after an instantaneous increase of the electronic temperature from $0$ (ground state) to $T_e$.
Results from both AIMD and DNNP-MD are reported, sampled in the microcanonical (NVE) ensemble.
 Due to the larger unit cell, the DNNP-MD trajectories display reduced statistical fluctuations and allow for a more accurate extrapolation to the steady-state which follows the initial transient behavior.
 At $T_e=1.0$~eV, both AIMD and DNNP-MD reach a steady-state $T_L$ very close to the notional initial value  (298~K). In fact, in the AIMD, the difference the average and initial $T_L$ are within the statistical uncertainty dictated by the finite-size microcanonical fluctuations.

At more elevated electronic temperatures, \emph{i.e.}, $T_e = 2.0$--$2.2$~eV, the intantaneous lattice temperature, $T_L(t)$, rises rapidly over a few 100s of fs, in noticeably oscillatory way. The observed steady-state values are $T_L=423.2\pm0.3$~K at $T_e=2.0$~eV, $T_L=518.2\pm0.4$~K at $T_e=2.1$~eV, and $T_L=641.0\pm0.4$~K at $T_e=2.2$~eV, where the errors represent the standard error of the mean.
The average steady-state values of $T_L$ from the DNNP-MD simulations are consistent with values from the AIMD simulations, although the for the latter the statistical error is much larger. At $T_e = 2.6$~eV, the instantaneous lattice temperature exhibits a even more abrupt initial jump, quickly exceeding 2,000~K -- a value larger than the $\alpha$-\ce{SiO2} melting temperature in full equilibrium (\emph{i.e.}, $T_e=T_L$) conditions.
A similar large increase of the steady-state ionic temperature stepping from $T_e\approx2.2$ to $T_e\approx2.6$~eV has been previously observed by Boero \emph{et al.} and attributed to the weakening and breaking of the \ce{Si-O} bonds.\cite{Boero2005}

It should be noticed that, although the instantaneous lattice temperature, $T_L(t)$, is a well-defined dynamical quantity, its definition does not imply that the lattice is in a well-defined thermodynamic equilibrium state. Indeed, we have have found that the atomic speeds initially do not follow a Maxwell-Boltzmann distribution, as shown in Fig.~S6 in The Supplementary Information. The best fit to a Maxwell-Boltzmann distribution done separately for \ce{Si} and \ce{O} atoms give rather different kinetic temperatures, indicated as $T_\mathrm{Si}$ and $T_\mathrm{O}$ in Fig.~S6. Even after 100 fs, the two values are still rather different, although they tend to converged toward the same kinetic average temperature, $T_L(t)$. In general, we find that $T_\mathrm{O}>T_\mathrm{Si}$, as expected since $m_\mathrm{Si}>m_\mathrm{O}$.
We also observe that $T_\mathrm{O}$ rises very quickly, initially ``overshooting'' and then falling back and converging toward the value of $T_\mathrm{Si}$, which has risen much slowly and more monotonically. This behavior is particular evident for $T_e\ge 2.0$~eV, in panels (c)--(e) of Fig.~S6. The oscillations of $T_L(t)$ displayed in Fig.~\ref{fig:lattice_temperature} can be ascribed to this lack of proper thermodynamic equilibration. Indeed, the magnitude of the ``overshoots" observed in the oscillatory dynamics of $T_L(t)$ correlates with that of the ``overshoots" of $T_\mathrm{O}$. 

The observed behavior can be rationalized using a simple kinetic model, in which collisions between atoms of the same species exchange momentum more efficiently than collisions between atoms of different species. As a consequences, atoms of the same species first ``thermalize'' among themselves and then fully equilibrate on a longer timescale. Assuming that a similar impulsive force is felt by all atoms upon raising of $T_e$, light species are also expected to accelerate more quickly, as observed. Although qualitative and simplified, this kinetic model is suggestive as a bounded-to-atomic fluid transition proposed in \ce{SiO2} at elevated pressures and temperatures\cite{Hicks2006,Green2018,Zhang2022b}. 

\begin{figure}[!htbp]
\centering

\begin{subfigure}{0.48\linewidth}
\centering
\includegraphics[width=\linewidth]{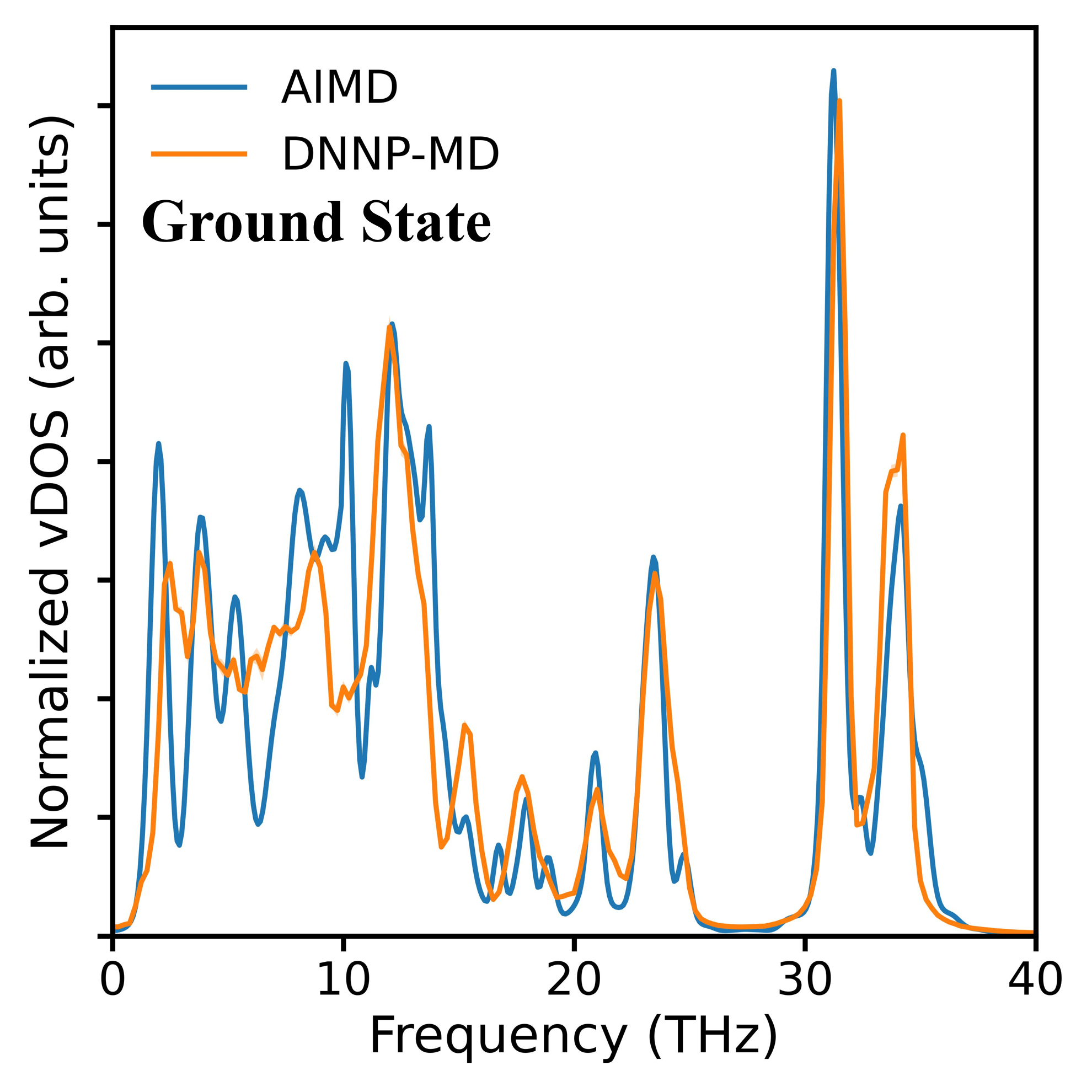}
\caption{Ground state}
\end{subfigure}\hfill
\begin{subfigure}{0.48\linewidth}
\centering
\includegraphics[width=\linewidth]{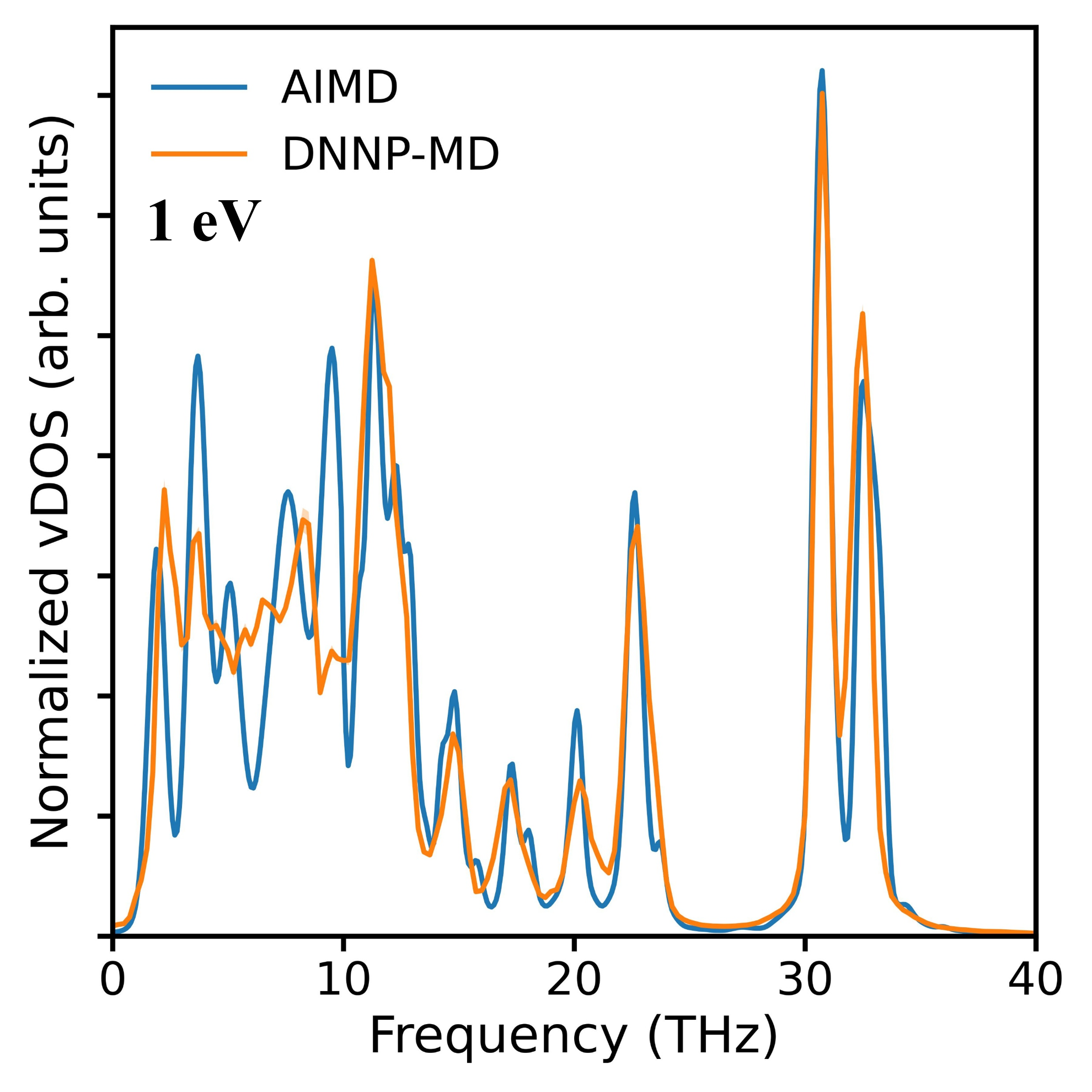}
\caption{$T_e = 1.0$~eV}
\end{subfigure}

\vspace{0.25cm}

\begin{subfigure}{0.48\linewidth}
\centering
\includegraphics[width=\linewidth]{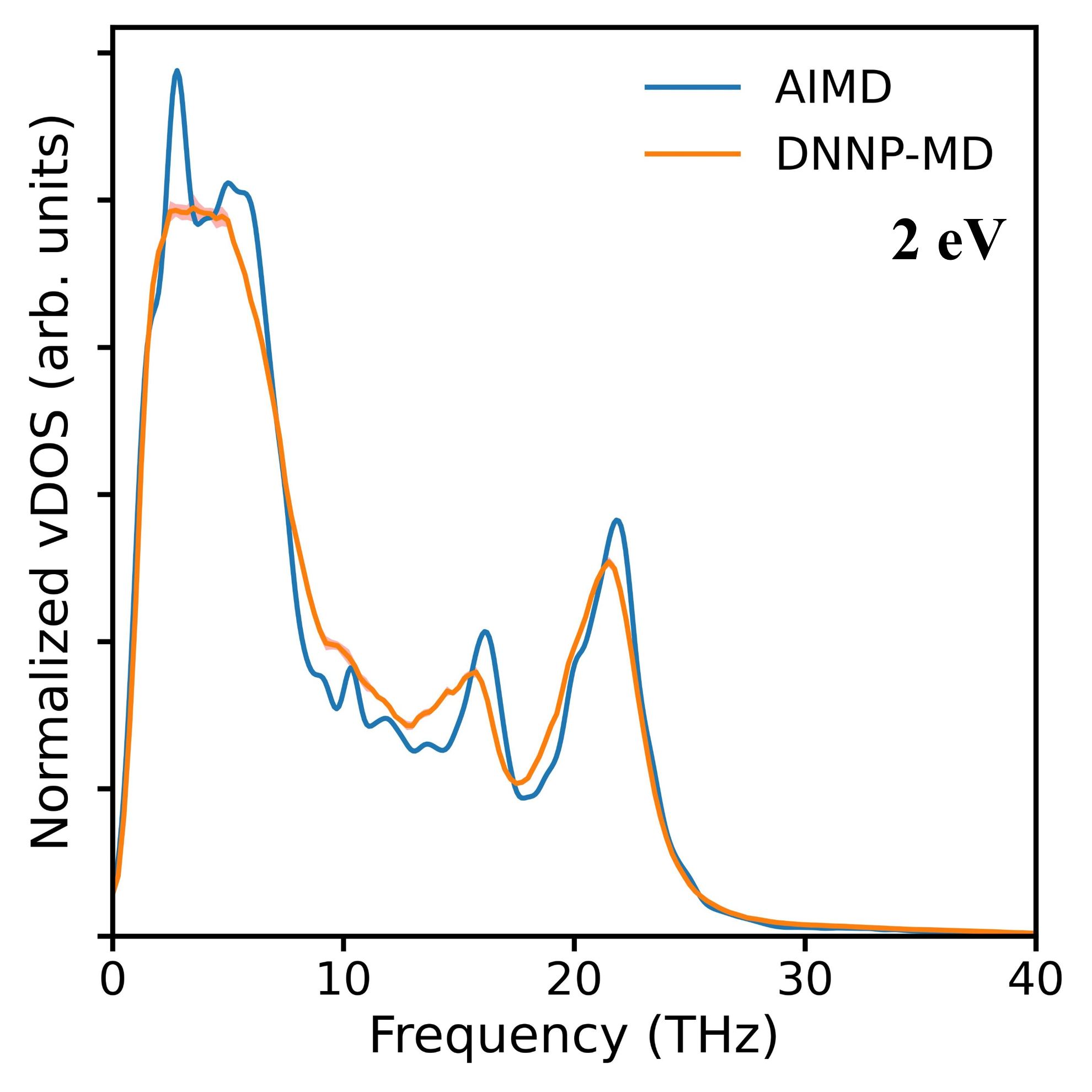}
\caption{$T_e = 2.0$~eV}
\end{subfigure}\hfill
\begin{subfigure}{0.48\linewidth}
\centering
\includegraphics[width=\linewidth]{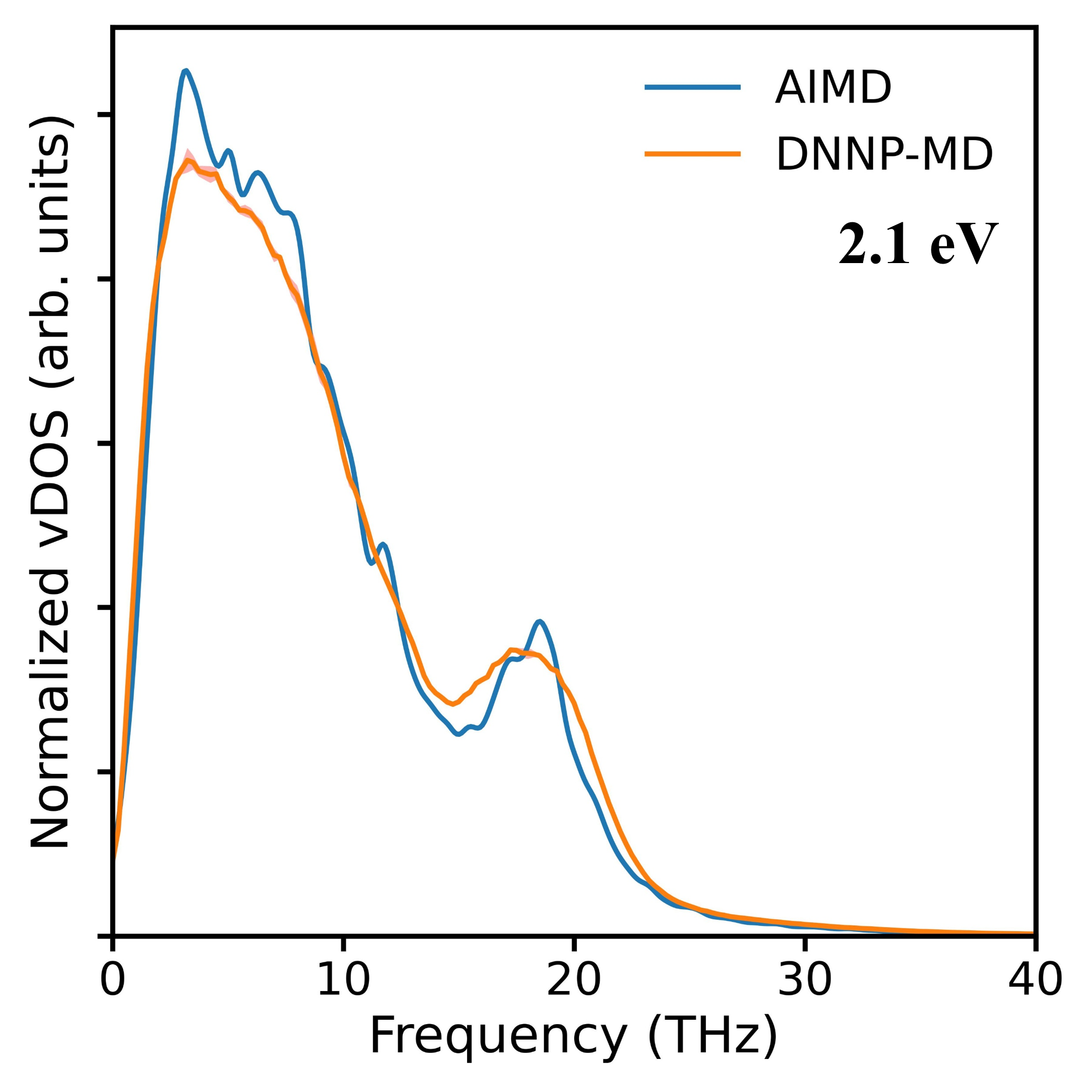}
\caption{$T_e = 2.1$~eV}
\end{subfigure}

\end{figure}

\begin{figure}[!htbp]
\ContinuedFloat
\centering

\begin{subfigure}{0.48\linewidth}
\centering
\includegraphics[width=\linewidth]{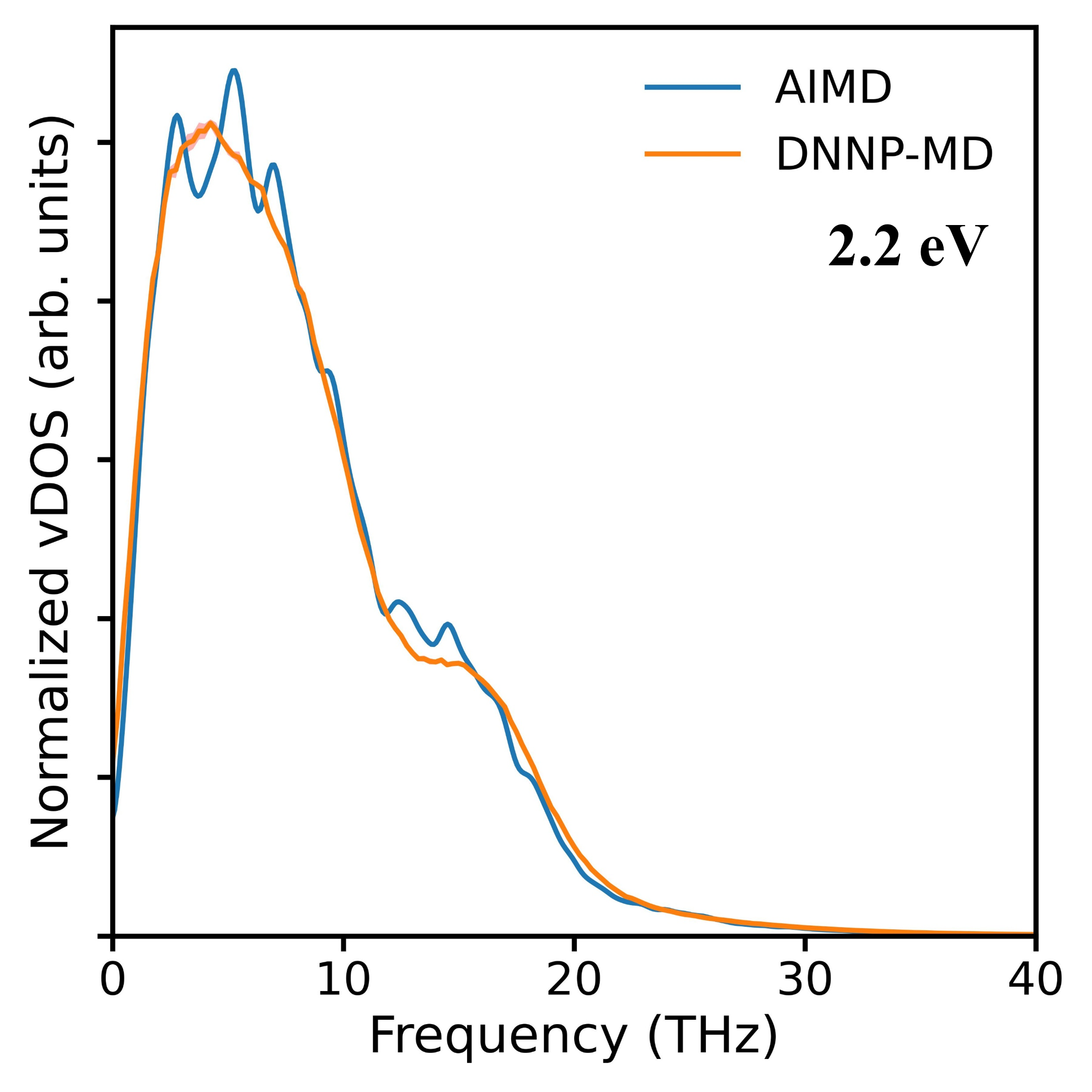}
\caption{$T_e = 2.2$~eV}
\end{subfigure}\hfill
\begin{subfigure}{0.48\linewidth}
\centering
\includegraphics[width=\linewidth]{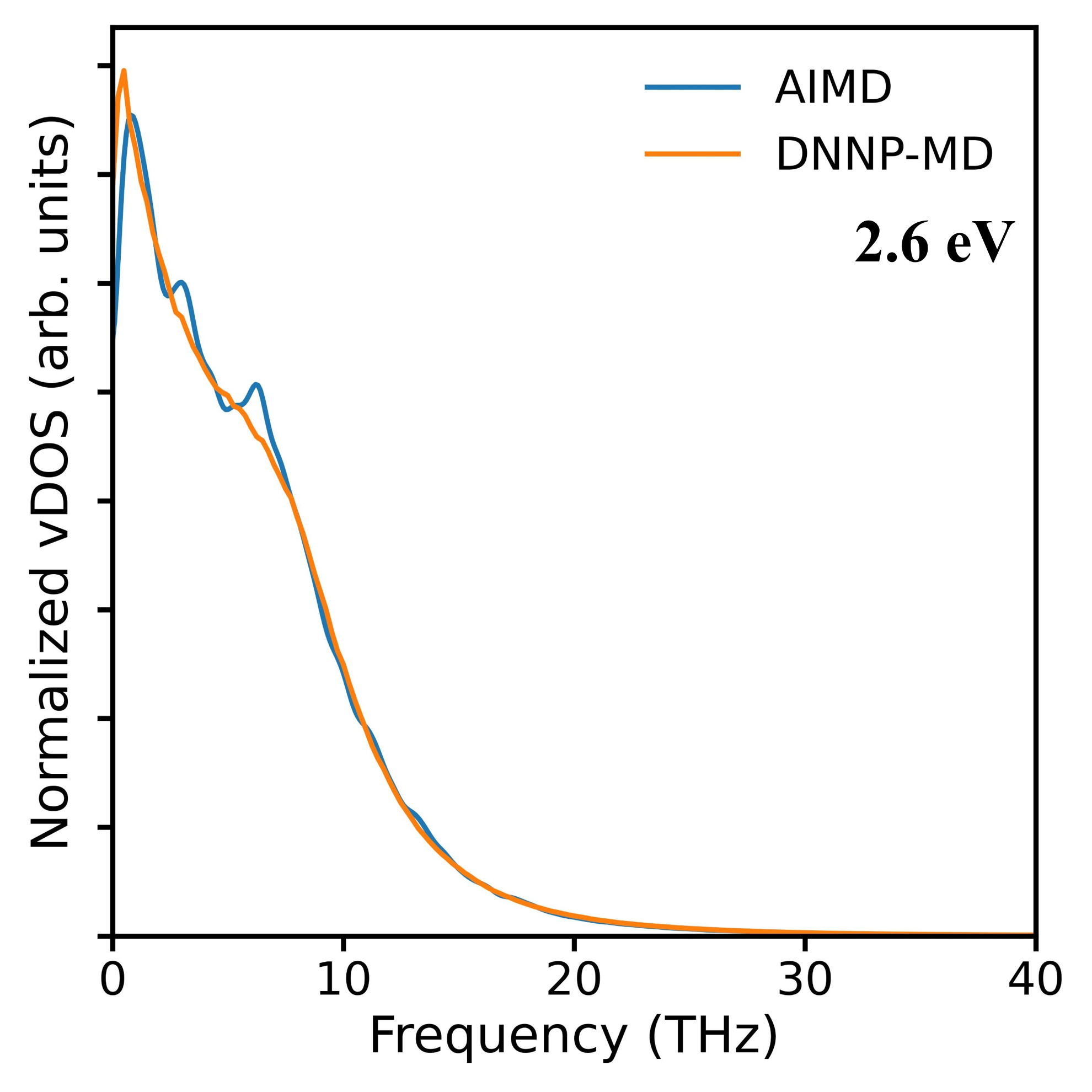}
\caption{$T_e = 2.6$~eV}
\end{subfigure}

\caption{Vibrational density of states (VDOS) of $\alpha$-SiO$_2$ obtained from AIMD and DNNP-MD simulations performed under NVE conditions for the ground state and for electronic temperatures up to $T_e = 2.6$~eV. The AIMD results correspond to a $2\times2\times2$ supercell of the $\alpha$-SiO$_2$ unit cell (72 atoms, 24 SiO$_2$ formula units), while the DNNP-MD results are obtained from a $7\times7\times7$ supercell (3087 atoms, 1029 SiO$_2$ formula units).}
\label{fig:VDOS_excited}
\end{figure}
\FloatBarrier

\subsection{Velocity Autocorrelation Function and Vibrational Density of States at Finite Electronic Temperature}

To further elucidate the nature of the steady-state achieved after a sudden increase of the electronic temperature, we analyze the velocity autocorrelation function (VACF) and the vibrational density of state (VDOS).
It is worth clarifying that the steady-state observed in our numerical simulation is not indicative of a long-lasting, thermodynamic phase, and that we interpret the present results primarily as lattice destabilization and ultrafast structural disordering and melting, in agreement with the recent literature.~\cite{Lian2016,Zier2019Comment,Lian2019Reply}. The electronically excited $\alpha$-\ce{SiO2} is expected to reach a fully equilibrated state through non-adiabatic electron-ion interactions which are not included in our AIMD simulations and beyond the scope of this work.

Figure~S5 presents the VACFs of $\alpha$-\ce{SiO2} obtained from AIMD and DNNP-MD NVE simulations described in Sec.~\ref{sec:lattice_temperature}.
At $T_e=0$ and $1$~eV, the VACF exhibits pronounced oscillations during the first few ps, reflecting well-preserved vibrational coherence characteristic of a crystalline solid. For $T_e = 2.0$-$2.2$~eV, the VACF decays rapidly to zero after a negative dip, indicating strong damping of vibrational motion and increased collisional back-scattering in a dynamically disordered lattice. At $T_e = 2.6$~eV, the VACF decays to zero within $\sim 10$-$20$~fs without developing pronounced oscillations or a negative dip, characteristic of a fluid phase above the Frenkel line\cite{Brazhkin2012,Trachenko2016}. Across all conditions, the close agreement between AIMD and DNNP-MD results demonstrates that the DNNPs accurately reproduce the dynamical behavior of the system at finite electronic temperature, while allowing the use of larger simulations cell for improved thermodynamical sampling.

Figure~\ref{fig:VDOS_excited} compares the vibrational density of states (VDOS) of $\alpha$-SiO$_2$ obtained from AIMD and DNNP-MD simulations NVE simulations described in Sec.~\ref{sec:lattice_temperature}. At $T_e=0$ and $1$~eV, the VDOS exhibits well-defined peaks across the full frequency range, with sharp features in the high-frequency \ce{Si-O} stretching modes ($\sim 25$--$35$~THz) and distinct low-frequency lattice modes below $10$~THz, characteristic of a crystalline solid. For $T_e = 1.0$~eV, these features remain largely preserved, although slight peak broadening indicates a reduction in phonon lifetimes.

With further increase in $T_e$ to $2.0$-$2.2$~eV, the VDOS undergoes substantial broadening accompanied by a pronounced suppression of high-frequency \ce{Si-O} stretching modes. This behavior is characterized by significant broadening of the vibrational density of states and a redistribution of spectral weight toward lower frequencies, indicating reduced phonon lifetimes and partial loss of long-range structural order, consistent with the onset of a fluid-like state. The finite limit for $\omega\to0$ is related to atomic diffusion, which increases with $T_e$. At $T_e = 2.6$~eV, the VDOS is almost completely structureless and monotonically decaying, as expected from a super-critical fluid.\cite{Trachenko2016}. This regime is consistent with proposed bounded-to-atomic fluid transition in \ce{SiO2} at elevated pressures and temperatures.\cite{Hicks2006,Green2018,Zhang2022b} 

\section{Conclusions}

In this work, we developed a multiscale first-principles-to-machine-learning framework to investigate ultrafast lattice dynamics in electronically excited $\alpha$-\ce{SiO2}. By training deep neural network potentials (DNNP) using finite-electronic-temperature, $T_e$, \emph{ab initio} molecular dynamics (AIMD) data, we performed molecular dynamics (MD) simulations at near first-principles accuracy over length scales hardly inaccessible to conventional first-principles simulations. 

In agreement with previous studies,\cite{Boero2005,Tsaturyan2022,Zhang2022b,Ono2025a} our results show that increasing electronic temperature induces pronounced lattice destabilization in $\alpha$-\ce{SiO2}, evidenced by violations of elastic stability criteria, substantial volumetric expansion, a sharp reduction of the bulk modulus, and progressive weakening of \ce{Si-O} bonding due to antibonding-state occupation. These indications were obtained by considering the projected Crystal Orbital Hamilton Population (pCOHP) analysis, the equation of state of $\alpha$-\ce{SiO2}, and phonon band structures, all at finite electronic temperatures.    

The use of DNNPs allowed us to obtain accurate phonon band structures, which  exhibit systematic softening with increasing electronic temperature, further elucidating the nature of the lattice destabilization at elevated $T_e$. 

From the knowledge of the electronic and phonon band structures, we can estimate the The Fr\"ohlich coupling constant, $\alpha$, which measures the long-range coupling between charge carriers and longitudinal optical phonons in polar materials. We found that $\alpha$ decreases as $T_e$ increases, leading to a cross-over to a non-polar phase of $\alpha$-\ce{SiO2} at elevated electronic temperature. This finding is corroborated by the Bader charge analysis. 

By means of DNNP-MD simulations, we obtained accurate estimates of the steady-state achieved after a sudden increase of the electronic temperature. In particular, we evaluated the average kinetic temperature and the velocity distribution functions, showing that a well-defined thermal equilibrium of the lattice degrees of freedom is not achieved over the first few hundreds of femtoseconds.

The analysis of velocity autocorrelation function (VACF) and vibrational density of states (VDOS) supports a bonded-to-atomic transition toward a supercritical phase of \ce{SiO2} at large $T_e$, as also recently suggested both experimentally and theoretically~\cite{Hicks2006,Green2018,Zhang2022b}.

\begin{acknowledgments}
The authors acknowledge financial support from the Engineering and Physical Sciences Research Council (EPSRC) through the Ultrafast Nanodosimetry grant (EP/W017245/1). 
We are grateful for use of the computing resources from the Northern Ireland High Performance Computing (NI-HPC) service funded by EPSRC (EP/T022175). 
LS thanks Dr Adrien Descamps for insightful discussions.
\end{acknowledgments}

\section*{Competing Interests}
The authors declare no competing interests.

\section*{Supporting Information}
Additional data and analyses relevant to this manuscript are provided in the Supporting Information.

\renewcommand{\bibsection}{\section*{References}}
\bibliography{references}

\end{document}


\begin{center}
{\Large\bfseries Supplemental Material}\\[0.4cm]

{\large\bfseries
Ultrafast Nonthermal Lattice Destabilization and Suppression of
Polar Optical Scattering in Electronically Excited
$\alpha$-SiO$_2$ from First-Principles and Deep Neural Network
Potential Modeling}\\[0.5cm]

Iyyappa Rajan Panneerselvam,
Mark Yeung,
Charlotte Palmer,
Brendan Dromey,
and Lorenzo Stella\\[0.2cm]

Centre for Light-Matter Interactions,
Department of Physics and Astronomy,
Queen's University Belfast,
Belfast BT7 1NN, United Kingdom
\end{center}

\vspace{0.5cm}

\begin{table}[H]
\centering
\footnotesize
\caption{Elastic stability criteria and the minimum eigenvalue of the elastic stiffness matrix for $\alpha$-SiO$_2$ at different electronic temperatures.}
\label{tab:SI_elastic}
\begin{tabular}{ccccccc}
\toprule
$T_e$ (eV) &
Condition (i) &
Condition (ii) &
Condition (iii) &
Condition (iv) &
$\lambda_{\min}$ (GPa) &
Stability \\
\midrule
Ground & $\checkmark$ & $\checkmark$ & $\checkmark$ & $\checkmark$ & 27.58  & Stable \\
1.0    & $\checkmark$ & $\checkmark$ & $\checkmark$ & $\checkmark$ & 22.94  & Stable \\
2.0    & $\checkmark$ & $\checkmark$ & $\checkmark$ & $\checkmark$ & 7.52   & Stable \\
2.1    & $\checkmark$ & $\checkmark$ & $\checkmark$ & $\checkmark$ & 2.97   & Stable \\
2.2    & $\checkmark$ & $\checkmark$ & $\times$     & $\times$     & -10.10 & Unstable \\
2.6    & $\times$     & $\checkmark$ & $\checkmark$ & $\times$     & -34.47 & Unstable \\
\bottomrule
\end{tabular}
\end{table}

\vspace{0.3cm}
\noindent
Conditions (i)--(iv) correspond to the elastic stability criteria for trigonal systems:
(i) $C_{11}-C_{12}>0$,
(ii) $C_{13}^2 < \tfrac{1}{2}C_{33}(C_{11}+C_{12})$,
(iii) $C_{14}^2 < \tfrac{1}{2}C_{44}(C_{11}-C_{12})$, and
(iv) $C_{44}>0$.
Only the minimum eigenvalue $\lambda_{\min}$ of the elastic stiffness matrix is reported here, as it provides a sufficient indicator of mechanical stability.

\vspace{0.6cm}

\begin{table}[H]
\centering
\setlength{\abovecaptionskip}{4pt}
\setlength{\belowcaptionskip}{6pt}
\caption{Third-order Birch--Murnaghan equation-of-state parameters for $\alpha$-SiO$_2$ at different electronic temperatures. 
$E_0$ denotes the HSE06-corrected electronic free energy at the equilibrium volume, $V_0$ is the equilibrium unit-cell volume, 
$B_0$ is the bulk modulus, and $B_0'$ is its pressure derivative. 
The energy per formula unit is obtained assuming three SiO$_2$ units per $\alpha$-quartz unit cell. 
For $T_e = 2.2$~eV, the reported equilibrium volume is obtained from an extrapolation of the equation-of-state fit beyond the sampled volume range.}
\label{tab:SI_EOS}

\begin{tabular}{cccccc}
\toprule
$T_e$ (eV) &
$E_0$ (eV/cell) &
$E_0$ (eV/f.u.) &
$V_0$ (\AA$^3$/cell) &
$B_0$ (GPa) &
$B_0'$ \\
\midrule
Ground state & $-87.673638$ & $-29.224546$ & $120.7356$ & $68.080$ & $-1.8983$ \\
1.0          & $-87.749389$ & $-29.249796$ & $122.1118$ & $61.433$ & $-1.0433$ \\
2.0          & $-91.251671$ & $-30.417224$ & $151.3179$ & $22.303$ & $2.5601$ \\
2.1          & $-92.128657$ & $-30.709552$ & $164.0036$ & $14.291$ & $3.2720$ \\
2.2          & $-93.453209$ & $-31.151070$ & $231.4832$ & $2.030$  & $5.3498$ \\
\bottomrule
\end{tabular}
\end{table}

\vspace{0.4cm}

\begin{figure}[H]
\centering
\begin{subfigure}[t]{0.48\linewidth}
\includegraphics[height=5.6cm]{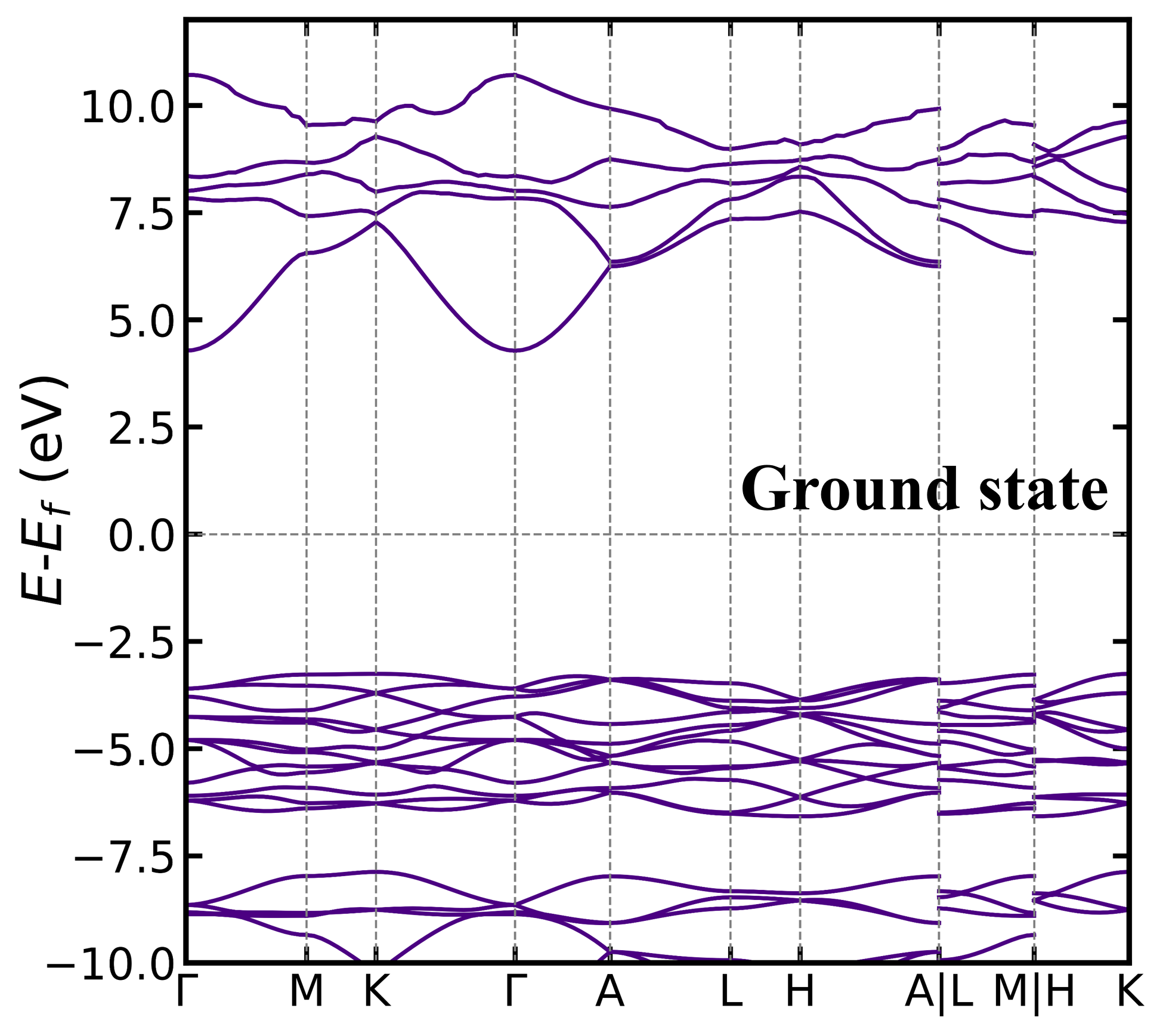}
\caption{Ground state}
\end{subfigure}\hfill
\begin{subfigure}[t]{0.48\linewidth}
\includegraphics[height=5.6cm]{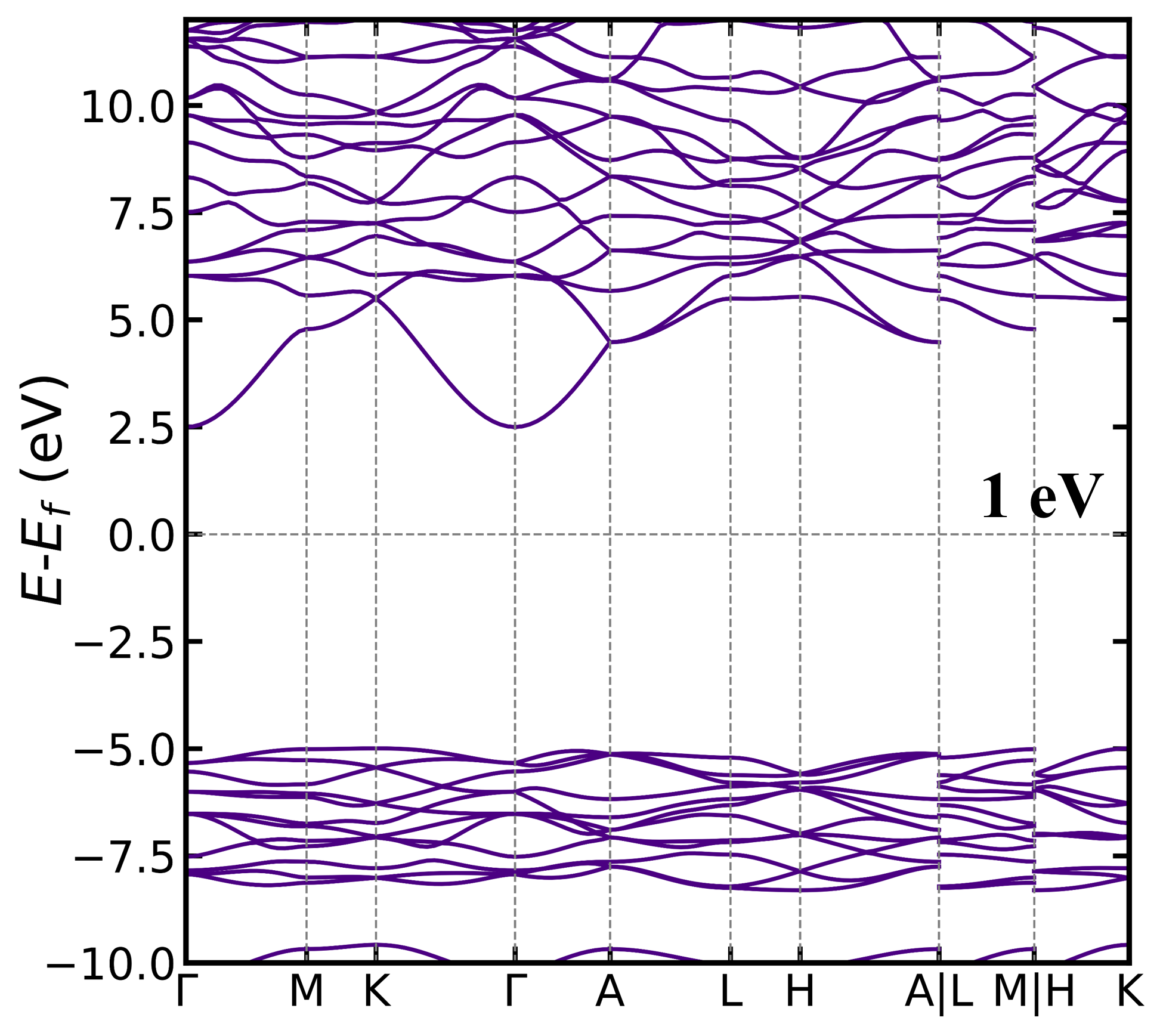}
\caption{$T_e = 1.0$~eV}
\end{subfigure}

\vspace{0.2cm}

\begin{subfigure}[t]{0.48\linewidth}
\includegraphics[height=5.6cm]{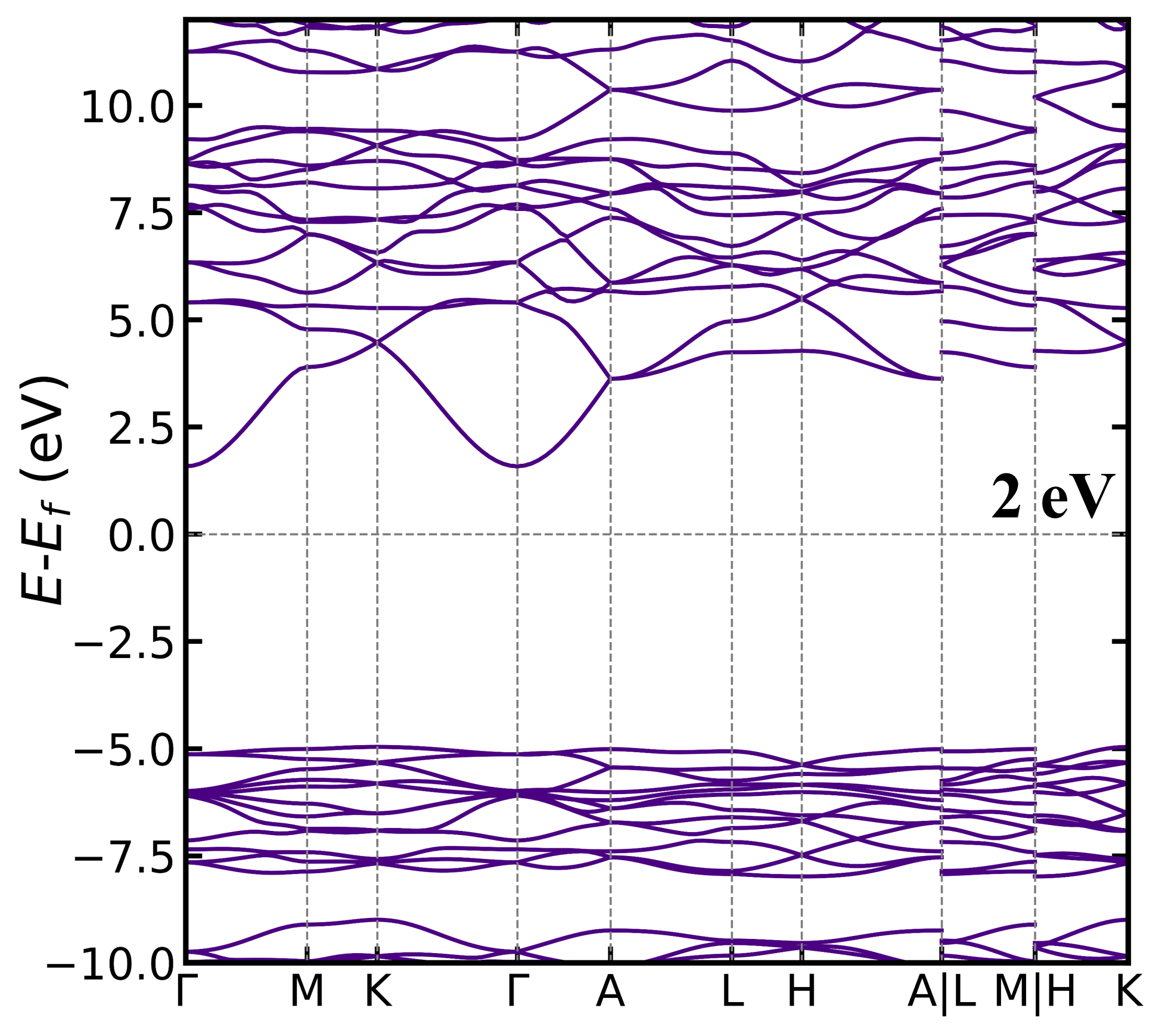}
\caption{$T_e = 2.0$~eV}
\end{subfigure}\hfill
\begin{subfigure}[t]{0.48\linewidth}
\includegraphics[height=5.6cm]{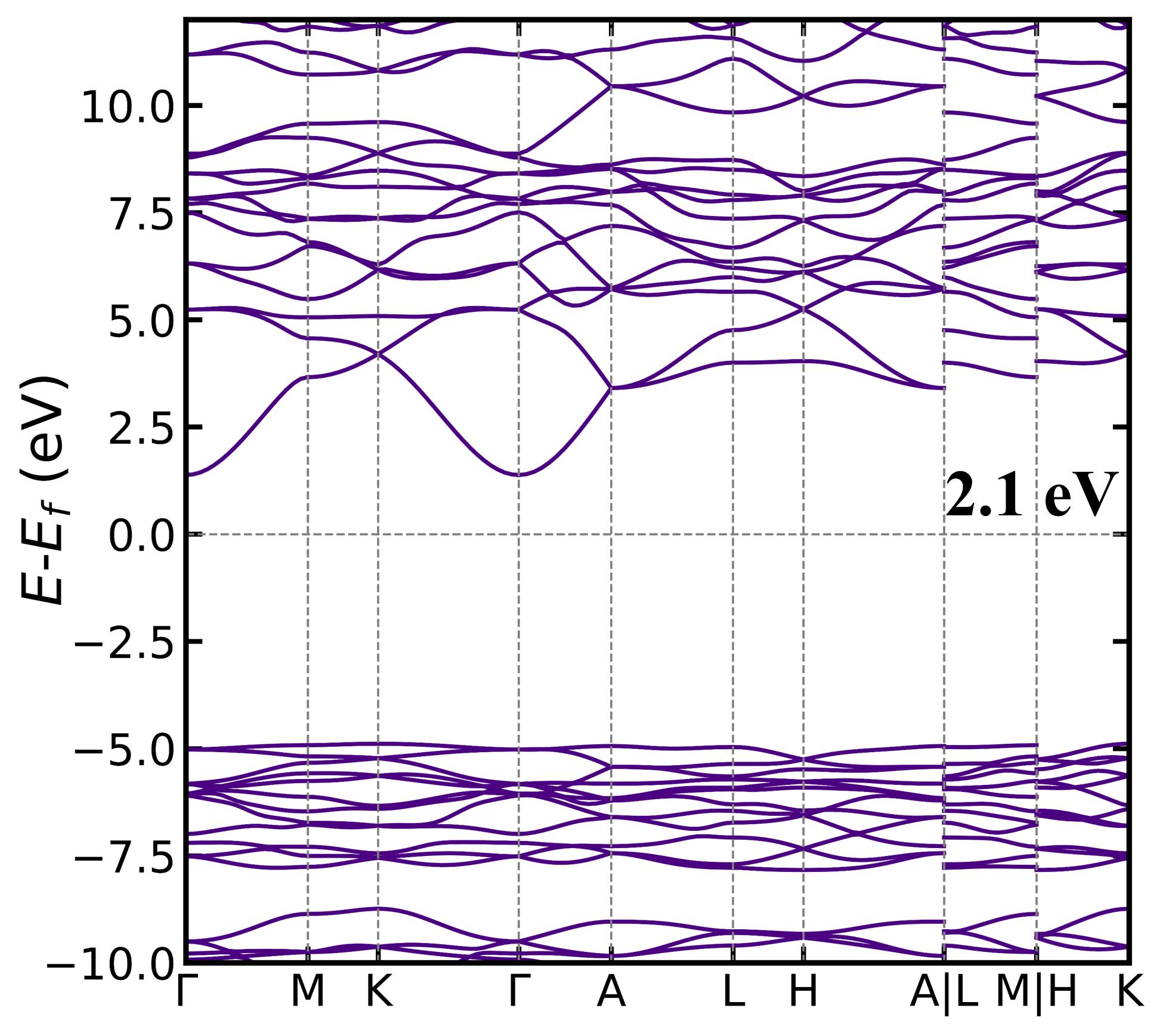}
\caption{$T_e = 2.1$~eV}
\end{subfigure}

\vspace{0.2cm}

\begin{subfigure}[t]{0.48\linewidth}
\includegraphics[height=5.6cm]{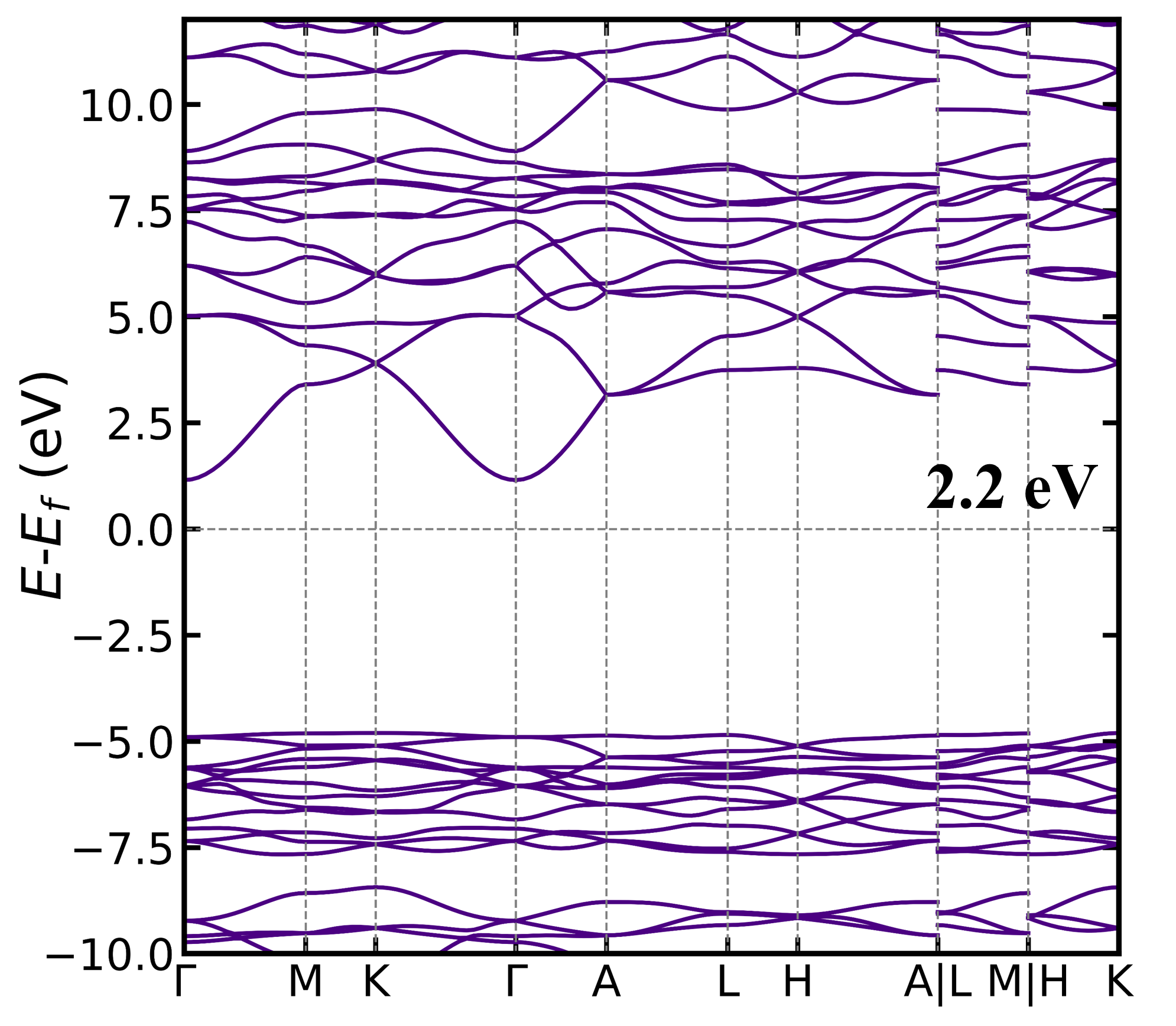}
\caption{$T_e = 2.2$~eV}
\end{subfigure}

\caption{HSE06 hybrid-functional band structures of $\alpha$-SiO$_2$ at the ground state and under electronic excitation corresponding to an electronic temperature range of $1.0$--$2.2$~eV.}
\label{fig:SI_band}
\end{figure}

\vspace{0.4cm}

\begin{figure}[H]
\centering
\begin{subfigure}[t]{0.48\linewidth}
\centering
\includegraphics[height=4.8cm]{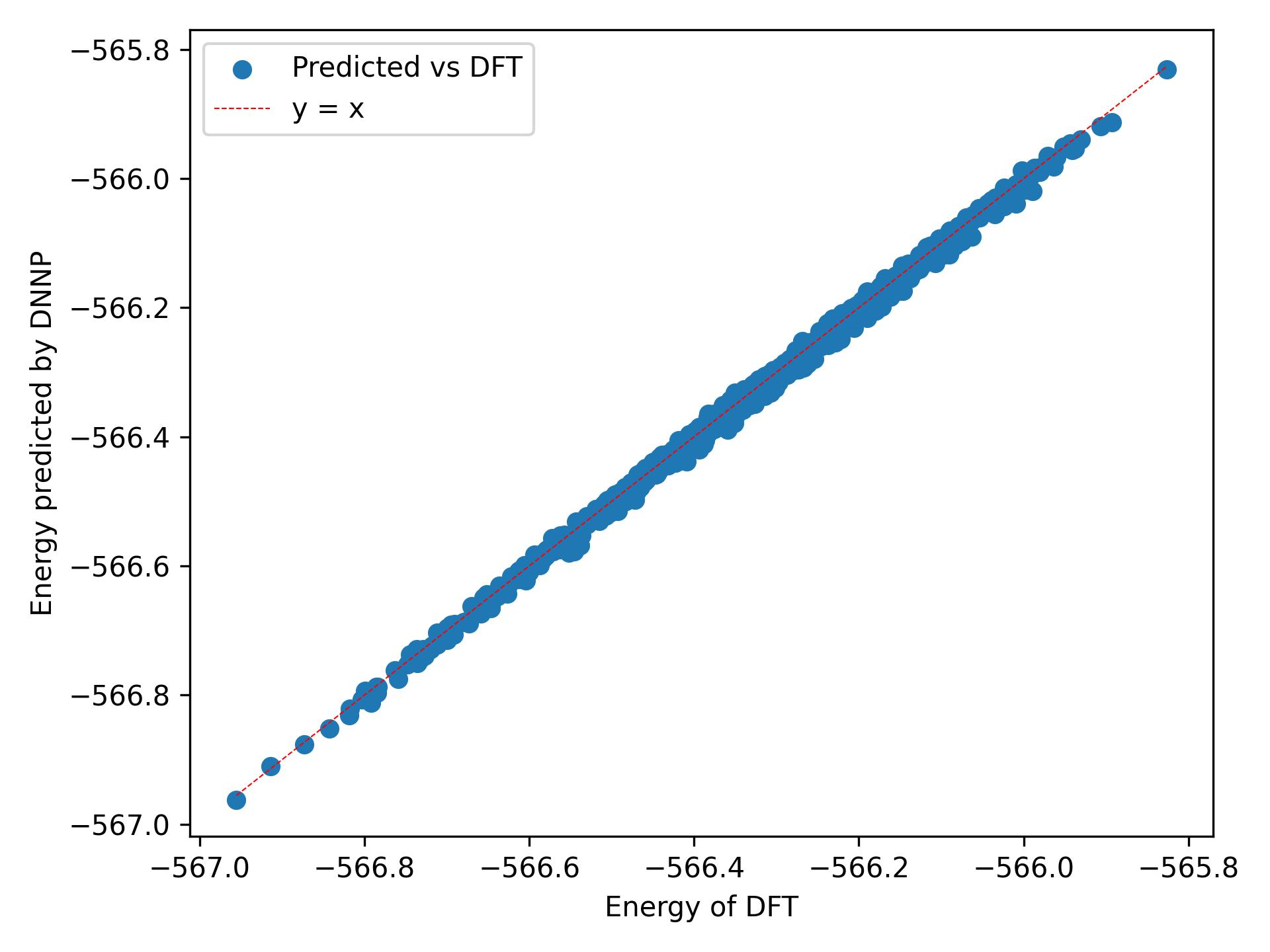}
\caption{Ground state}
\end{subfigure}\hfill
\begin{subfigure}[t]{0.48\linewidth}
\centering
\includegraphics[height=4.8cm]{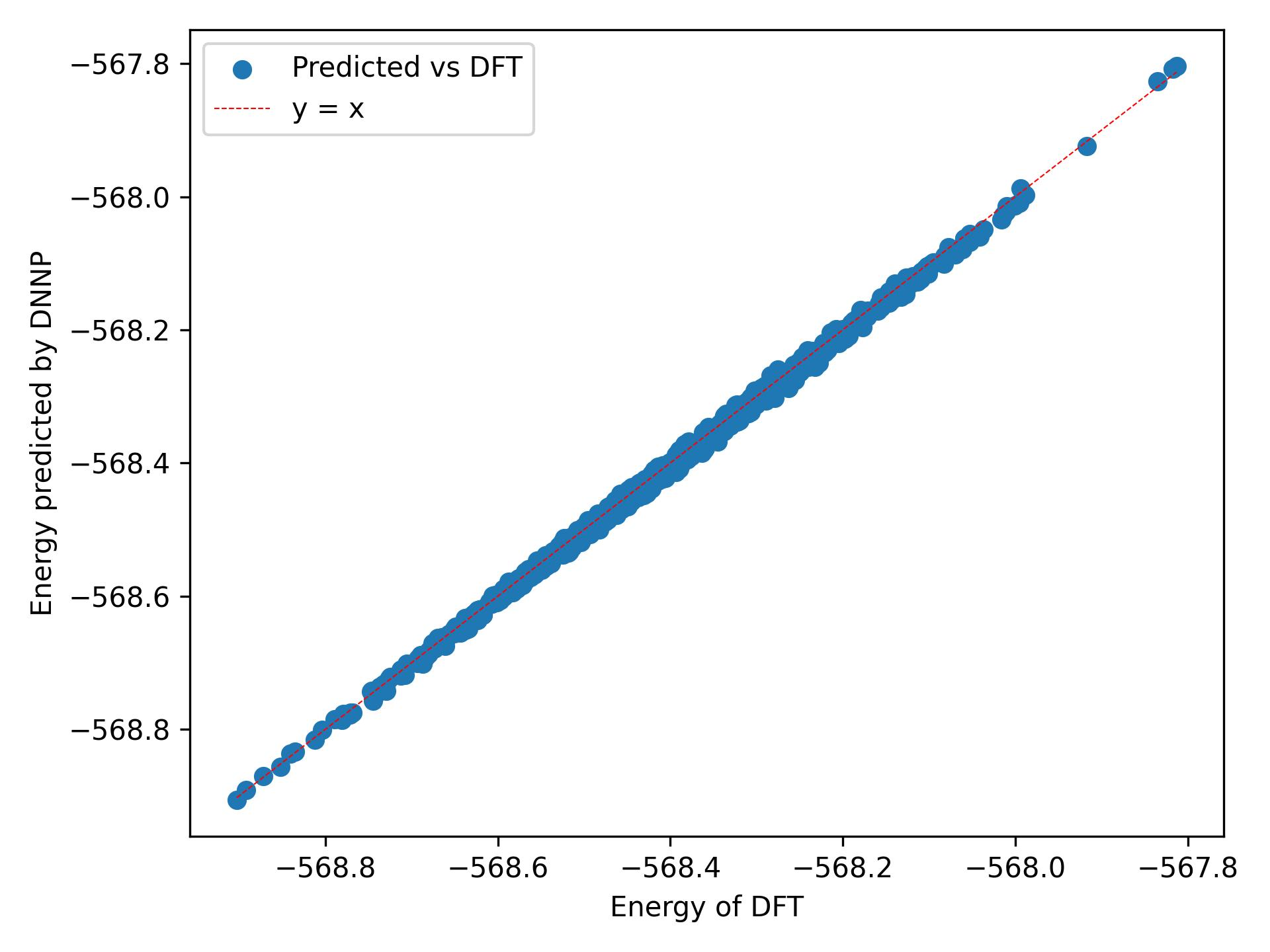}
\caption{$T_e = 1.0$~eV}
\end{subfigure}

\vspace{0.2cm}

\begin{subfigure}[t]{0.48\linewidth}
\centering
\includegraphics[height=4.8cm]{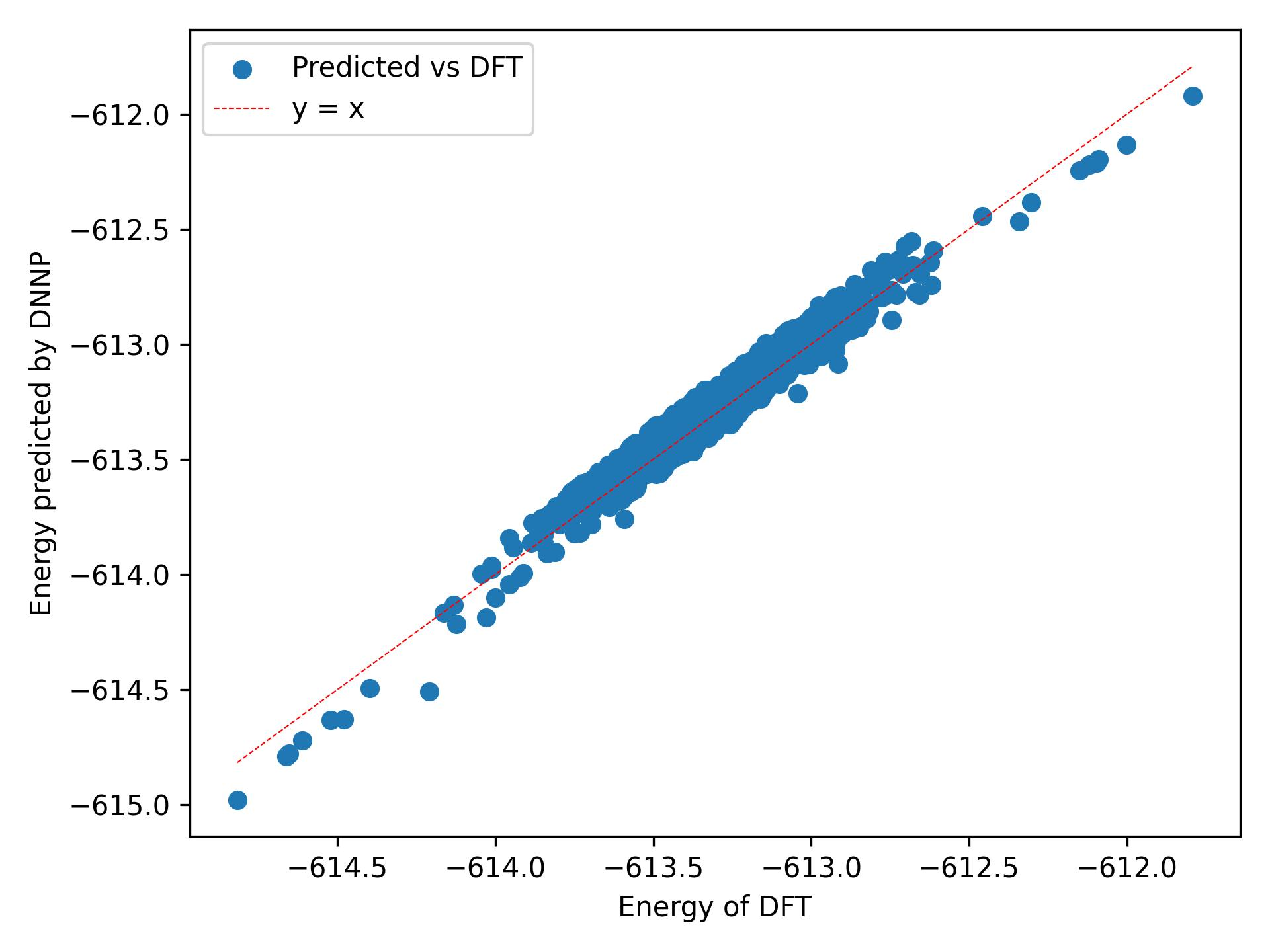}
\caption{$T_e = 2.0$~eV}
\end{subfigure}\hfill
\begin{subfigure}[t]{0.48\linewidth}
\centering
\includegraphics[height=4.8cm]{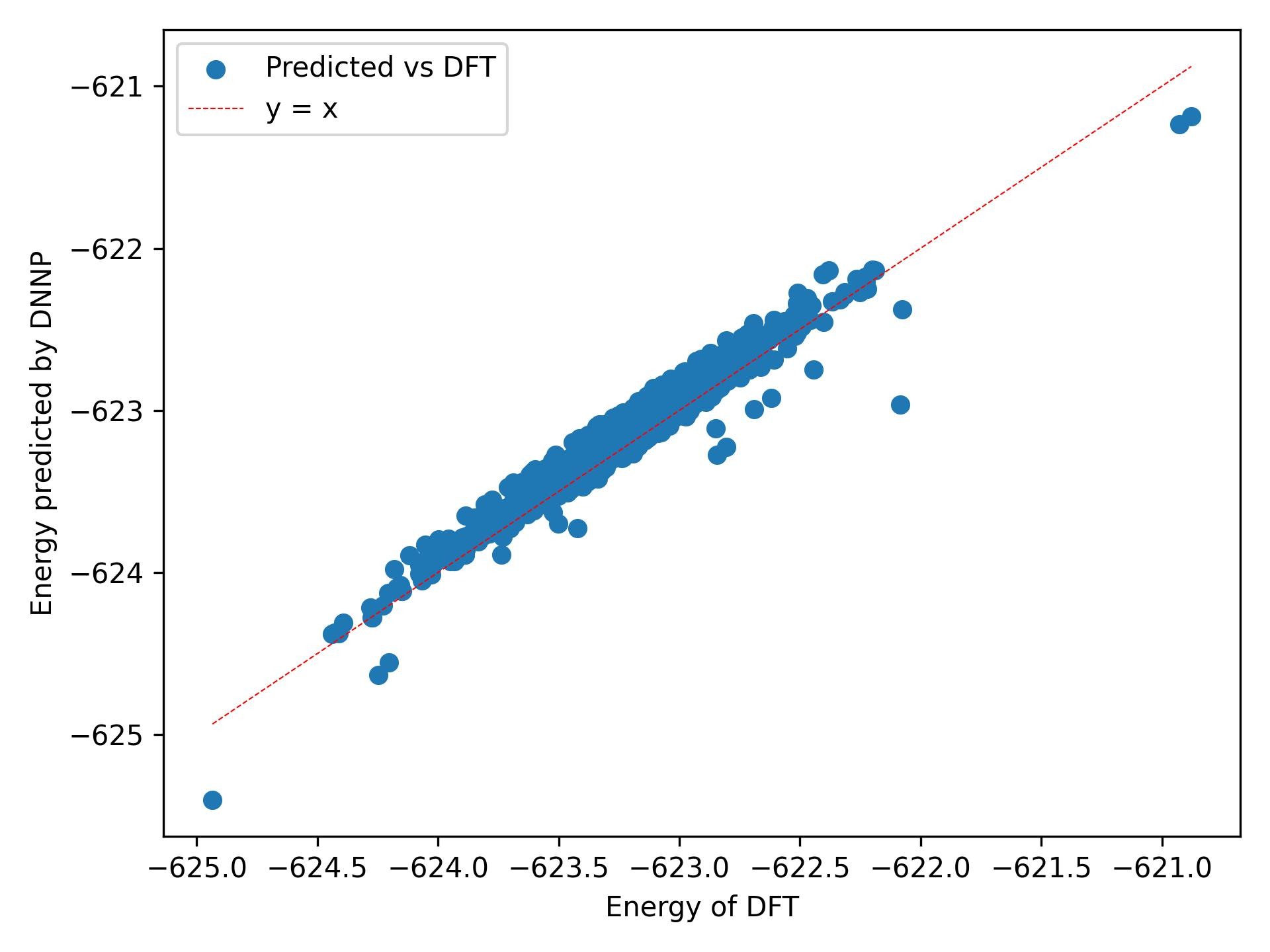}
\caption{$T_e = 2.1$~eV}
\end{subfigure}

\vspace{0.2cm}

\begin{subfigure}[t]{0.48\linewidth}
\centering
\includegraphics[height=4.8cm]{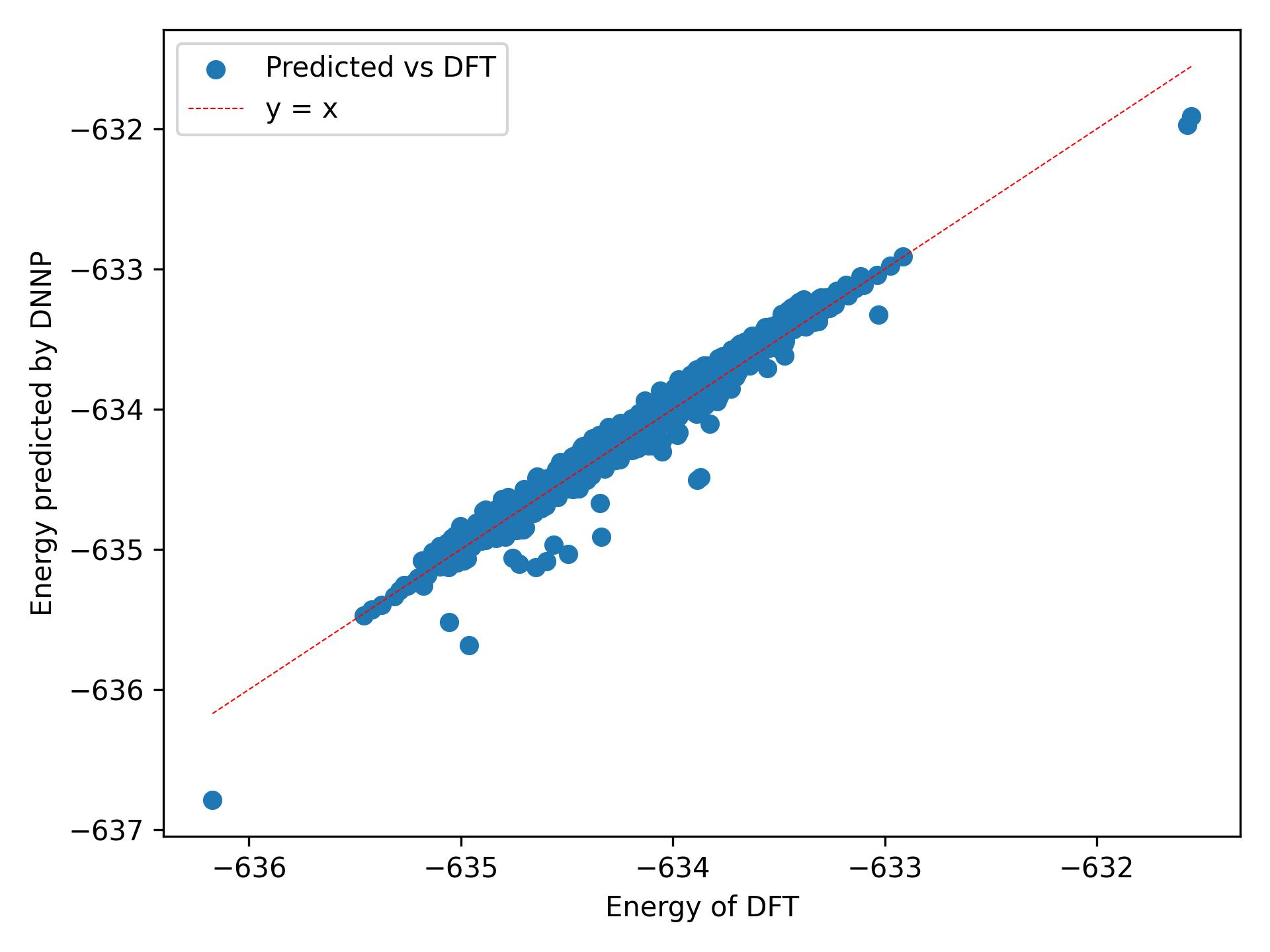}
\caption{$T_e = 2.2$~eV}
\end{subfigure}\hfill
\begin{subfigure}[t]{0.48\linewidth}
\centering
\includegraphics[height=4.8cm]{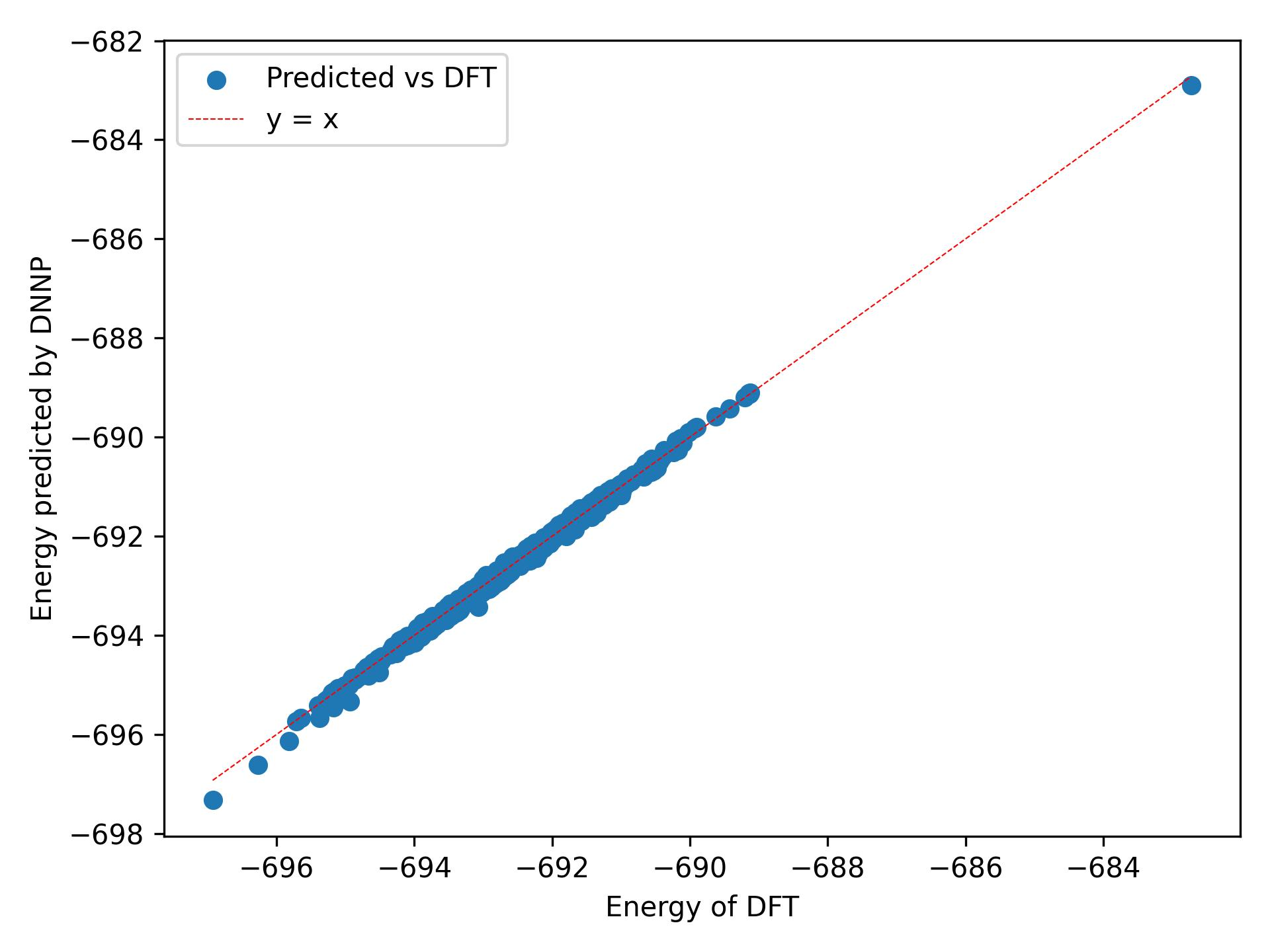}
\caption{$T_e = 2.6$~eV}
\end{subfigure}

\caption{Energy validation of the deep neural network potentials (DNNP) against DFT reference data for $\alpha$-SiO$_2$ at the ground state and under electronic excitation. Each panel compares DNN-predicted energies with corresponding DFT energies at the indicated electronic temperature.}
\label{fig:SI_energy_validation}
\end{figure}

\begin{figure}[H]
\centering

\begin{subfigure}{\linewidth}
\centering
\includegraphics[width=0.9\linewidth]{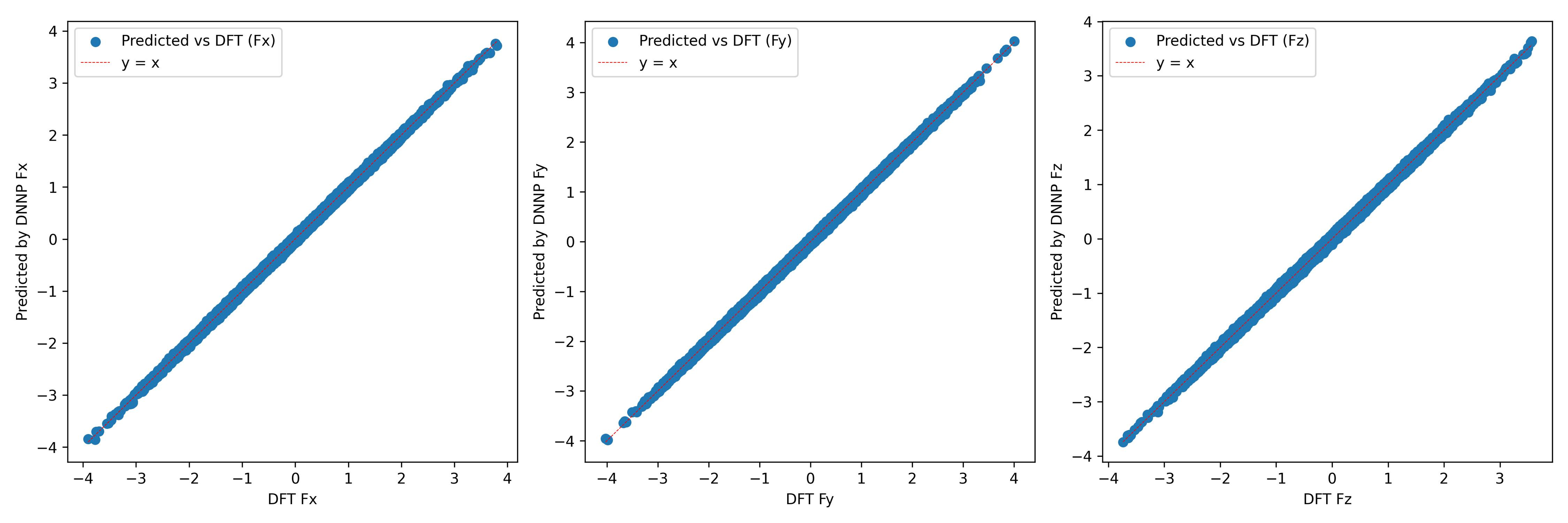}
\caption{Ground state}
\end{subfigure}

\vspace{0.25cm}

\begin{subfigure}{\linewidth}
\centering
\includegraphics[width=0.9\linewidth]{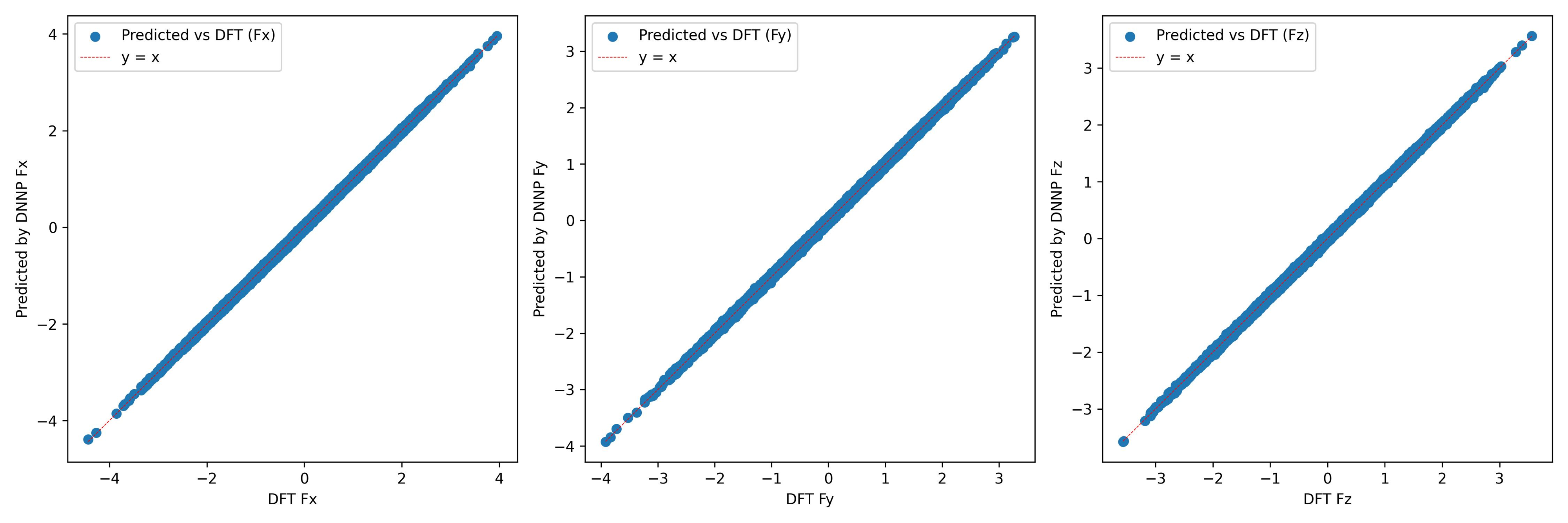}
\caption{$T_e = 1.0$~eV}
\end{subfigure}

\vspace{0.25cm}

\begin{subfigure}{\linewidth}
\centering
\includegraphics[width=0.9\linewidth]{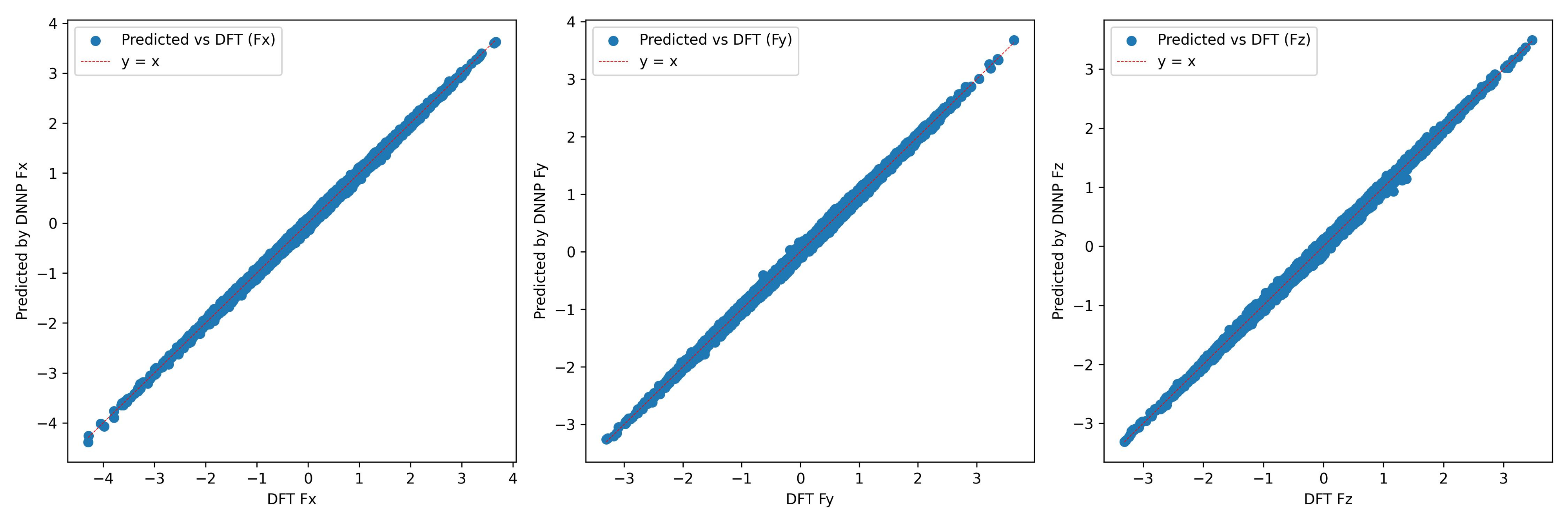}
\caption{$T_e = 2.0$~eV}
\end{subfigure}

\end{figure}

\addtocounter{figure}{-1}

\begin{figure}[H]
\centering

\begin{subfigure}{\linewidth}
\centering
\includegraphics[width=0.9\linewidth]{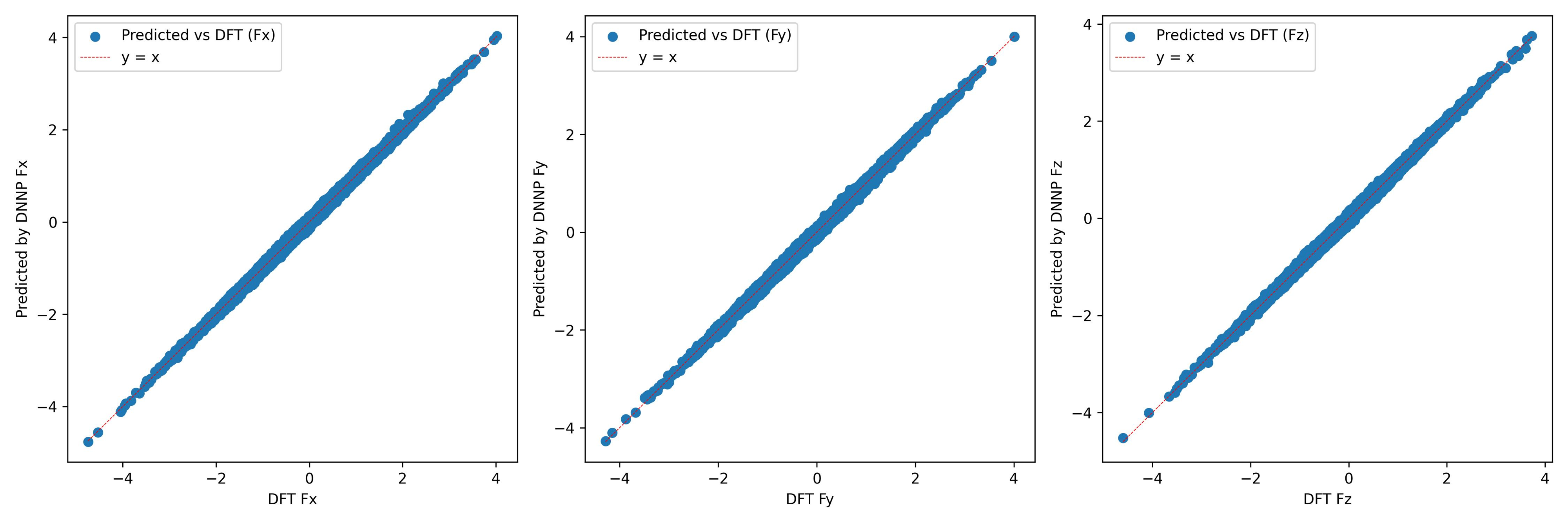}
\caption{$T_e = 2.1$~eV}
\end{subfigure}

\vspace{0.25cm}

\begin{subfigure}{\linewidth}
\centering
\includegraphics[width=0.9\linewidth]{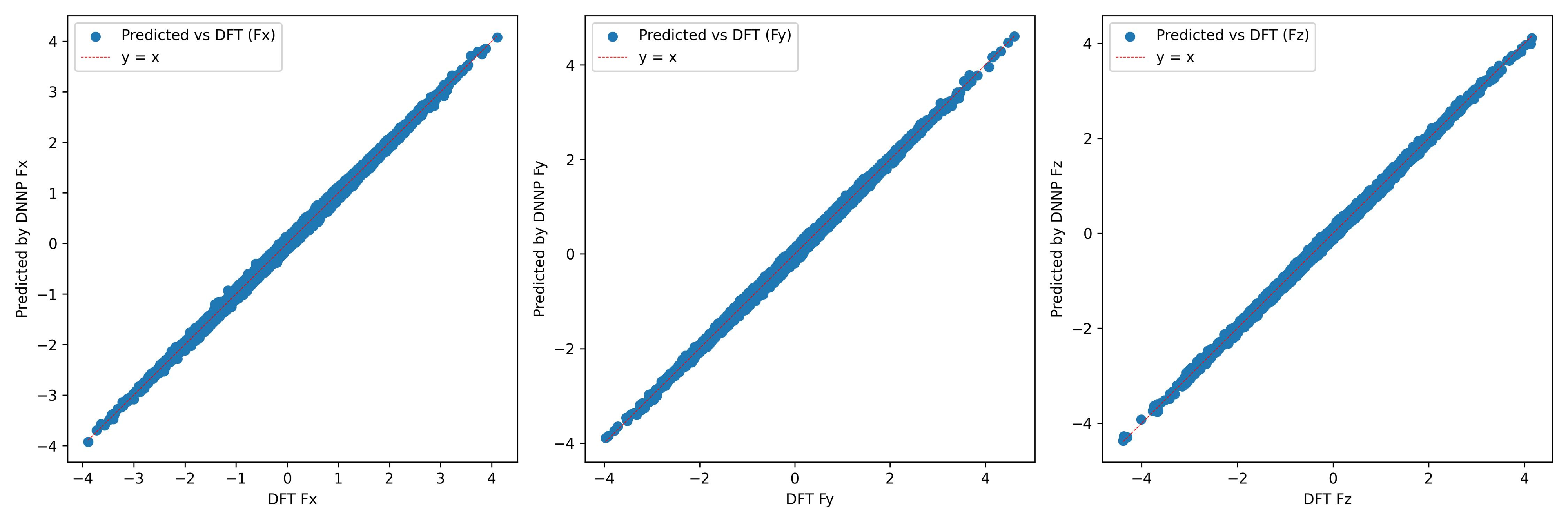}
\caption{$T_e = 2.2$~eV}
\end{subfigure}

\vspace{0.25cm}

\begin{subfigure}{\linewidth}
\centering
\includegraphics[width=0.9\linewidth]{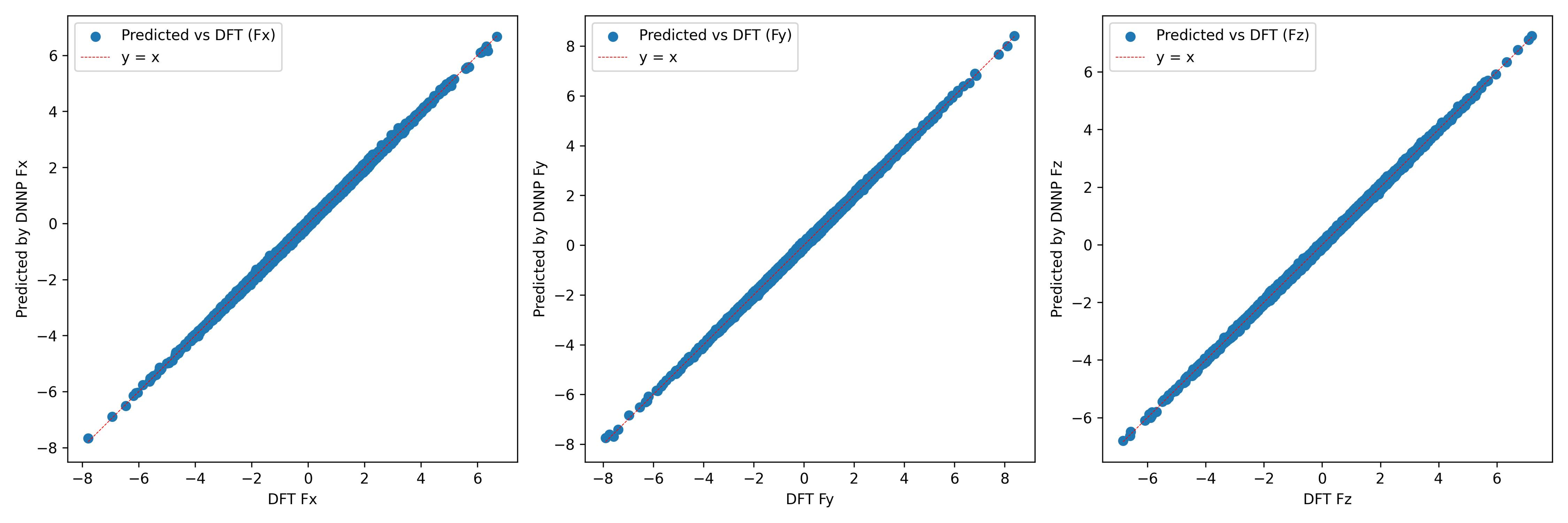}
\caption{$T_e = 2.6$~eV}
\end{subfigure}

\caption{Force validation of the DNNPs against DFT reference data for $\alpha$-SiO$_2$. Each panel shows parity plots comparing DNN-predicted and DFT forces ($F_x$, $F_y$, and $F_z$).}
\label{fig:SI_force_validation}
\end{figure}

\begin{table}[H]
\centering
\caption{Mean absolute error (MAE) and root-mean-square error (RMSE) of the DNNPs with respect to DFT reference data for $\alpha$-SiO$_2$ at different electronic temperatures. Energy and virial errors are reported both per simulation cell and per atom, while force errors are reported per Cartesian component.}
\label{tab:SI_MAE_RMSE}
\begin{tabular}{lcccccc}
\toprule
$T_e$ (eV) & Property & MAE & RMSE & MAE / atom & RMSE / atom \\
\midrule
Ground 
& Energy (eV) & $7.21\times10^{-3}$ & $9.33\times10^{-3}$ & $1.00\times10^{-4}$ & $1.30\times10^{-4}$ \\
& Force (eV/\AA) & $1.81\times10^{-2}$ & $2.32\times10^{-2}$ & -- & -- \\
& Virial (eV) & $9.22\times10^{-2}$ & $1.19\times10^{-1}$ & $1.28\times10^{-3}$ & $1.65\times10^{-3}$ \\
\midrule
1.0 
& Energy (eV) & $6.37\times10^{-3}$ & $7.87\times10^{-3}$ & $8.84\times10^{-5}$ & $1.09\times10^{-4}$ \\
& Force (eV/\AA) & $1.36\times10^{-2}$ & $1.74\times10^{-2}$ & -- & -- \\
& Virial (eV) & $6.91\times10^{-2}$ & $8.97\times10^{-2}$ & $9.59\times10^{-4}$ & $1.25\times10^{-3}$ \\
\midrule
2.0 
& Energy (eV) & $5.97\times10^{-2}$ & $7.05\times10^{-2}$ & $8.29\times10^{-4}$ & $9.79\times10^{-4}$ \\
& Force (eV/\AA) & $2.21\times10^{-2}$ & $2.86\times10^{-2}$ & -- & -- \\
& Virial (eV) & $1.65\times10^{-1}$ & $2.16\times10^{-1}$ & $2.30\times10^{-3}$ & $3.00\times10^{-3}$ \\
\midrule
2.1 
& Energy (eV) & $8.78\times10^{-2}$ & $1.12\times10^{-1}$ & $1.22\times10^{-3}$ & $1.56\times10^{-3}$ \\
& Force (eV/\AA) & $2.77\times10^{-2}$ & $3.51\times10^{-2}$ & -- & -- \\
& Virial (eV) & $2.23\times10^{-1}$ & $2.99\times10^{-1}$ & $3.10\times10^{-3}$ & $4.15\times10^{-3}$ \\
\midrule
2.2 
& Energy (eV) & $6.33\times10^{-2}$ & $9.43\times10^{-2}$ & $8.79\times10^{-4}$ & $1.31\times10^{-3}$ \\
& Force (eV/\AA) & $2.80\times10^{-2}$ & $3.58\times10^{-2}$ & -- & -- \\
& Virial (eV) & $2.31\times10^{-1}$ & $3.07\times10^{-1}$ & $3.21\times10^{-3}$ & $4.26\times10^{-3}$ \\
\midrule
2.6 
& Energy (eV) & $5.24\times10^{-2}$ & $6.95\times10^{-2}$ & $7.28\times10^{-4}$ & $9.65\times10^{-4}$ \\
& Force (eV/\AA) & $2.21\times10^{-2}$ & $2.88\times10^{-2}$ & -- & -- \\
& Virial (eV) & $2.01\times10^{-1}$ & $2.60\times10^{-1}$ & $2.79\times10^{-3}$ & $3.61\times10^{-3}$ \\
\bottomrule
\end{tabular}
\end{table}

\vspace{0.4cm}

\begin{figure}[H]
\centering

\begin{subfigure}[t]{0.48\linewidth}
\centering
\includegraphics[height=5.6cm]{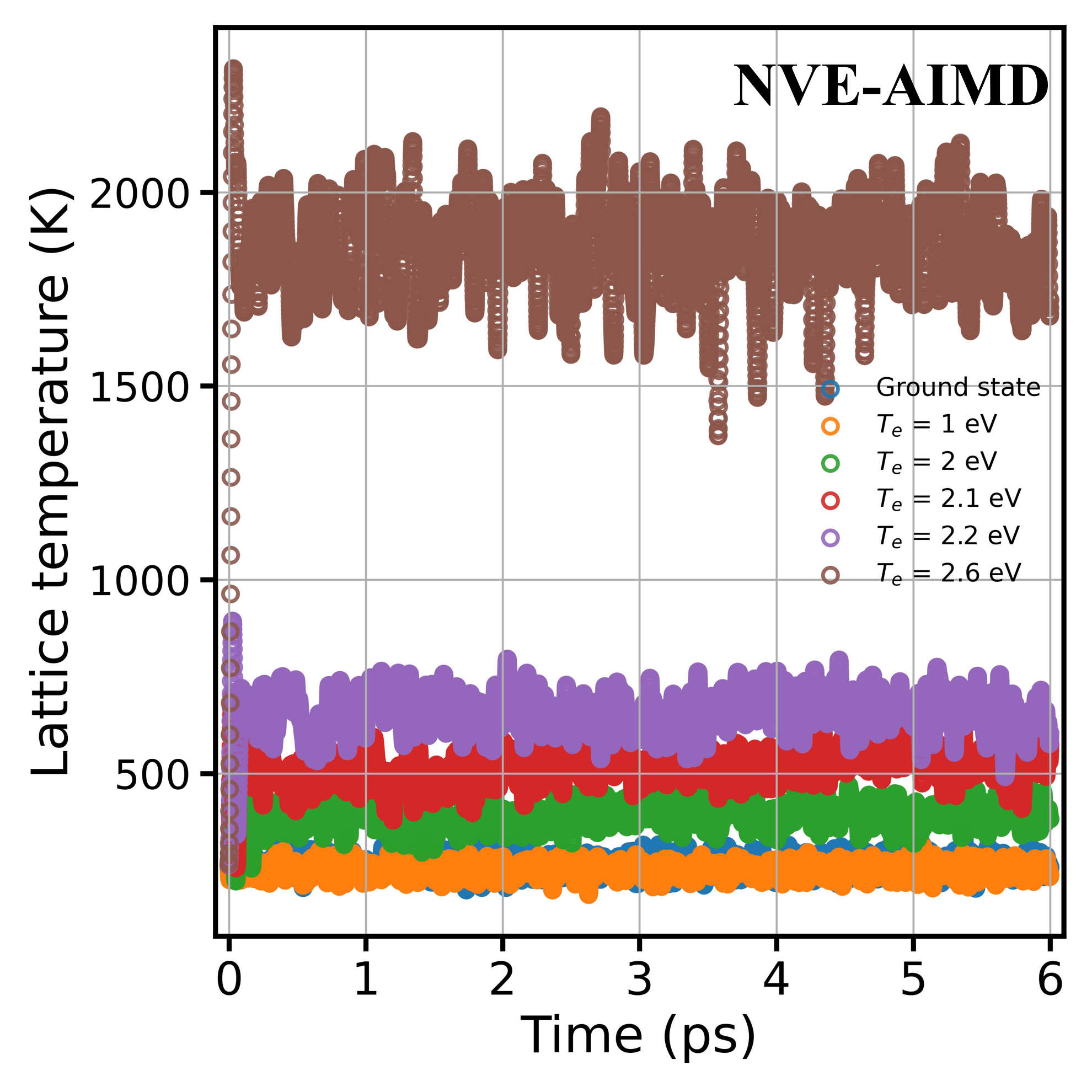}
\caption{NVE-AIMD (up to 6~ps)}
\end{subfigure}\hfill
\begin{subfigure}[t]{0.48\linewidth}
\centering
\includegraphics[height=5.6cm]{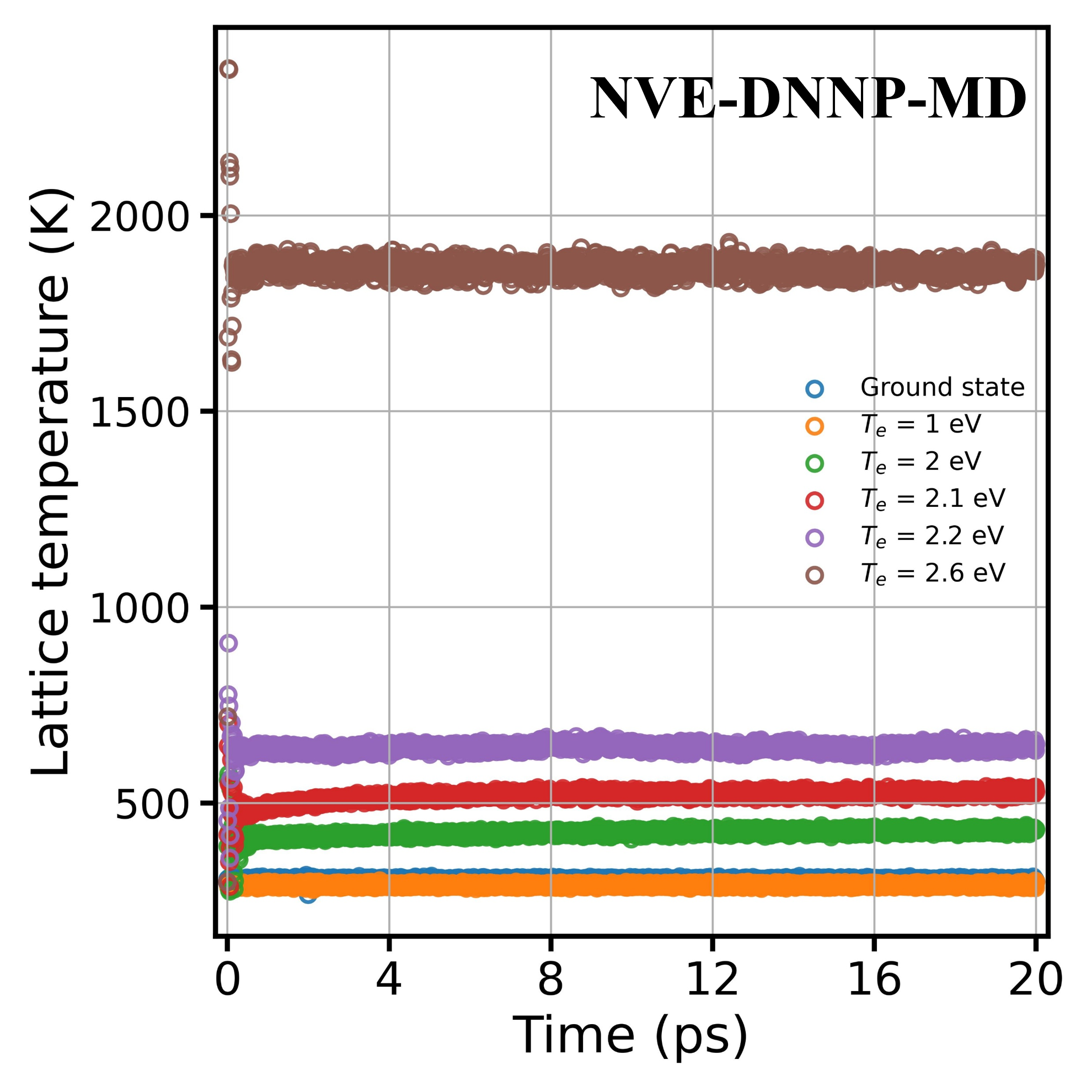}
\caption{NVE-DNNP-MD (up to 20~ps)}
\end{subfigure}

\caption{Extended-time lattice temperature evolution in $\alpha$-SiO$_2$ under electronic excitation obtained from NVE-AIMD (a) and NVE-DNNP-MD (b) simulations. The longer simulation times highlight the subsequent relaxation and fluctuation behavior beyond the initial ultrafast heating regime discussed in the main text.}
\label{fig:SI_lattice_temperature_long}
\end{figure}

\vspace{0.4cm}

\begin{figure}[H]
\centering

\begin{subfigure}[t]{0.48\linewidth}
\centering
\includegraphics[height=5.6cm]{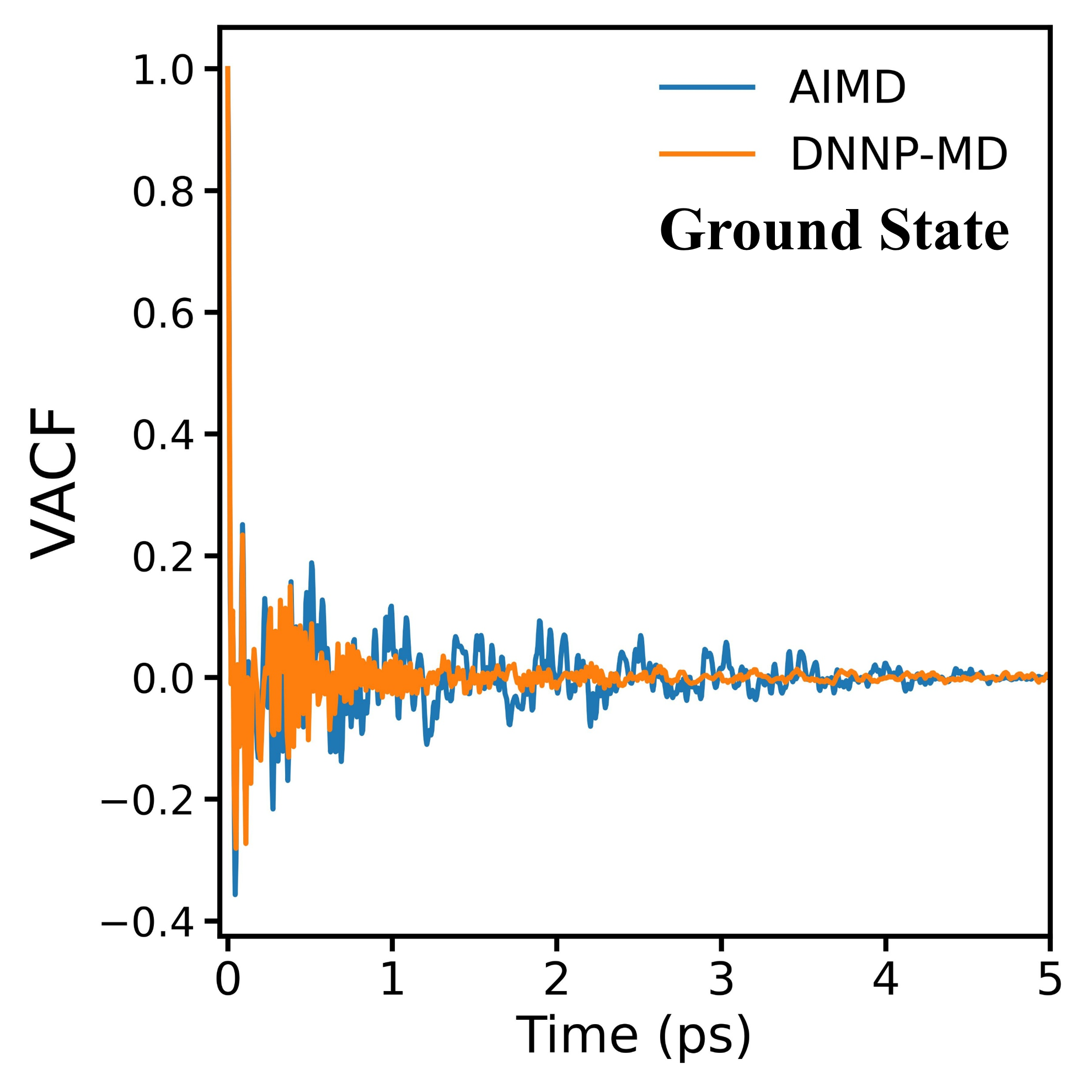}
\caption{Ground state}
\end{subfigure}\hfill
\begin{subfigure}[t]{0.48\linewidth}
\centering
\includegraphics[height=5.6cm]{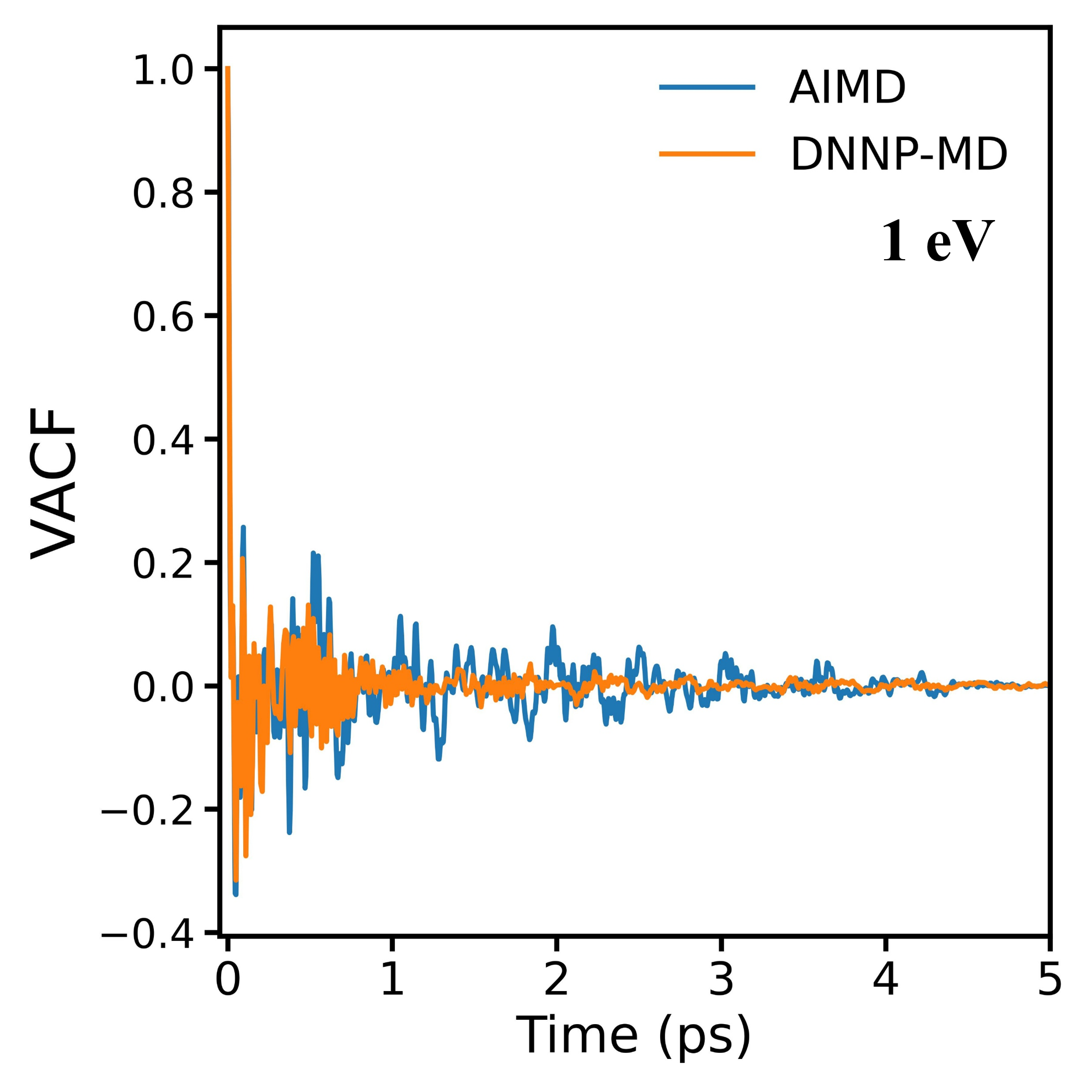}
\caption{$T_e = 1.0$~eV}
\end{subfigure}

\vspace{0.2cm}

\begin{subfigure}[t]{0.48\linewidth}
\centering
\includegraphics[height=5.6cm]{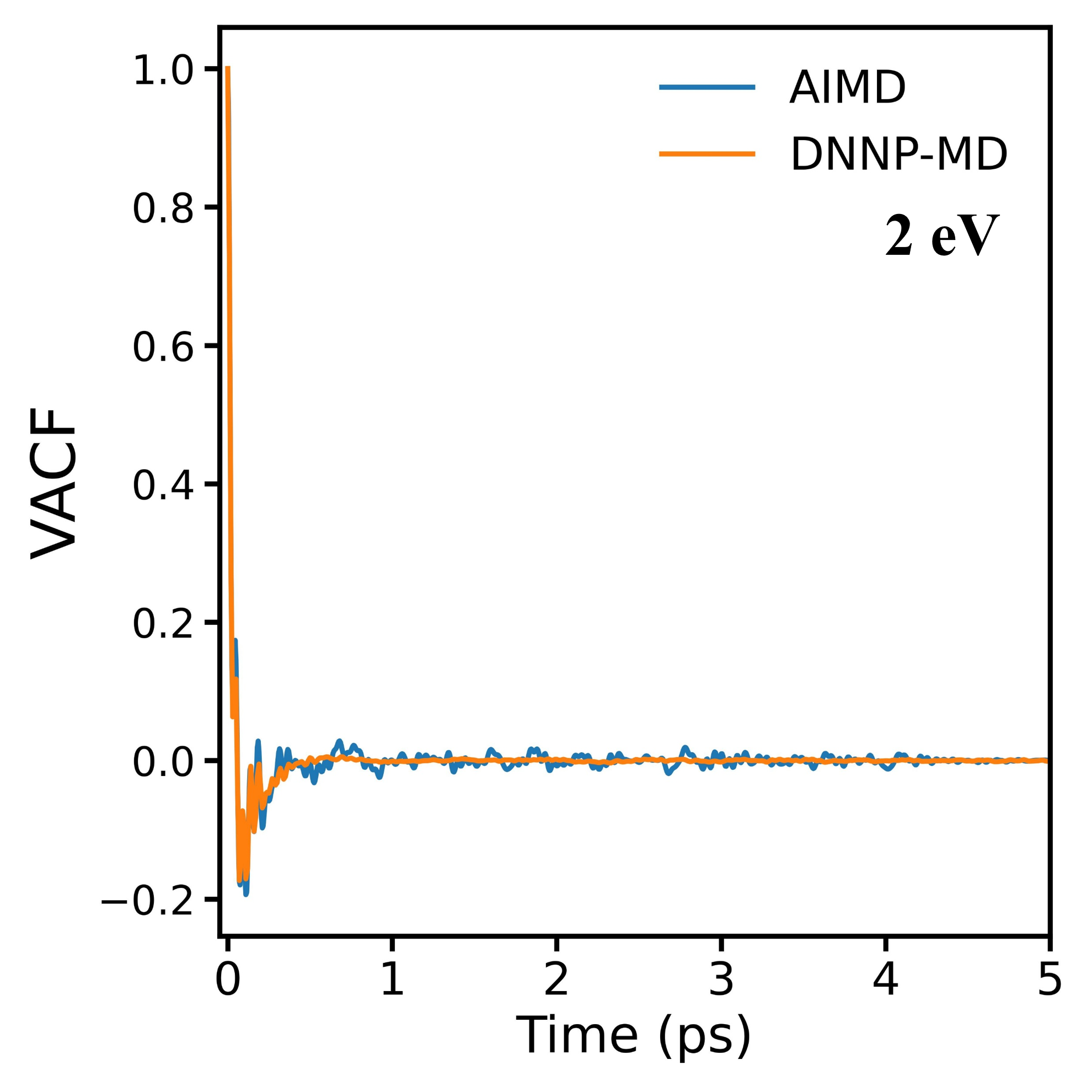}
\caption{$T_e = 2.0$~eV}
\end{subfigure}\hfill
\begin{subfigure}[t]{0.48\linewidth}
\centering
\includegraphics[height=5.6cm]{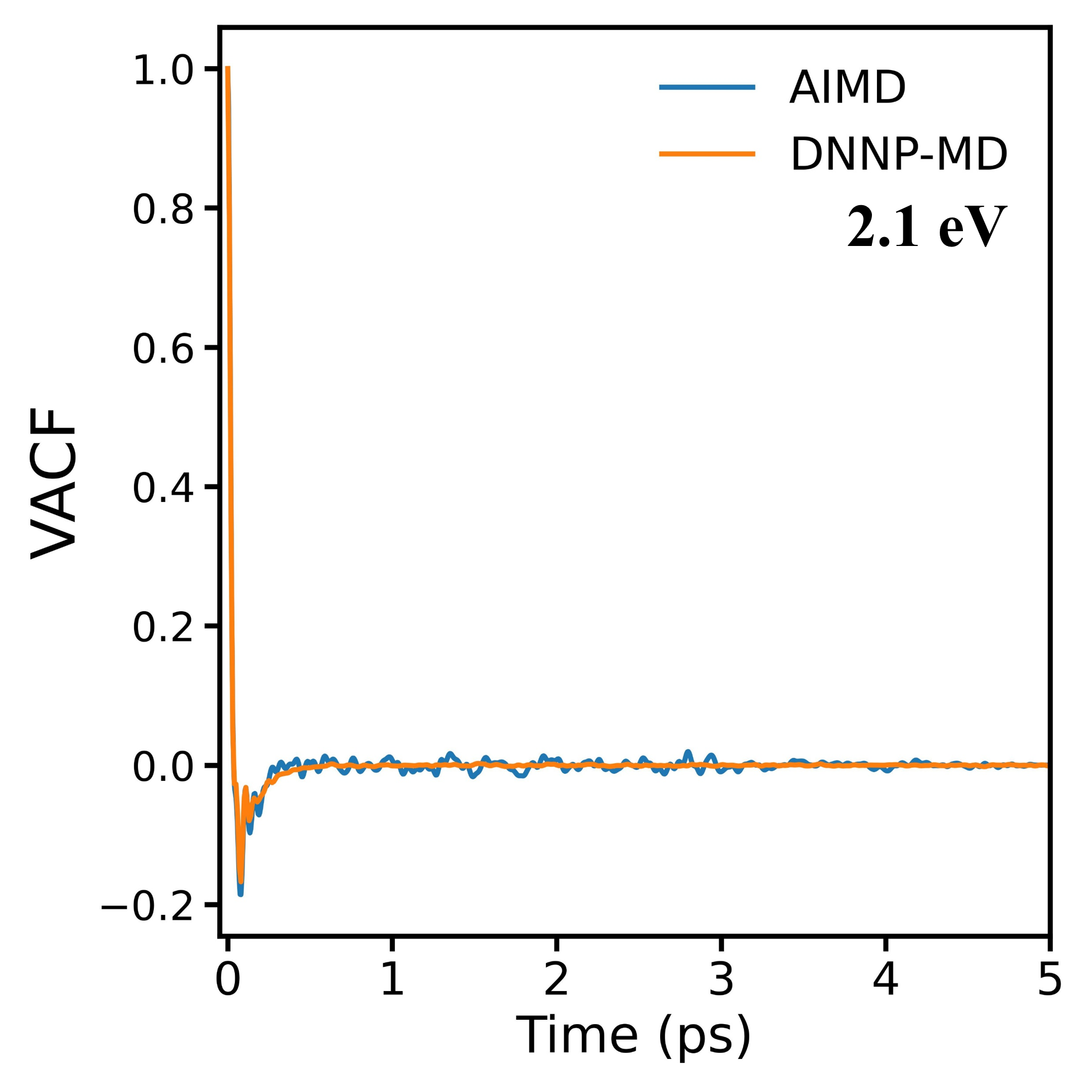}
\caption{$T_e = 2.1$~eV}
\end{subfigure}

\vspace{0.2cm}

\begin{subfigure}[t]{0.48\linewidth}
\centering
\includegraphics[height=5.6cm]{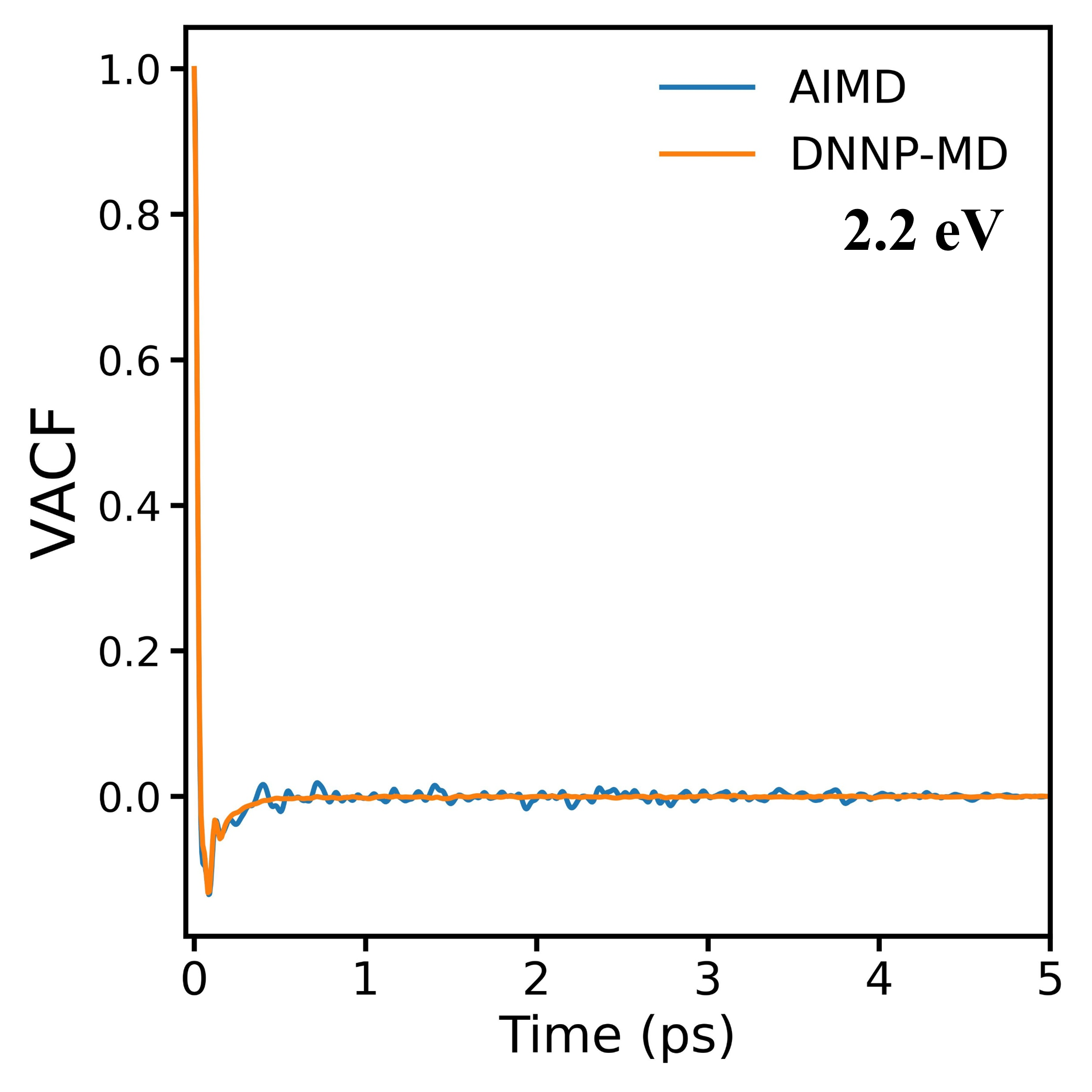}
\caption{$T_e = 2.2$~eV}
\end{subfigure}\hfill
\begin{subfigure}[t]{0.48\linewidth}
\centering
\includegraphics[height=5.6cm]{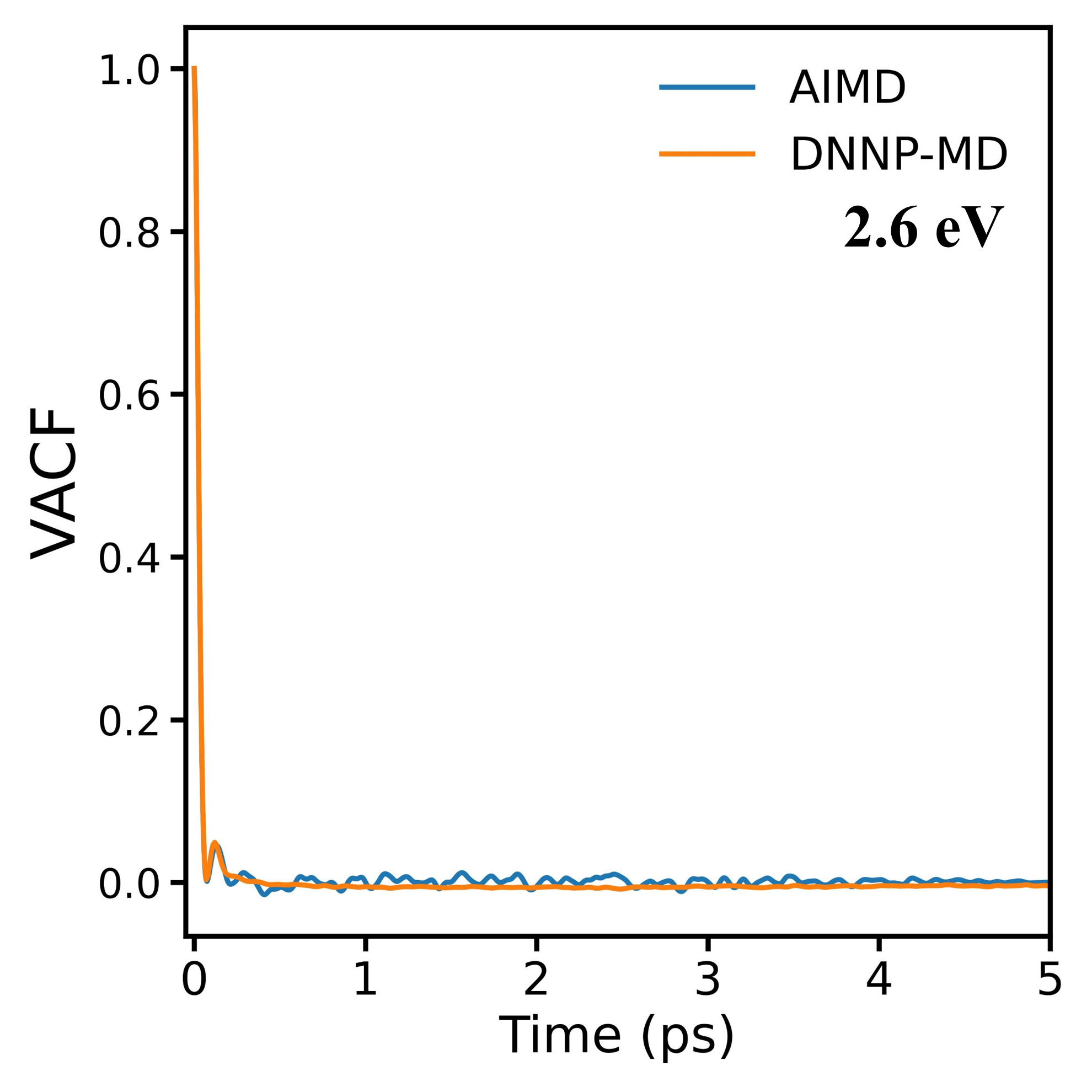}
\caption{$T_e = 2.6$~eV}
\end{subfigure}

\caption{Velocity autocorrelation functions (VACFs) of $\alpha$-SiO$_2$ at the ground state and under electronic excitation corresponding to an electronic temperature range of $1.0$--$2.6$~eV.}
\label{fig:SI_VACF}
\end{figure}

\subsection*{Maxwell-Boltzmann velocity distribution analysis}

To analyze the nonequilibrium lattice response under electronic excitation, atomic velocity distributions of Si and O atoms were extracted from the molecular dynamics trajectories generated using the DeePMD-LAMMPS framework. The simulations were performed with a timestep of 1~fs, and trajectory snapshots were written every 10 MD steps (10~fs interval).

For each selected trajectory frame, the atomic speed magnitude was calculated as

\begin{equation}
v_i = \sqrt{v_{x,i}^2 + v_{y,i}^2 + v_{z,i}^2},
\end{equation}

where $v_{x,i}$, $v_{y,i}$, and $v_{z,i}$ denote the Cartesian velocity components of atom $i$.

The instantaneous temperature associated with each atomic species was estimated from the equipartition relation,

\begin{equation}
T = \frac{m \langle v^2 \rangle}{3 k_{\mathrm{B}}},
\end{equation}

where $m$ is the atomic mass, $k_{\mathrm{B}}$ is the Boltzmann constant, and $\langle v^2 \rangle$ represents the ensemble-averaged squared speed.

The theoretical Maxwell-Boltzmann speed distribution was then evaluated using

\begin{equation}
f(v) =
4\pi
\left(
\frac{m}{2\pi k_{\mathrm{B}} T}
\right)^{3/2}
v^2
\exp
\left(
-\frac{m v^2}{2 k_{\mathrm{B}} T}
\right),
\end{equation}

where $v$ is the atomic speed. For visualization purposes, both the molecular dynamics velocity histograms and the corresponding Maxwell-Boltzmann distributions were normalized and vertically shifted to compare the temporal evolution of the velocity distributions during the initial nonequilibrium lattice dynamics following electronic excitation.

\vspace{0.4cm}

\renewcommand{\thefigure}{S6a}
\begin{figure}[p]
\centering
\includegraphics[width=0.95\linewidth]{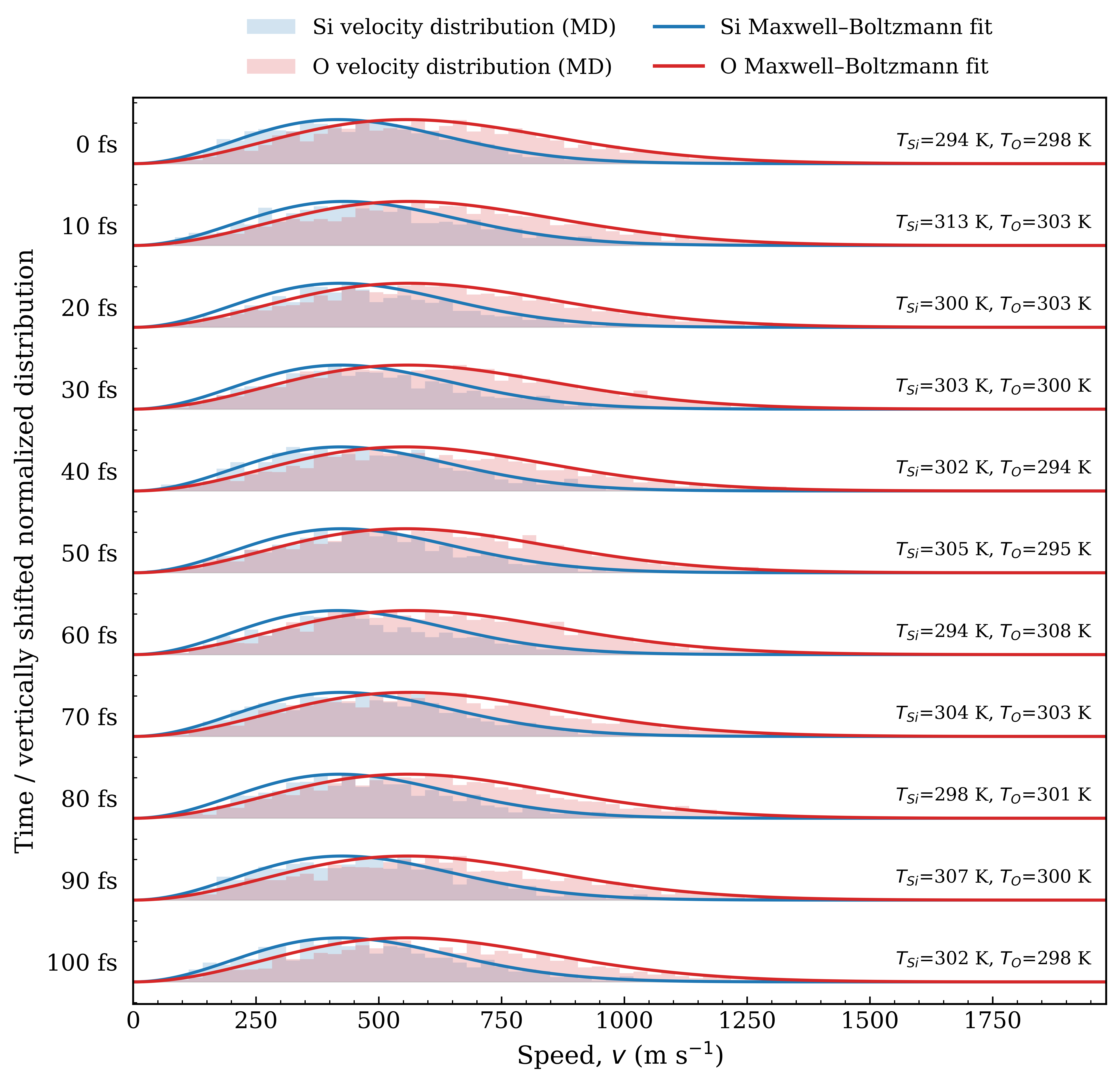}
\caption{Ground state}
\label{fig:S6a}
\end{figure}

\clearpage

\renewcommand{\thefigure}{S6b}
\begin{figure}[p]
\centering
\includegraphics[width=0.95\linewidth]{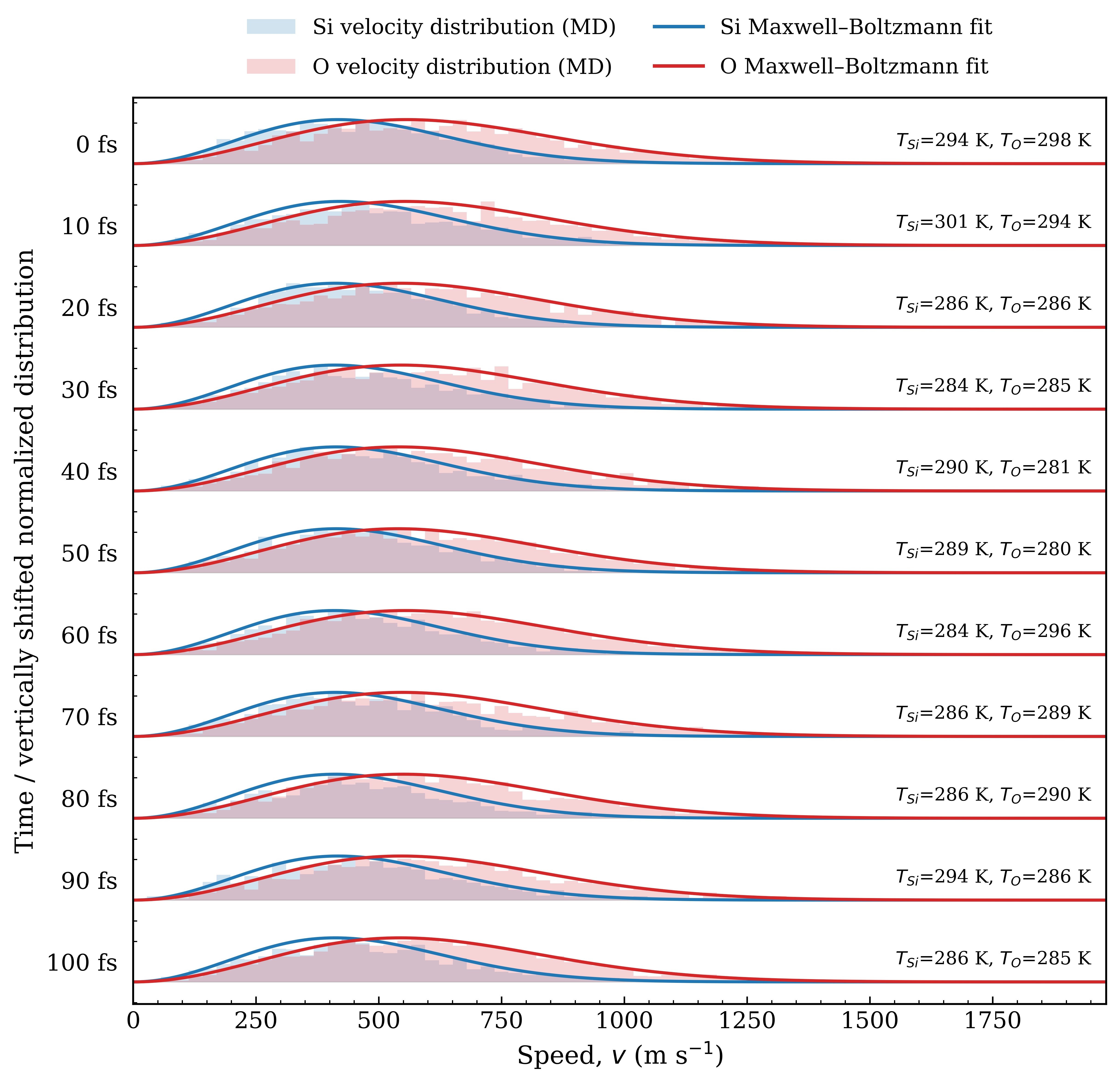}
\caption{$T_e = 1.0$~eV}
\label{fig:S6b}
\end{figure}

\clearpage

\renewcommand{\thefigure}{S6c}
\begin{figure}[p]
\centering
\includegraphics[width=0.95\linewidth]{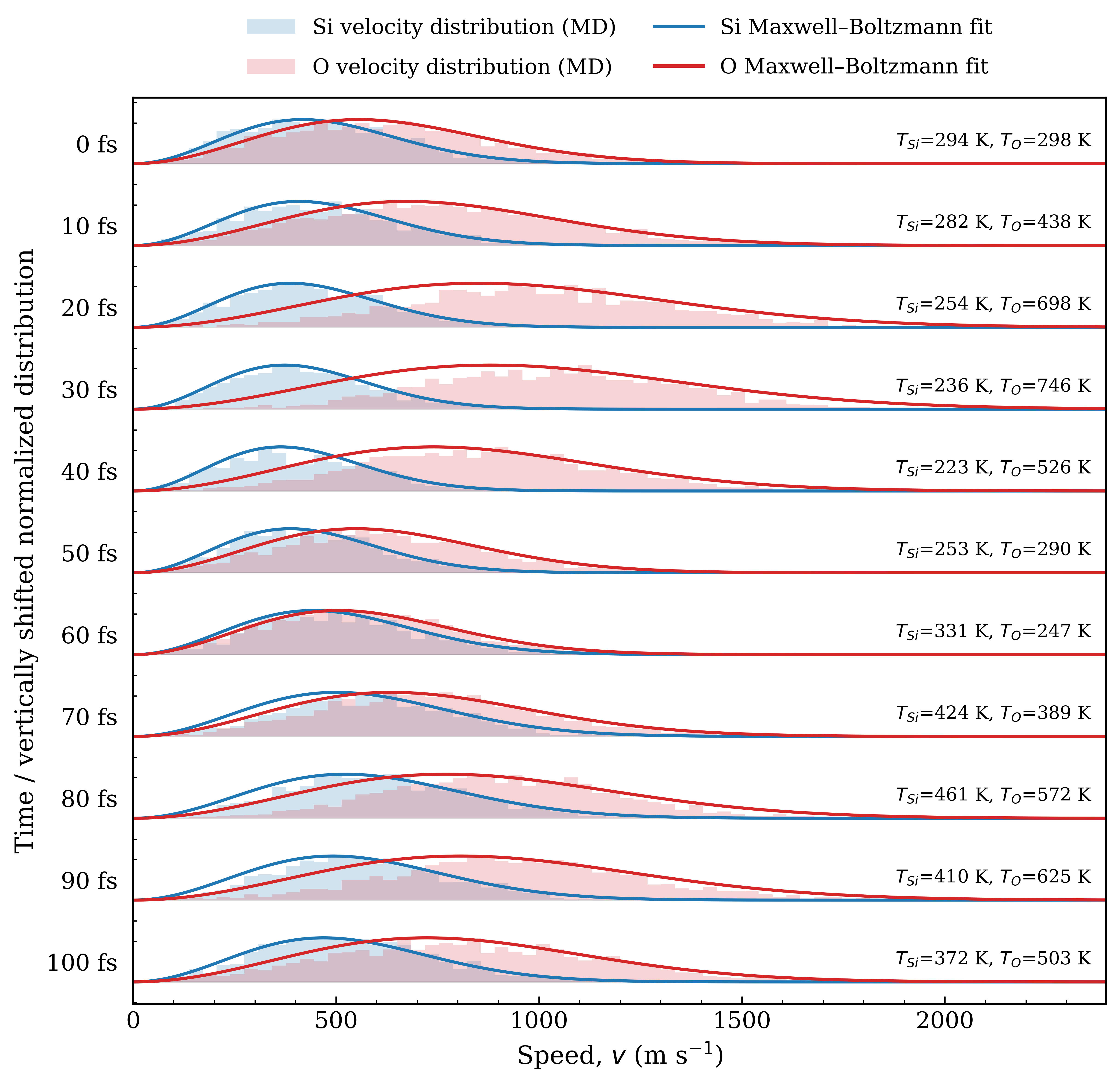}
\caption{$T_e = 2.0$~eV}
\label{fig:S6c}
\end{figure}

\clearpage

\renewcommand{\thefigure}{S6d}
\begin{figure}[p]
\centering
\includegraphics[width=0.95\linewidth]{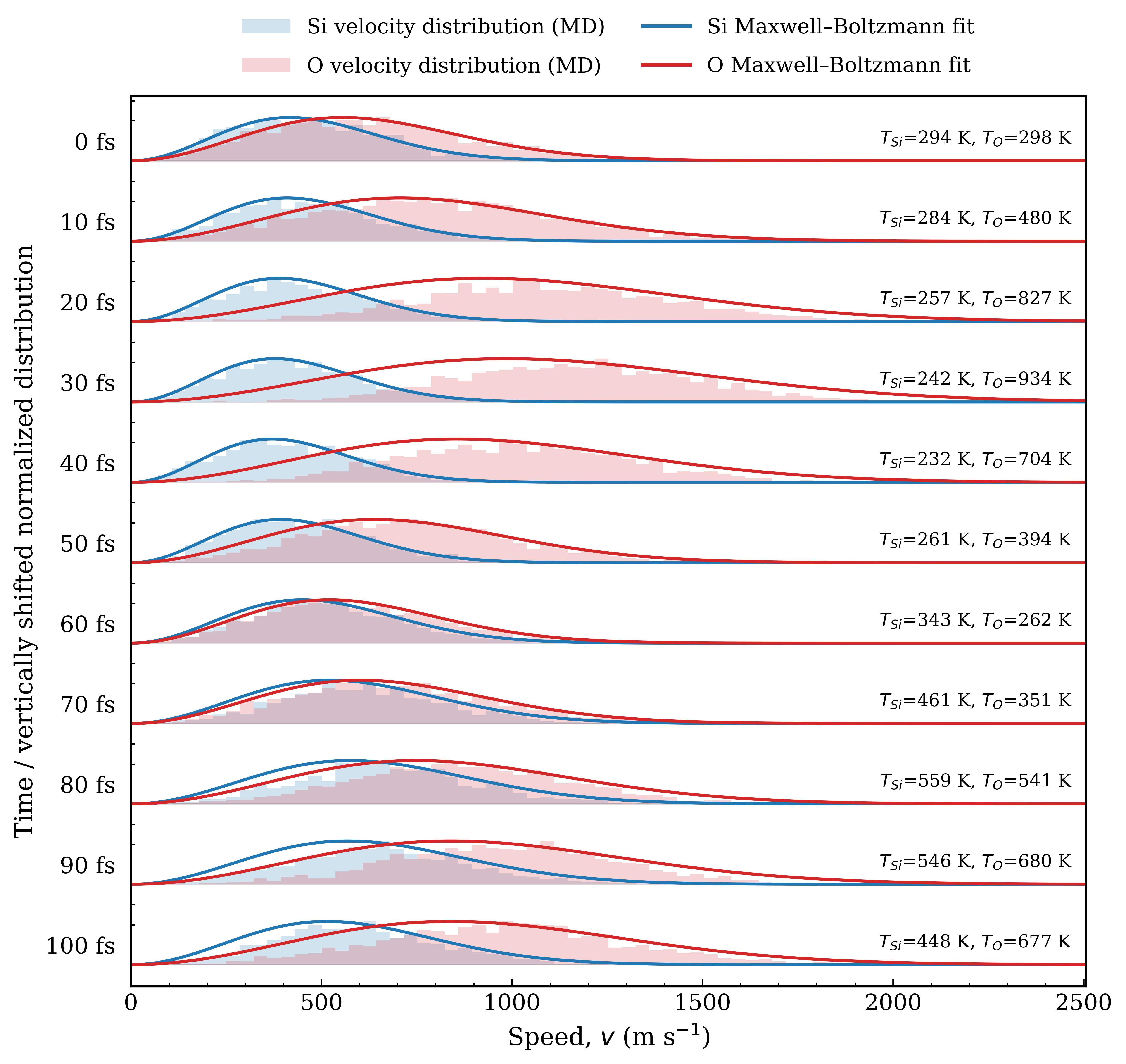}
\caption{$T_e = 2.1$~eV}
\label{fig:S6d}
\end{figure}

\clearpage

\renewcommand{\thefigure}{S6e}
\begin{figure}[p]
\centering
\includegraphics[width=0.95\linewidth]{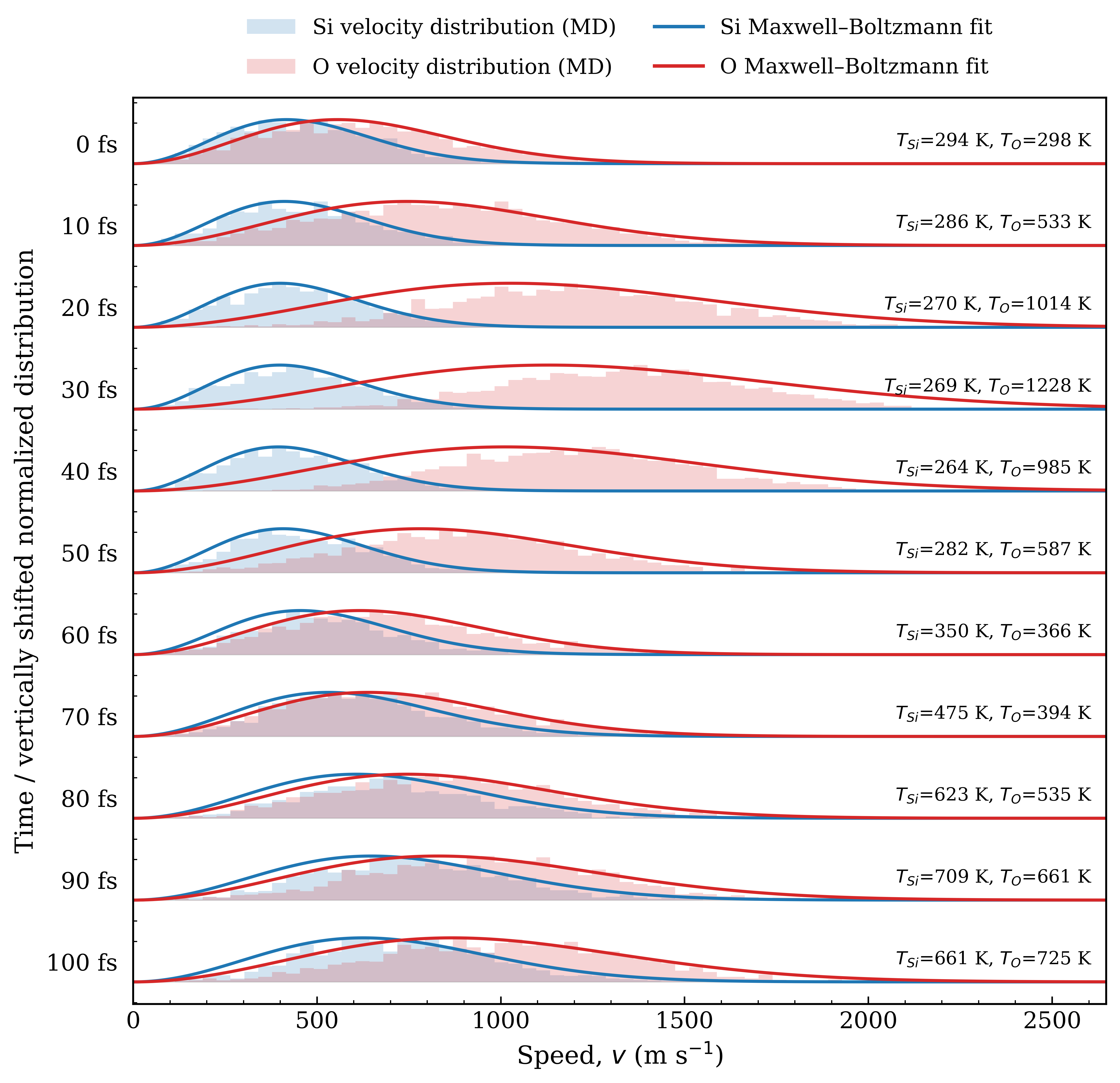}
\caption{$T_e = 2.2$~eV}
\label{fig:S6e}
\end{figure}

\clearpage

\renewcommand{\thefigure}{S6f}
\begin{figure}[p]
\centering
\includegraphics[width=0.95\linewidth]{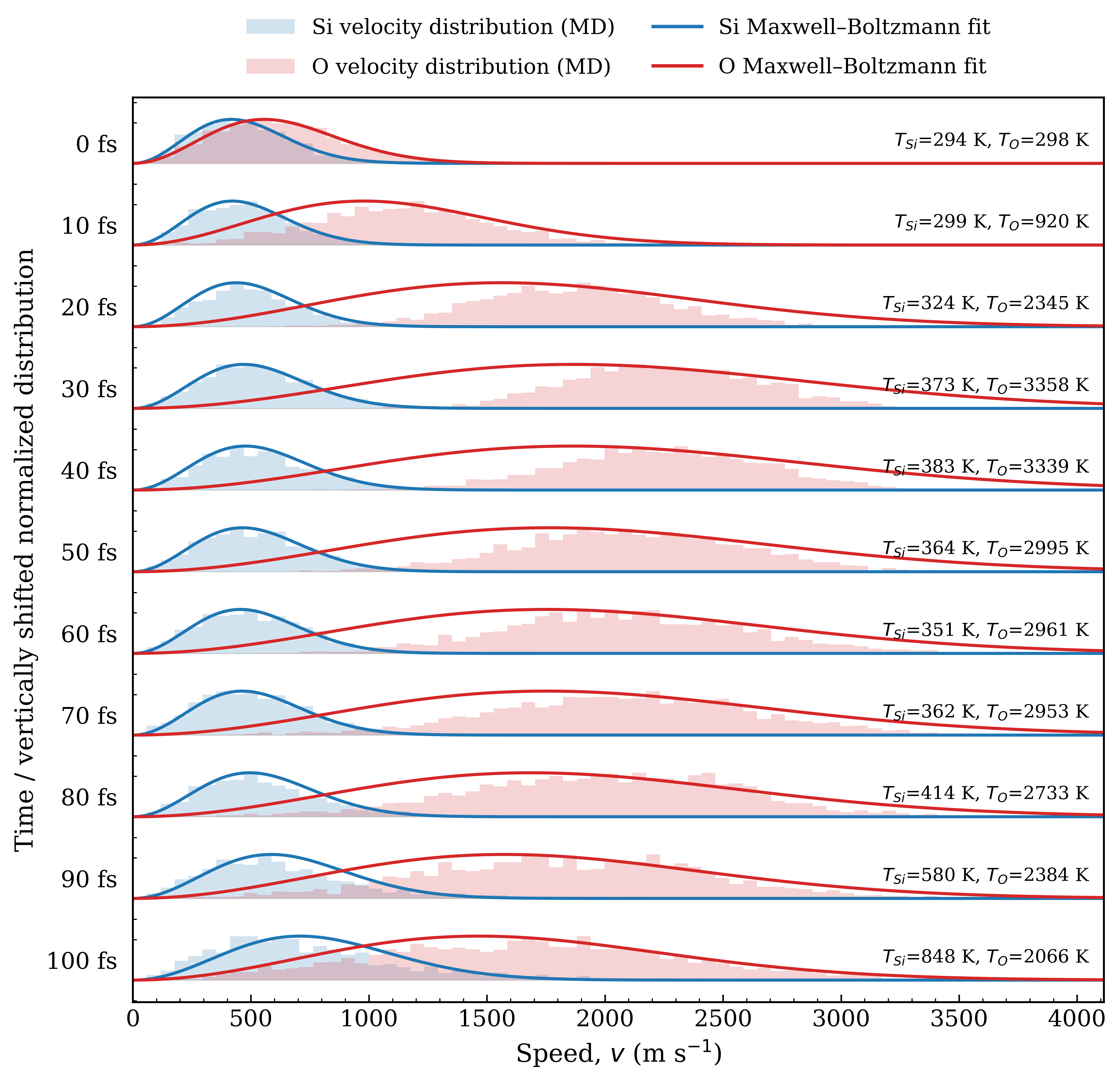}
\caption{$T_e = 2.6$~eV}
\label{fig:S6f}

\vspace{0.2cm}

\caption*{\textbf{Figure S6:} Time-resolved Maxwell-Boltzmann velocity distributions of Si and O atoms in $\alpha$-SiO$_2$ at the ground state and under electronic excitation corresponding to electronic temperatures ranging from $1.0$ to $2.6$~eV. The shaded distributions represent molecular dynamics velocity distributions, while the solid curves denote Maxwell-Boltzmann fits for selected trajectory frames sampled every 10~fs over the initial 100~fs of nonequilibrium lattice dynamics following electronic excitation.}
\end{figure}

\clearpage

\renewcommand{\thefigure}{S\arabic{figure}}